\pdfoutput=1
\documentclass[12pt,a4paper]{article}

\usepackage{ifthen} 
\newboolean{pdflatex}
\setboolean{pdflatex}{true} 

\newboolean{articletitles}
\setboolean{articletitles}{true} 

\newboolean{uprightparticles}
\setboolean{uprightparticles}{false} 

\def\paperauthors{Giacomo Graziani$^1$, Lucio Anderlini$^1$, Saverio Mariani$^{1,2,3}$, Edoardo Franzoso$^{4,5}$, Luciano Libero Pappalardo$^{4,5}$, Pasquale di Nezza$^{6}$} 
\def\papertitle{A Neural-Network-defined Gaussian Mixture Model for particle identification applied to the LHCb fixed-target programme}
\def\paperasciititle{\papertitle}
\def\paperkeywords{{High Energy Physics}, {LHCb}, {Particle identification}, {Machine Learning}, {Neural Network}} 
\def\papercopyright{\the\year\ CERN for the benefit of the LHCb collaboration} 
\def\paperlicence{CC BY 4.0 licence}
\def\paperlicenceurl{https://creativecommons.org/licenses/by/4.0/}

\usepackage[top=1in, bottom=1.25in, left=1in, right=1in]{geometry}

\usepackage{microtype}
\usepackage{lineno}  
\usepackage{xspace} 
\usepackage{caption} 
\usepackage{import}

\usepackage{graphicx}  
\usepackage{color}
\usepackage{colortbl}
\graphicspath{{./figs/}} 

\usepackage{amsmath} 
\usepackage{amssymb}
\usepackage{amsfonts}
\usepackage{upgreek} 

\newcommand*\patchAmsMathEnvironmentForLineno[1]{%
\expandafter\let\csname old#1\expandafter\endcsname\csname #1\endcsname
\expandafter\let\csname oldend#1\expandafter\endcsname\csname
end#1\endcsname
 \renewenvironment{#1}%
   {\linenomath\csname old#1\endcsname}%
   {\csname oldend#1\endcsname\endlinenomath}%
}
\newcommand*\patchBothAmsMathEnvironmentsForLineno[1]{%
  \patchAmsMathEnvironmentForLineno{#1}%
  \patchAmsMathEnvironmentForLineno{#1*}%
}
\AtBeginDocument{%
\patchBothAmsMathEnvironmentsForLineno{equation}%
\patchBothAmsMathEnvironmentsForLineno{align}%
\patchBothAmsMathEnvironmentsForLineno{flalign}%
\patchBothAmsMathEnvironmentsForLineno{alignat}%
\patchBothAmsMathEnvironmentsForLineno{gather}%
\patchBothAmsMathEnvironmentsForLineno{multline}%
\patchBothAmsMathEnvironmentsForLineno{eqnarray}%
}

\usepackage{hyperxmp}

\usepackage[pdftex,
            pdfauthor={\paperauthors},
            pdftitle={\paperasciititle},
            pdfkeywords={\paperkeywords},
            pdfcopyright={Copyright (C) \papercopyright},
            pdflicenseurl={\paperlicenceurl}]{hyperref}

\usepackage[colorinlistoftodos,textsize=scriptsize]{todonotes}

\usepackage[bottom,flushmargin,hang,multiple]{footmisc}

\usepackage[all]{hypcap} 

\usepackage{xspace} 
\usepackage{upgreek}

\def\lhcb   {\mbox{LHCb}\xspace}

\def\lhc    {\mbox{LHC}\xspace}

\def\rich   {RICH\xspace}

\def\spd    {SPD\xspace}

\def\MagUp {\mbox{\em Mag\kern -0.05em Up}\xspace}

\ifthenelse{\boolean{uprightparticles}}%
{

 \def\Ppi         {\ensuremath{\uppi}\xspace}

 \def\Pphi        {\ensuremath{\upphi}\xspace}

 \def\PDelta      {\ensuremath{\Delta}\xspace}                 
 \def\PXi         {\ensuremath{\Xi}\xspace}                 
 \def\PLambda     {\ensuremath{\Lambda}\xspace}                 
 \def\PSigma      {\ensuremath{\Sigma}\xspace}                 
 \def\POmega      {\ensuremath{\Omega}\xspace}                 
 \def\PUpsilon    {\ensuremath{\Upsilon}\xspace}

 \def\PB      {\ensuremath{\mathrm{B}}\xspace}                 
                  
 \def\PD      {\ensuremath{\mathrm{D}}\xspace}

 \def\PK      {\ensuremath{\mathrm{K}}\xspace}

 \def\Pi      {\ensuremath{\mathrm{i}}\xspace}

 \def\Pp      {\ensuremath{\mathrm{p}}\xspace}

 \def\Ps      {\ensuremath{\mathrm{s}}\xspace}

 \def\thebaroffset{0.0em}
}
{

 \def\Ppi         {\ensuremath{\pi}\xspace}

 \def\Pphi        {\ensuremath{\phi}\xspace}

 \mathchardef\PDelta="7101
 \mathchardef\PXi="7104
 \mathchardef\PLambda="7103
 \mathchardef\PSigma="7106
 \mathchardef\POmega="710A
 \mathchardef\PUpsilon="7107
                  
 \def\PB      {\ensuremath{B}\xspace}                 
                  
 \def\PD      {\ensuremath{D}\xspace}

 \def\PK      {\ensuremath{K}\xspace}

 \def\Pi      {\ensuremath{i}\xspace}

 \def\Pp      {\ensuremath{p}\xspace}

 \def\Ps      {\ensuremath{s}\xspace}

 \def\thebaroffset{0.18em}
}
\newcommand{\offsetoverline}[2][\thebaroffset]{\kern #1\overline{\kern -#1 #2}}%

\makeatletter
\ifcase \@ptsize \relax
  \newcommand{\miniscule}{\@setfontsize\miniscule{4}{5}}
\or
  \newcommand{\miniscule}{\@setfontsize\miniscule{5}{6}}
\or
  \newcommand{\miniscule}{\@setfontsize\miniscule{5}{6}}
\fi
\makeatother

\DeclareRobustCommand{\optbar}[1]{\shortstack{{\miniscule (\rule[.5ex]{1.25em}{.18mm})}
  \\ [-.7ex] $#1$}}

\def\squark    {{\ensuremath{\Ps}}\xspace}

\def\pion   {{\ensuremath{\Ppi}}\xspace}

\def\pip    {{\ensuremath{\pion^+}}\xspace}
\def\pim    {{\ensuremath{\pion^-}}\xspace}

\def\kaon    {{\ensuremath{\PK}}\xspace}

\def\KorKbar {\kern \thebaroffset\optbar{\kern -\thebaroffset \PK}{}\xspace}

\def\Kp      {{\ensuremath{\kaon^+}}\xspace}
\def\Km      {{\ensuremath{\kaon^-}}\xspace}

\def\KS      {{\ensuremath{\kaon^0_{\mathrm{S}}}}\xspace}

\newcommand{\phiz}{\ensuremath{\Pphi}\xspace}

\def\D       {{\ensuremath{\PD}}\xspace}

\def\DorDbar {\kern \thebaroffset\optbar{\kern -\thebaroffset \PD}\xspace}
\def\Dz      {{\ensuremath{\D^0}}\xspace}

\def\Dp      {{\ensuremath{\D^+}}\xspace}
\def\Dm      {{\ensuremath{\D^-}}\xspace}

\def\DpDm    {\ensuremath{\Dp {\kern -0.16em \Dm}}\xspace}

\def\Dstarp  {{\ensuremath{\D^{*+}}}\xspace}

\def\B       {{\ensuremath{\PB}}\xspace}

\def\BorBbar {\kern \thebaroffset\optbar{\kern -\thebaroffset \PB}\xspace}

\def\Bd      {{\ensuremath{\B^0}}\xspace}

\def\BdorBdbar {\kern \thebaroffset\optbar{\kern -\thebaroffset \Bd}\xspace}

\def\Bs      {{\ensuremath{\B^0_\squark}}\xspace}

\def\BsorBsbar {\kern \thebaroffset\optbar{\kern -\thebaroffset \Bs}\xspace}

\def\Y#1S{\ensuremath{\PUpsilon{(#1S)}}\xspace}

\def\proton      {{\ensuremath{\Pp}}\xspace}
\def\antiproton  {{\ensuremath{\overline \proton}}\xspace}

\def\Lz          {{\ensuremath{\PLambda}}\xspace}
\def\Lbar        {{\ensuremath{\offsetoverline{\PLambda}}}\xspace}
\def\LorLbar     {\kern \thebaroffset\optbar{\kern -\thebaroffset \PLambda}\xspace}

\newcommand{\decay}[2]{\ensuremath{#1\!\to #2}\xspace} 

\def\to                 {\ensuremath{\rightarrow}\xspace}

\def\AT#1     {\ensuremath{A_{\mathrm{T}}^{#1}}\xspace}           

\def\C#1      {\ensuremath{\mathcal{C}_{#1}}\xspace}                       
\def\Cp#1     {\ensuremath{\mathcal{C}_{#1}^{'}}\xspace}                    
\def\Ceff#1   {\ensuremath{\mathcal{C}_{#1}^{\mathrm{(eff)}}}\xspace}        
\def\Cpeff#1  {\ensuremath{\mathcal{C}_{#1}^{'\mathrm{(eff)}}}\xspace}       
\def\Ope#1    {\ensuremath{\mathcal{O}_{#1}}\xspace}                       
\def\Opep#1   {\ensuremath{\mathcal{O}_{#1}^{'}}\xspace}                    

\newcommand{\aunit}[1]{\ensuremath{\text{\,#1}}}       

\newcommand{\tev}{\aunit{Te\kern -0.1em V}\xspace}
\newcommand{\gev}{\aunit{Ge\kern -0.1em V}\xspace}
\newcommand{\mev}{\aunit{Me\kern -0.1em V}\xspace}
\newcommand{\kev}{\aunit{ke\kern -0.1em V}\xspace}
\newcommand{\ev}{\aunit{e\kern -0.1em V}\xspace}
 
\newcommand{\mevc}{\ensuremath{\aunit{Me\kern -0.1em V\!/}c}\xspace}
\newcommand{\gevc}{\ensuremath{\aunit{Ge\kern -0.1em V\!/}c}\xspace}
\newcommand{\mevcc}{\ensuremath{\aunit{Me\kern -0.1em V\!/}c^2}\xspace}
\newcommand{\gevcc}{\ensuremath{\aunit{Ge\kern -0.1em V\!/}c^2}\xspace}

\def\mm   {\aunit{mm}\xspace}

\newcommand{\chisq}{\ensuremath{\chi^2}\xspace}
\newcommand{\chisqndf}{\ensuremath{\chi^2/\mathrm{ndf}}\xspace}

\def\gsim{{~\raise.15em\hbox{$>$}\kern-.85em
          \lower.35em\hbox{$\sim$}~}\xspace}
\def\lsim{{~\raise.15em\hbox{$<$}\kern-.85em
          \lower.35em\hbox{$\sim$}~}\xspace}

\def\sPlot{\mbox{\em sPlot}\xspace}

\def\pt         {\ensuremath{p_{\mathrm{T}}}\xspace}

\def\ptot       {\ensuremath{p}\xspace}

\def\dllkpi     {\ensuremath{\mathrm{DLL}_{\kaon,\pion}}\xspace}
\def\dllppi     {\ensuremath{\mathrm{DLL}_{\proton,\pion}}\xspace}

\def\geant      {\mbox{\textsc{Geant4}}\xspace}

\def\tell1  {TELL1\xspace}
\def\ukl1   {UKL1\xspace}

\newcommand{\ie}{\mbox{\itshape i.e.}\xspace}

\usepackage{cite} 
\usepackage{mciteplus}

\def\pHe{\ensuremath{\proton\text{He}}\xspace}
\def\pAr{\ensuremath{\proton\text{Ar}}\xspace}
\def\xPID        {{\ensuremath{x}}\xspace} 
\def\xsPID        {{\ensuremath{\underline{x}}}\xspace} 
\def\WV          {{\ensuremath{\underline{\theta}}}\xspace}
\def\pNe{\ensuremath{\proton\text{Ne}}\xspace}
\def\pp{\ensuremath{\proton\proton}\xspace}
\def\pbar{\antiproton}
\def\dllpk {\ensuremath{\mathrm{DLL}_{\proton,\kaon}}\xspace}
\def\pdf{\textit{pdf}\xspace}

\usepackage{longtable} 
\usepackage{comment}
\usepackage{multirow}

\begin{document}

\renewcommand{\thefootnote}{\fnsymbol{footnote}}
\setcounter{footnote}{1}


\begin{titlepage}

\vspace*{-1.5cm}

\noindent
\begin{tabular*}{\linewidth}{lc@{\extracolsep{\fill}}r@{\extracolsep{0pt}}}
\ifthenelse{\boolean{pdflatex}}
{\vspace*{-1.2cm}\mbox{\!\!\!\includegraphics[width=.14\textwidth]{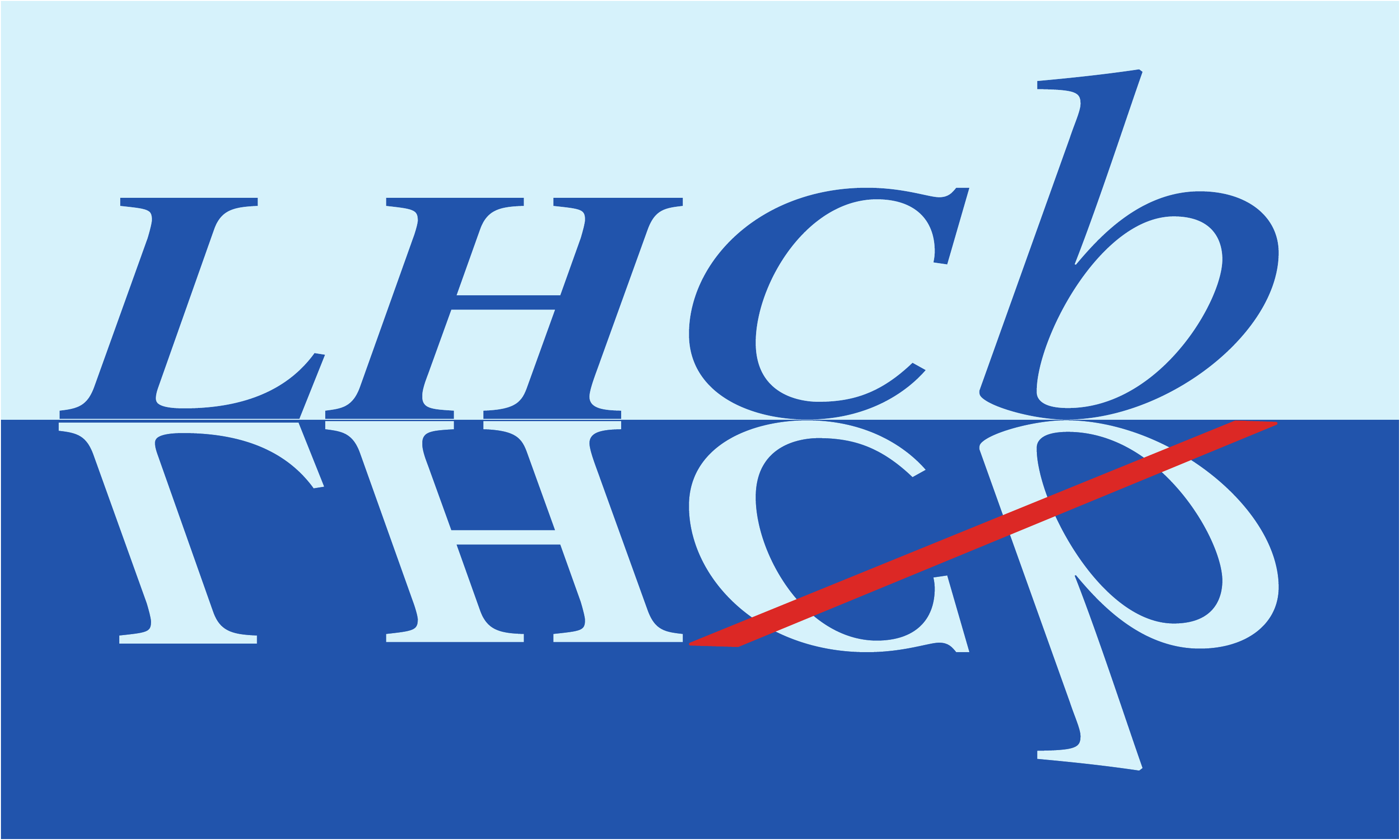}} & &}%
{\vspace*{-1.2cm}\mbox{\!\!\!\includegraphics[width=.12\textwidth]{figs/lhcb-logo.eps}} & &}
 \\
 & & LHCb-DP-2021-007\\  
 & & \today \\ 
 & & \\
\hline
\end{tabular*}


\vspace*{1.0cm}

{\normalfont\bfseries\boldmath\huge
\begin{center}
  \papertitle
\end{center}
}

\vspace*{1.2cm}

\begin{center}
Giacomo Graziani$^1$, Lucio Anderlini$^1$, Saverio Mariani $^{1,2,3}$, \\
Edoardo Franzoso$^{4,5}$, Luciano Libero Pappalardo$^{4,5}$,\\
Pasquale di Nezza$^{6}$
\bigskip\\
{\normalfont\itshape\footnotesize
$ ^1$INFN Sezione di Firenze, Florence, Italy\\
$ ^2$Universit\`a degli studi di Firenze, Florence, Italy\\
$ ^3$European Organization for Nuclear Research (CERN), Geneva, Switzerland\\
$ ^4$INFN Sezione di Ferrara, Ferrara, Italy\\
$ ^5$Universit\`a degli studi di Ferrara, Ferrara, Italy\\
$ ^6$INFN Laboratori Nazionali di Frascati, Frascati, Italy
}
\end{center}

\vspace{\fill}

\begin{abstract}
\noindent
Particle identification in large high-energy physics experiments typically relies on
classifiers obtained by combining many experimental observables. Predicting the probability density function (\pdf) of such classifiers in the multivariate space covering the relevant experimental features is usually challenging. 
The detailed simulation of the detector response from first principles
cannot provide the reliability needed for the most precise physics measurements.
Data-driven modelling is usually preferred, though sometimes limited by the available data size and different coverage of the feature space by the control channels. 
In this paper, we discuss a novel approach to the modelling of particle
identification classifiers using machine-learning techniques. The marginal \pdf of the classifiers is described with a Gaussian Mixture Model, whose parameters are predicted by Multi Layer Perceptrons trained on calibration data. As a proof of principle, the method is applied to the data acquired by the LHCb experiment in its fixed-target configuration. The model is trained on a data sample of proton-neon collisions and applied to smaller data samples of proton-helium and proton-argon collisions collected at different centre-of-mass energies. The method is shown to perform better than a detailed simulation-based approach, to be fast and suitable to be applied to a large variety of use cases.
\end{abstract}

\vspace*{0.6cm}

\begin{center}
  Published in JINST 17 (2022) P02018
\end{center}

\vspace*{0.6cm}
\vspace{\fill}
{\footnotesize
\centerline{\copyright~\papercopyright. \href{\paperlicenceurl}{\paperlicence}.}}
\vspace*{2mm}

\end{titlepage}

\pagestyle{empty}  


\newpage
\setcounter{page}{2}
\mbox{~}


\renewcommand{\thefootnote}{\arabic{footnote}}
\setcounter{footnote}{0}

\cleardoublepage


\pagestyle{plain} 
\setcounter{page}{1}
\pagenumbering{arabic}


\section{Introduction}
\label{sec:Introduction}
In large high-energy physics experiments, particles deposit energy in several detectors, some of which are specialized in charged or neutral particle identification (PID).
Usually a combination of techniques among Cherenkov and transition radiations, ionization
loss, time-of-flight measurements and calorimetry are simultaneously employed to
guarantee redundancy and a wide kinematic coverage.
To exploit all the available information, multivariate classification techniques are typically used, combining the response of all involved detectors into variables,
called PID classifiers in the following, optimizing the separation among different particle species.
Once a PID classifier has been built, predicting its performance is crucial to physics analysis. In general, this implies the knowledge of its probability density function (\pdf) for each particle species in the multivariate space spanning all the features that can affect the detector response. These include the kinematic
properties of the particle under scrutiny, but also the underlying event, \ie the signals due to other particles in the same event. As the number of these features increases, the problem quickly becomes intractable with simple methods. The detailed simulation of the detector response for the physical process of interest provides a way to predict the \pdf from first principles. However, the amount of simulated data needed to accurately cover the multivariate feature space is often prohibitive and the imperfection of the simulation needs to be determined through a comparison to real events. Data-driven modelling is thus preferred: calibration channels where the particle species is determined by their kinematics, independently on the response of the detectors used for PID, are used.
An example is the \decay{\Lz}{\proton\pim} decay, where a high-purity sample of events can be selected thanks to the isolated decay vertex and the narrow mass peak, by which the proton and the pion can be distinguished by their kinematic properties.
For a given decay mode, the distribution of the selected particles in the feature space will differ from that of the calibration channels. Ideally one would like to determine from the control sample, for each particle species, the marginal \pdf of the PID classifier(s) as a function of the relevant experimental features, in order to predict it for the physics channel under study.
In this paper, we propose a method to approach this problem based on machine-learning techniques. Control channels in data are used to identify  the experimental features affecting the PID classifier and to derive a statistical model of its \pdf.
The method is conceived to be generic, not relying on a specific
set of experimental feature variables, to be trainable in a relatively short time, allowing the analyst to easily modify the feature space and tune the hyperparameters for best performance, and to use a state-of-the-art neural network (NN) architecture.
After discussing the general idea of this approach, a concrete implementation is discussed
as a benchmark case, namely the modelling of charged particle identification in the
fixed-target data collected by the \lhcb experiment.
We conclude by summarizing the performance and general applicability of the proposed method.

\section{The method}
\label{sec:method}
The basic  idea of the  proposed approach is to empirically describe the marginal \pdf of a PID classifier \xPID, depending on a set of features \WV, through a Gaussian Mixture Model (GMM) as
\begin{equation}
 \xPID_p \sim \sum_{j=1}^{N_{g,p}} \alpha_{j,p}(\WV)  \mathcal{G} ( \xPID , \mu_{j,p}(\WV), \sigma_{j,p}(\WV))
\label{eq:model}
\end{equation}
where the index $p$ refers to the particle species, $\mathcal{G}$ is a Gaussian distribution
with mean $\mu$ and standard deviation $\sigma$, and $N_g$ is the number of Gaussian functions in the model.
The parameters $\alpha, \mu, \sigma$, whose values can have a non-linear dependence on \WV,
are estimated from a maximum likelihood fit to a set of $n_p$ training events for each particle species. The initialization is performed using a set of Multi Layer Perceptron (MLP) neural networks, where the loss function is the negative logarithm of the likelihood function. This definition is part of the novelty of the presented approach and, in contrast to other generative models used in similar applications such as GANs and VAEs, aims to explicitly model a probability density function. Training events from the calibration channels can be contaminated by unwanted background, which can be subtracted using the \sPlot method~\cite{Pivk:2004ty}, evaluating a weight $w$ for each event to quantify its probability to belong to the considered signal. Weighting the training events can also be useful to modify the distribution in feature space of the training sample, when there are large differences with respect to the target physics sample. In this way the model will be more accurate in the feature space regions where it is more useful to the application. When weights are used, the loss function becomes~\cite{MLFit_err, sFit, ML_sPlot}
\begin{equation}
 \mathcal{L} = - \sum_{i=1}^{n_p}  w_i \log\left[ \sum_{j=1}^{N_{g,p}} \alpha_{j,p}(\WV_i)  \mathcal{G} ( \xPID_i , \mu_{j,p}(\WV_i), \sigma_{j,p}(\WV_i))
\right].
\end{equation}
The model can be readily extended, as in the case of our benchmark example, to a set \xsPID of (approximately) linearly correlated target variables, by replacing the one-dimensional Gaussian functions with multivariate normal distributions. With $N_g$ large enough, the model can reproduce any smooth function with exponential-like tails, which is the typical behaviour of PID classifiers. The main advantage of the chosen functional form is the speed of normalization when performing the fit. This offers the possibility to use a simple NN design with relatively short training time and to tune the hyperparameters of the model in a reasonable time. We make use of state-of-the-art open-source machine-learning libraries to implement the model, namely scikit-learn~\cite{scikit-learn} and Keras~\cite{Keras} with TensorFlow~\cite{TF} backend. These offer support for the NN backward error propagation and auto-differentiation and can exploit GPU acceleration. Input variables are preprocessed to ease the training phase. A linear transformation is applied to the target variables \xsPID, to map the range of their values to the interval $[0,1)$, using the MinMaxScaler algorithm in scikit-learn. For the other features, whose \pdf is not aimed to be reproduced by the model, a transformation is applied converting their \pdf to a Gaussian distribution with $\mu=0$ and $\sigma=1$ implemented by the QuantileTransformer algorithm in scikit-learn. Equalizing the range and, for the features, the functional form of the \pdf, allows the NN layer for a standardized initialization and speeds up the numerical estimation of the derivatives in the loss function minimization procedure.
The chosen NN architecture, implemented using Keras/TensorFlow, consists in three tanh-activated layers with a constant number of nodes and an output layer matched to the shape of the predicted parameters. Each training requires a sensible choice for the following hyperparameters:
\begin{itemize}
	\item[$\bullet$] the number of Gaussian distributions $N_g$;
	\item[$\bullet$] the number of nodes in the NN hidden layers;
	\item[$\bullet$] the number of epochs, namely the iterations in the training process;
	\item[$\bullet$] the batch size, namely the number of training events to be considered for each iteration of the minimization procedure;
	\item[$\bullet$] the learning rate, namely the initial size of the step on the parameter values to be applied on each epoch (which is then linearly decreased with the epoch to ease the convergence).
\end{itemize}
To speed up the training, it is performed in two steps and
a crude estimation of the parameters is firstly obtained by
considering them independent on the features \WV. This first
fit is performed with a reduced number of epochs, and its result is used to initialize the second one. During this second step, the parameters are monitored along the iterations to check that their values converge smoothly. This is indeed a first cross-check against a possible inappropriate value of the learning rate, leading to an oscillating behaviour, or to effects of overtraining, leading to abrupt variations of the parameters.
The training dataset is then split in ranges of the features with similar population and the corresponding \xsPID histograms are compared to the fitted \pdf, as obtained separately for each feature interval. A good agreement between the data and predicted distributions for all the feature values validates the trained model. A concrete implementation of the proposed approach on data collected by the \lhcb experiment
is presented in the following Section and available on GitLab~\cite{pid4smog_module}.

\section{Charged PID in the LHCb fixed-target data}
\label{sec:LhcbIntro}
\begin{figure}
\centering
	\includegraphics[width=0.9\textwidth]{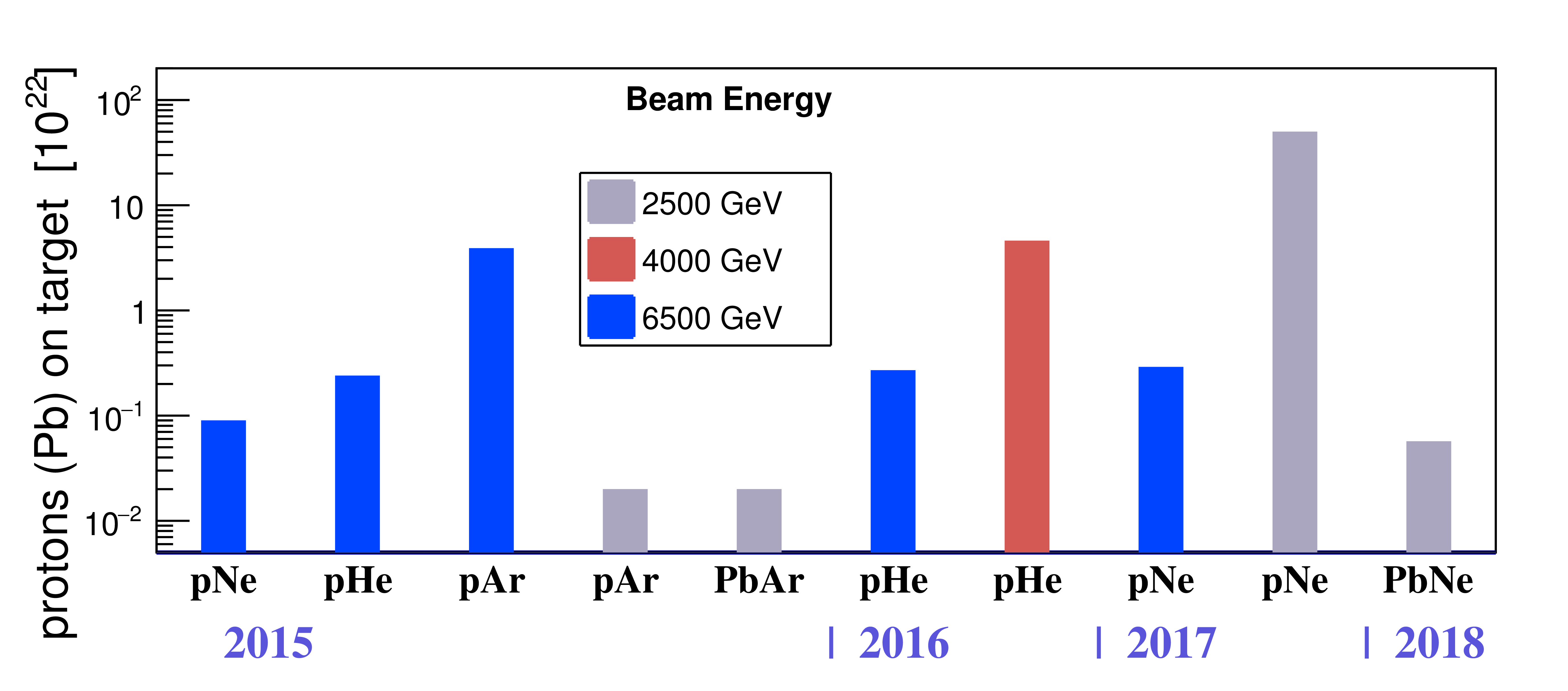}
	\caption{List of samples collected by the \lhcb experiment in fixed-target mode. The data size is expressed by the number of beam particles (protons or lead nuclei) traversing the gas volume around the interaction point. Figure from \cite{LHCb-TDR-020}.}
	\label{fig:SMOG_samples}
\end{figure}

\begin{figure}[bt]
  \centering
  \includegraphics[width=0.49\linewidth]{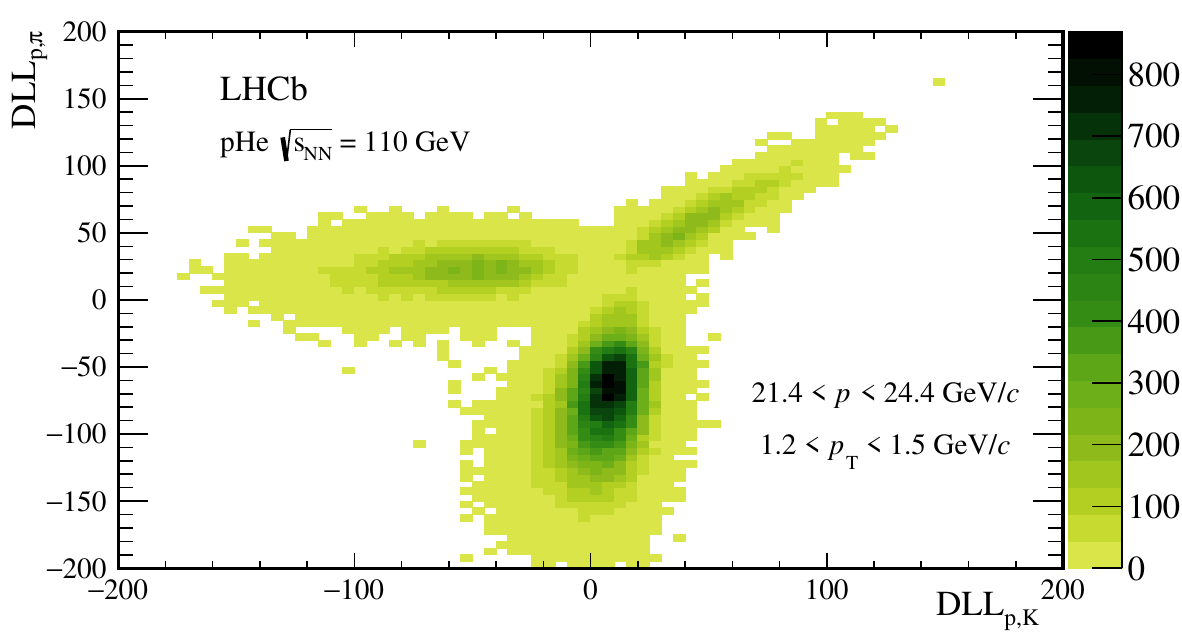}
  \includegraphics[width=0.49\linewidth]{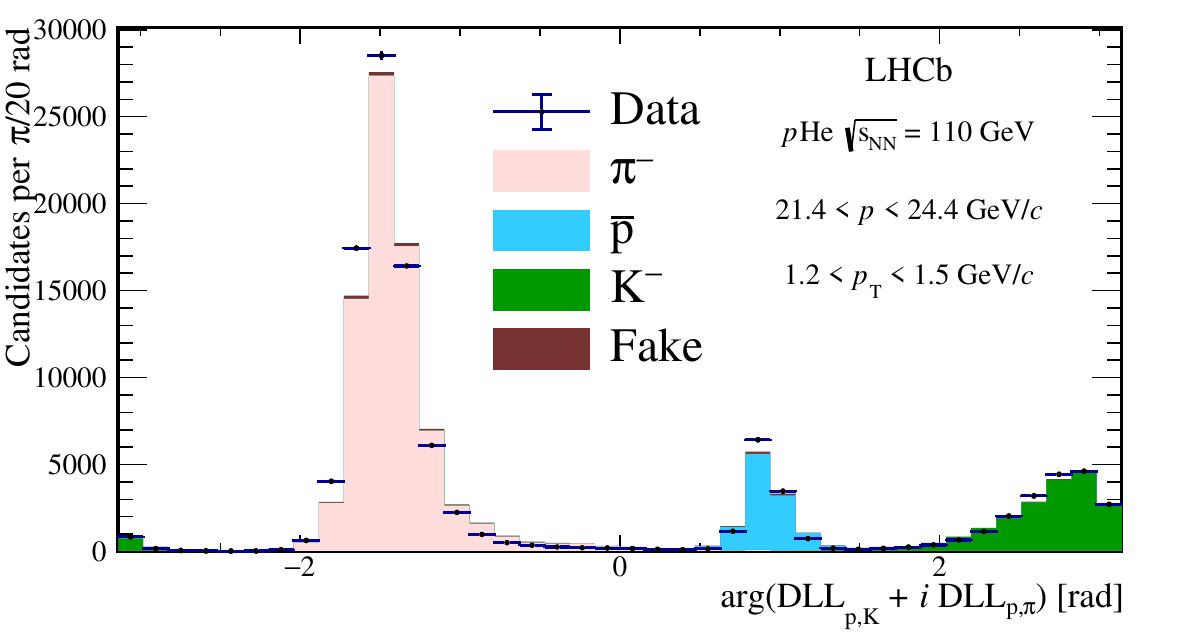}
  \caption{
    Two-dimensional template fit to the PID distribution of negatively charged tracks for a particular bin (\mbox{$ 21.4 < \ptot < 24.4\gevc$}, \mbox{$1.2 < \pt < 1.5\gevc$}).
    The (\dllpk, \dllppi) distribution, shown in the left plot, is fitted to determine the relative
    contribution of \pim, \Km and \pbar particles, using simulation to determine the template distributions and
    the fraction of fake tracks (which are barely visible).
    In the right plot, the result of the fit is projected into the variable \mbox{$\arg{(\dllpk + i\, \dllppi)}$}~\cite{LHCb-PAPER-2018-031}.}
   \label{fig:pHePID}
\end{figure}

The \lhcb detector, described in
detail in Refs.~\cite{LHCb-DP-2008-001,LHCb-DP-2014-002}, is a single-arm forward spectrometer conceived for heavy-flavour
physics in \pp collisions at the CERN LHC.
Charged particles reconstructed by the spectrometer are dominated by three hadron species: pions, kaons and protons\footnote{Charge conjugation is implied in all the paper,
thus particle species include both electric charge signs.}.
Charged particle identification is provided by
two Ring Imaging Cherenkov (\rich) detectors~\cite{LHCb-DP-2012-003} for momenta ranging from 2 to 100~\gevc. The RICH1 detector is optimized for the lower momentum region, covering the angular acceptance of [25,300] mrad, while the RICH2 was designed for the forward particles ([15,120] mrad) with a higher momentum. The momentum thresholds for the Cherenkov light emission are 2.5, 9.3 and 17.7 \gevc in RICH1 and 4.4, 15.6 and 29.6 \gevc in RICH2 for pions, kaons and protons, respectively. For each track, the yields and positions of the recorded  photons, whose emission angle depends on the incident particle velocity and thus on the particle mass for a given momentum, are checked against the three particle species hypotheses. Two PID classifiers are built as logarithms of the likelihood ratio between the proton and pion (\dllppi), the proton and kaon (\dllpk) or the kaon and pion (\dllkpi) hypotheses. A finely segmented scintillating detector (SPD) is positioned downstream of the RICH2 to provide a measurement of the detector occupancy already at the earlier stages of
the reconstruction procedure.
Since 2015, the \lhcb experiment also operates in fixed-target mode, recording collisions
between the \lhc beams and gas targets injected into the \lhc beam pipe~\cite{FerroLuzzi:2005em}. Fixed-target runs of limited duration with different beam energy and target gas species have been collected,
as summarized in Fig.~\ref{fig:SMOG_samples}.
The first physics result obtained in fixed-target configuration was the measurement of antiproton production from a proton-helium (\pHe) sample acquired in 2016~\cite{LHCb-PAPER-2018-031}. The antiprotons were distinguished
from the negatively charged pions and kaons through a template fit to the two-dimensional
space (\dllppi, \dllpk), as illustrated in Fig.~\ref{fig:pHePID}. The templates were obtained from a detailed simulation of the detector and from the \decay{\Lbar}{\pbar\pip}, \decay{\KS}{\pim\pip} and \decay{\phiz}{\Km\Kp} calibration channels in data for antiprotons, pions and kaons, respectively.
The data available statistics was limited, while the abundant samples recorded in \pp collisions, which are routinely used to determine the PID performance in \lhcb results~\cite{LHCb-DP-2018-001}, only presented a small overlap in feature space with \pHe data.
Indeed, the RICH detectors response is strongly affected by the detector occupancy, which is much larger on average in \pp than \pHe collisions and by the particle kinematics, which also differs between the two samples. Moreover, while the origin of \pp collision is defined within a few centimetres, the recorded beam-gas collisions spread over about one metre along the beam line.
The limited accuracy of the PID response modelling turned out to be one of
the dominant uncertainties in the antiproton production measurement~\cite{LHCb-PAPER-2018-031}.\\
A larger sample of proton-neon (\pNe) collisions with a nucleon-nucleon centre-of-mass energy of $\sqrt{s_{NN}} = 68 \gev$ was recorded in 2017, providing a source
of calibration events acquired in conditions more similar to the other fixed-target samples.
We apply here the method proposed in Section~\ref{sec:method} to model the PID response
using the \pNe dataset and to predict the classifier \pdf for track selections performed on smaller \pHe and \pAr data samples, taking into account the different feature distributions due to the different particle selection criteria and experimental conditions such as the gas target and beam energy. The underlying hypothesis is that the response of the RICH detectors is stable across the different data takings.

\subsection{Calibration channels}
\begin{figure}[bt]
  \centering
	\includegraphics[width = 0.48\textwidth]{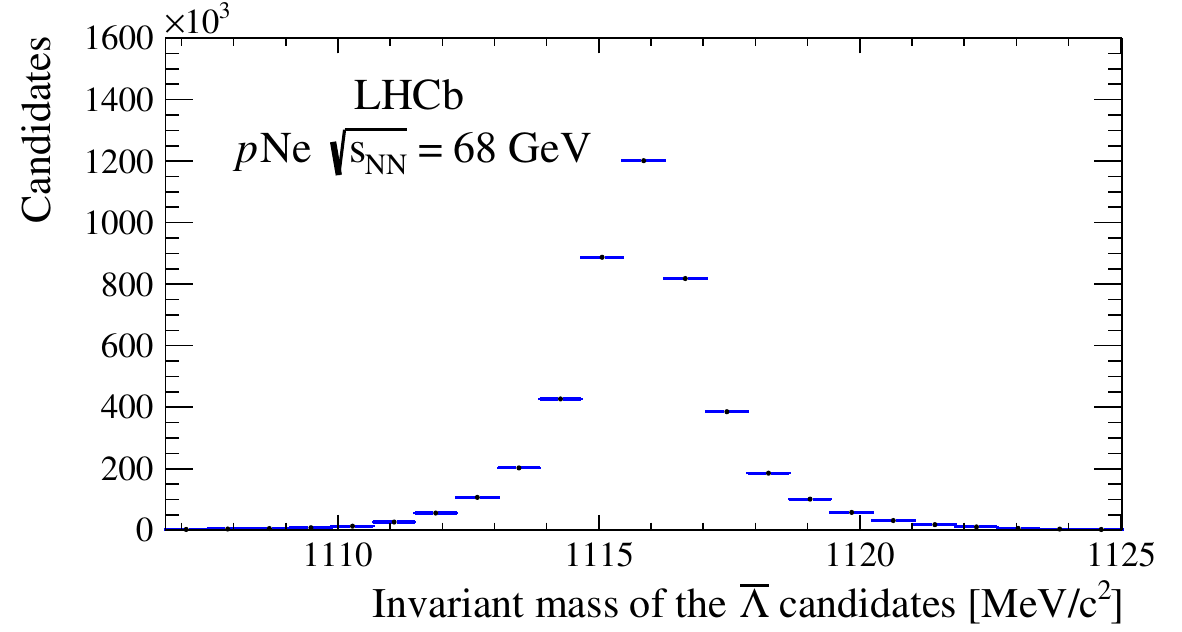}
	\includegraphics[width = 0.48\textwidth]{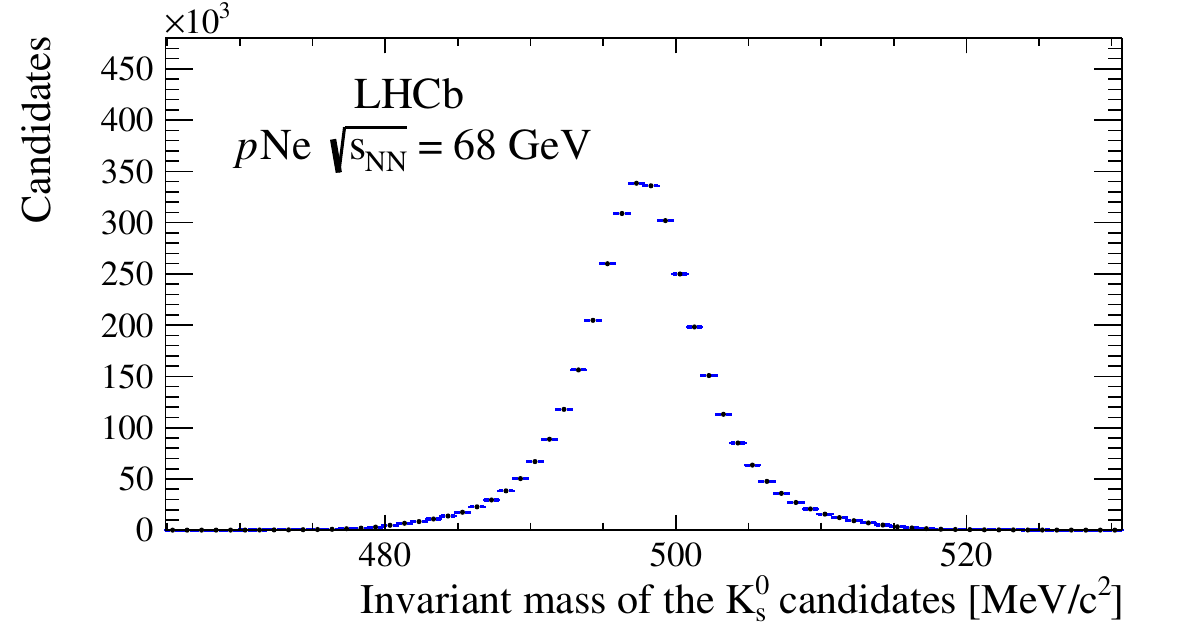}
	\includegraphics[width = 0.48\textwidth]{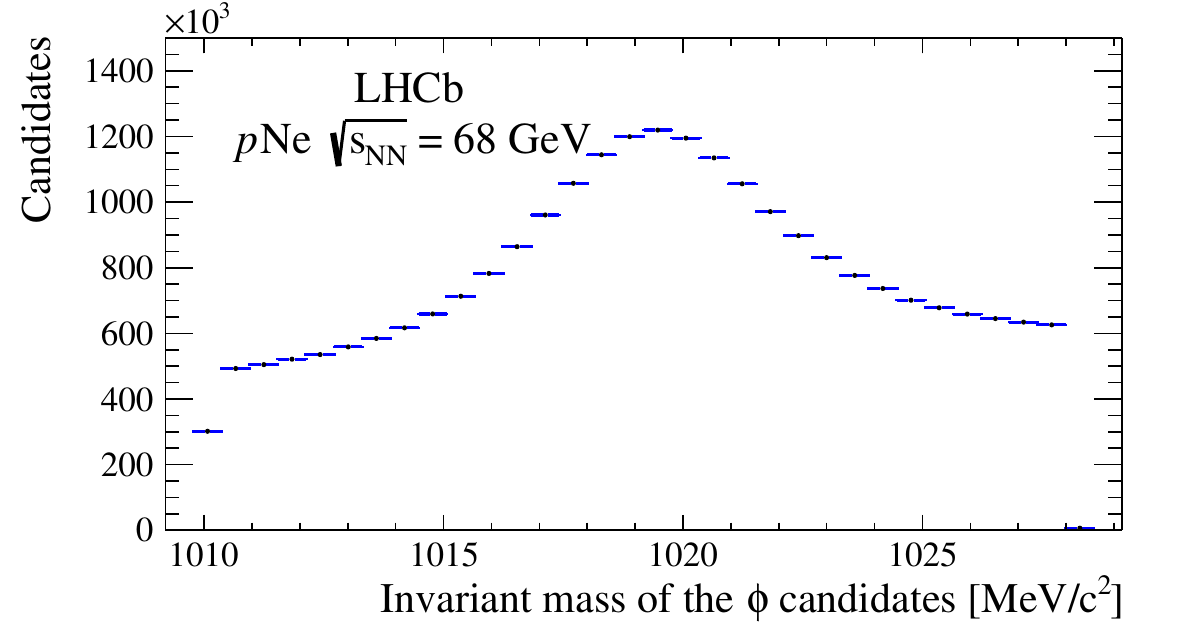}
  \caption{Invariant mass distributions for the $\decay{\Lbar}{\antiproton \pip}$ (top left), $\decay{\KS}{\pim\pip}$ (top right) and $\decay{\phiz}{\Km\Kp}$ (bottom) calibration channels.}
   \label{fig:calmasses}
\end{figure}

\begin{figure}
\centering
\includegraphics[width=0.48\textwidth]{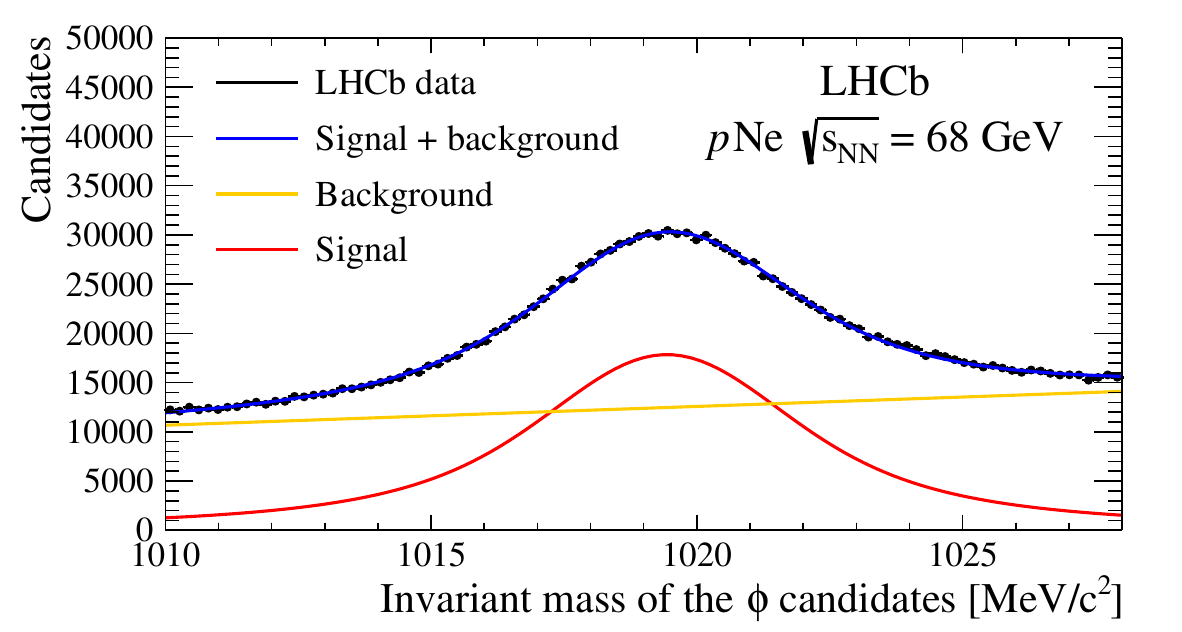}
\includegraphics[width=0.48\textwidth]{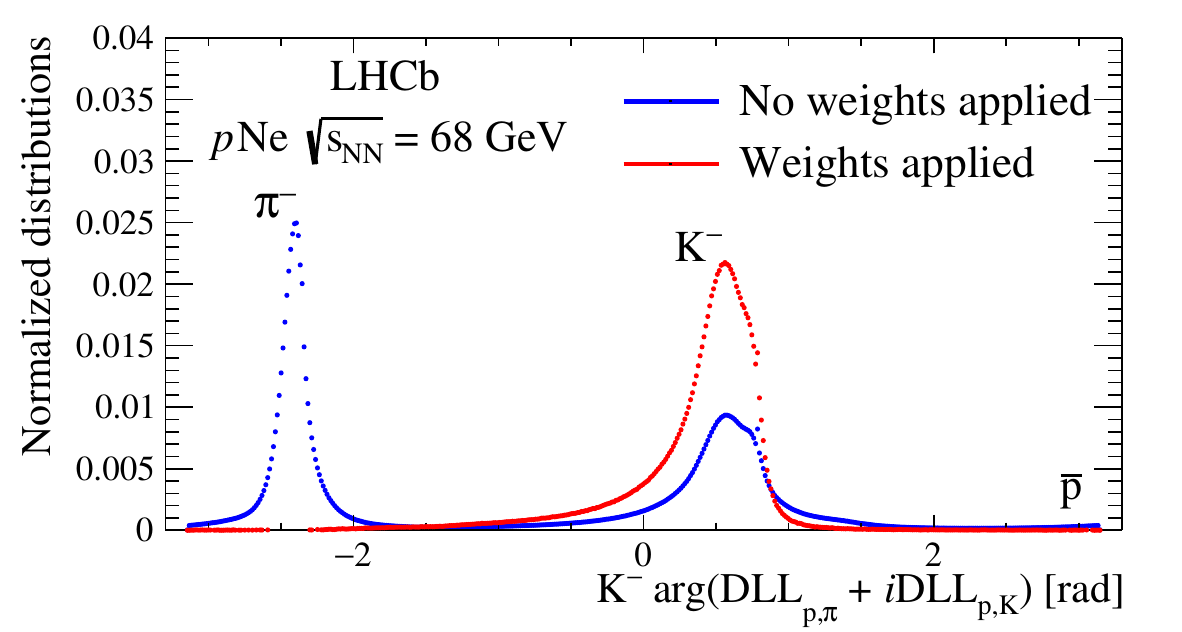}
\caption{Application of the \sPlot technique to the $\decay{\phiz}{\Km \Kp}$ decay. The left plot shows the fit of the \phiz candidates mass distribution; the right one the validation of the evaluated weights by comparing a combination of the PID variables where the three particle species can be easily distinguished before (in blue) and after (in red) the weights application.}
\label{fig_calib:phi_sPlot}
\end{figure}

Three decays of light hadrons abundantly produced in the recorded collisions are used to model the PID response to particles of known species:  \decay{\Lbar}{\antiproton \pip} for antiprotons, \decay{\KS}{\pim\pip} for pions, \decay{\phiz}{\Km\Kp} for kaons.
Antiprotons in \Lbar decays are identified solely from kinematics, as they always have higher  momentum than the pions. In all cases, the control decays are distinguished from the background using a set of requirements on the quality and the kinematics of the reconstructed tracks and requiring that the reconstructed invariant mass of the two-track combination is compatible with the decaying particle mass. The association of the final-state particle with signals in the \rich subdetectors is also imposed. For \decay{\phiz}{\Km\Kp} decays, where background contamination is more abundant, one of the two candidate final-state particles is required to be positively identified as a kaon, and only the other one is used in the control sample.\\
The invariant mass distributions for the three channels after selection are shown in Fig.~\ref{fig:calmasses}.
The purity for the \Lbar and \KS decay samples is measured to be larger than 99\%, while a significant residual background is present in the \phiz sample. The \sPlot method is used to compute a weight for each event representing the probability it belongs to the signal category based on a fit describing the invariant mass distribution as the sum of a signal and a background component, as  illustrated on the left plot of Fig.~\ref{fig_calib:phi_sPlot}. The signal component is parametrized as  a convolution of a Breit-Wigner and a Gaussian distribution, while a first-order polynomial function is used to model the background.
The weights obtained with the \sPlot technique can be then applied to all variables uncorrelated with the invariant mass, as is the case, to an excellent approximation, of the PID classifiers, in order to remove the background. To verify the procedure, the variable  $\arg{(\dllppi + i\, \dllpk)}$, which, as shown in the right plot of Fig.~\ref{fig_calib:phi_sPlot}, exhibits three distinct peaks corresponding to the particle species, is plotted before and after the \sPlot weighting. The pion contamination from background events is clearly suppressed.

\subsection{Feature selection}
\begin{figure}
\centering
\includegraphics[width = 0.48\textwidth]{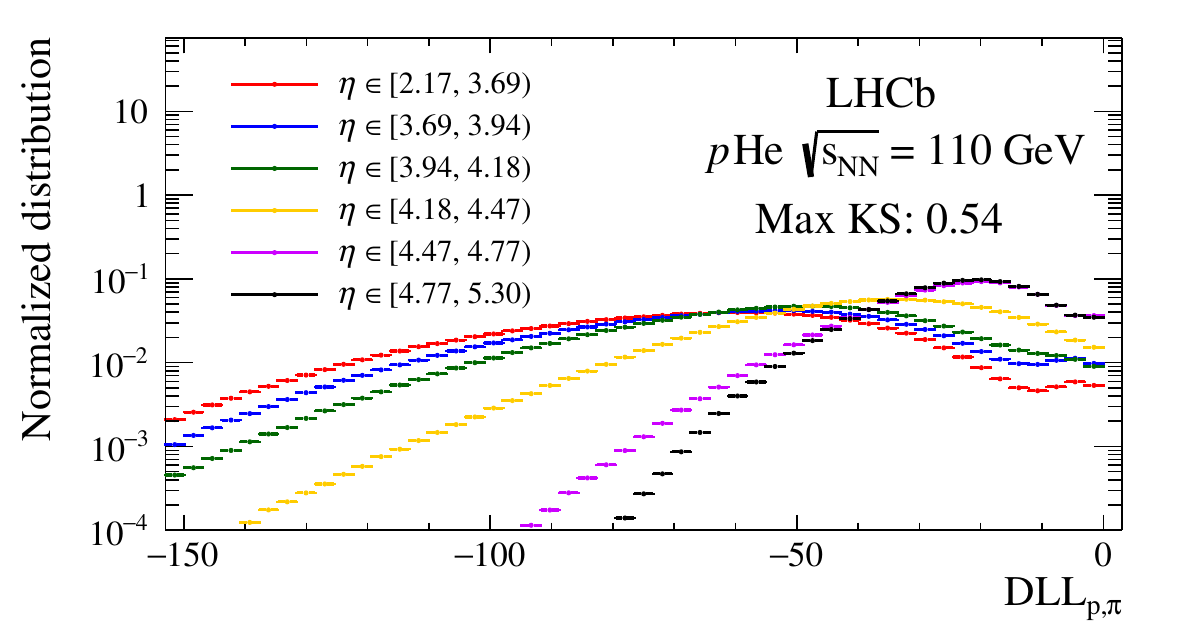}
\includegraphics[width = 0.48\textwidth]{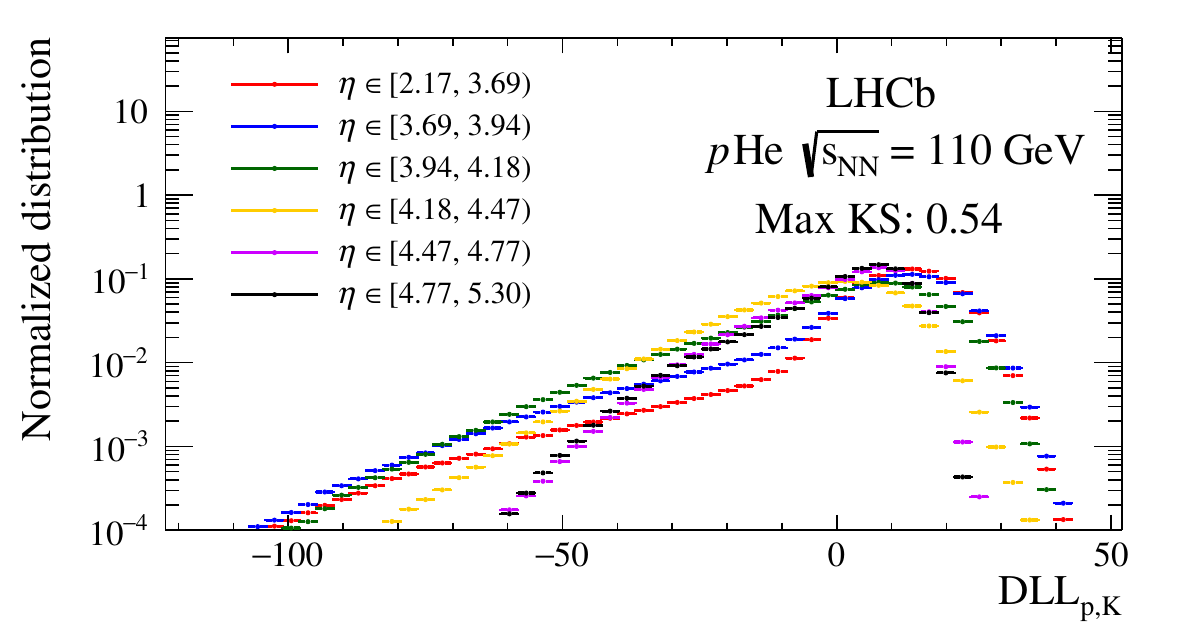}
\includegraphics[width = 0.48\textwidth]{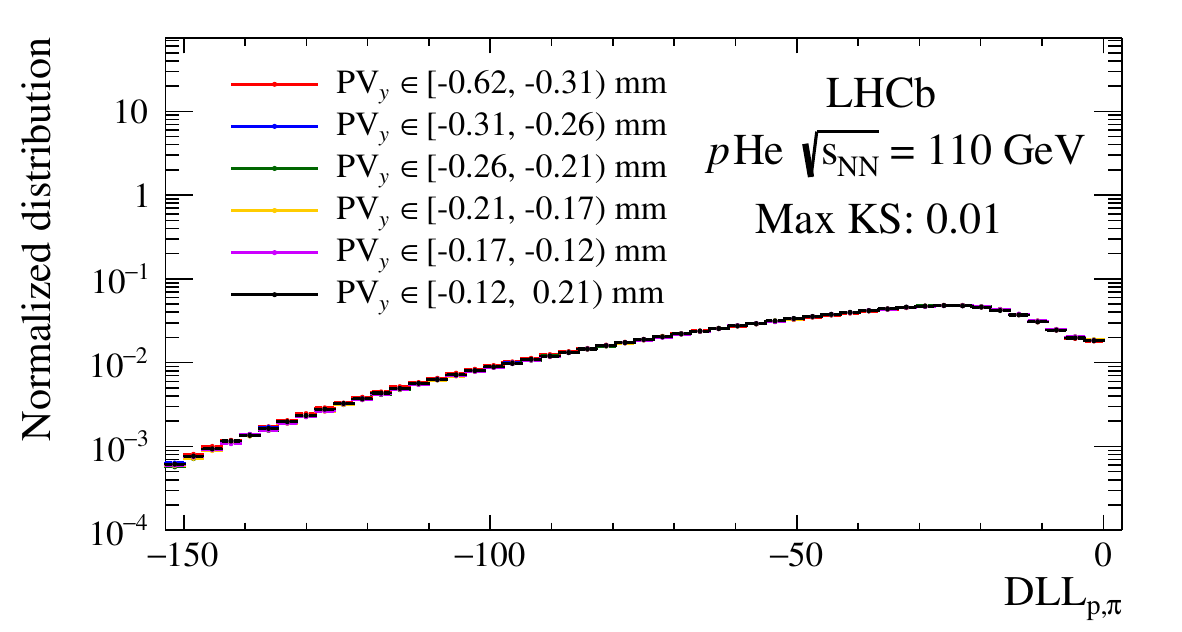}
\includegraphics[width = 0.48\textwidth]{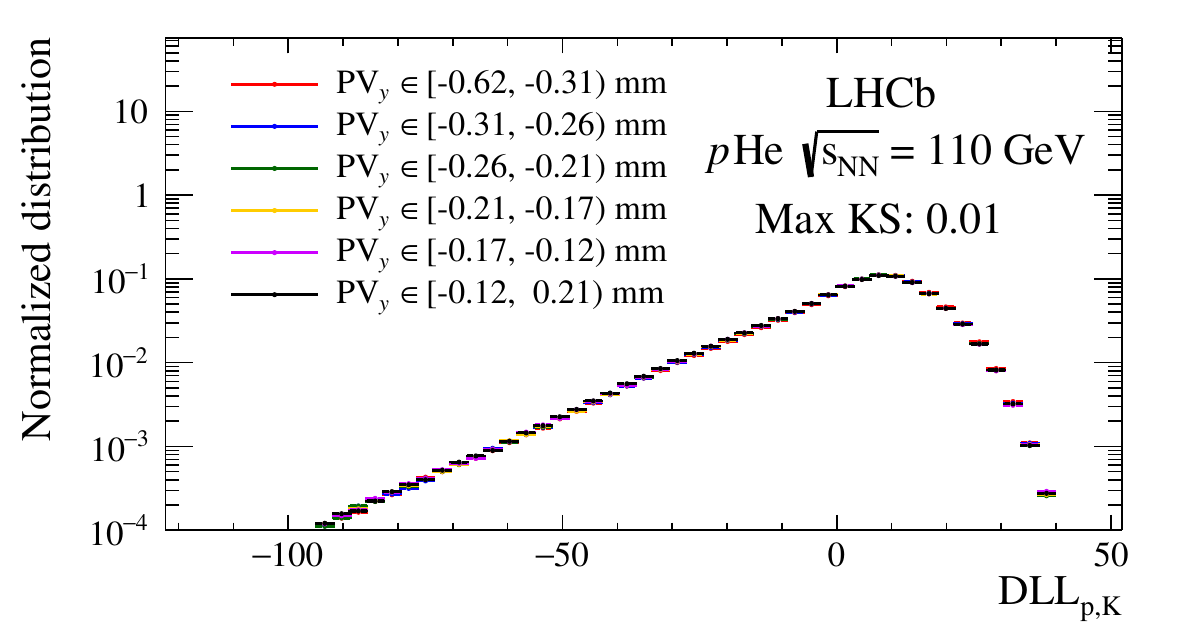}
\caption{Pion \dllppi (left) and \dllpk (right) distributions in the \pHe sample in bins of its pseudorapidity (top) or \textit{y} coordinate of the primary vertex (bottom). The maximum Kolmogorov-Smirnov distance values are given to illustrate the procedure explained in the text.}
\label{fig:feature_imp}
\end{figure}

The RICH performance strongly depends on the particle kinematics, with fast variations when crossing the Cherenkov threshold values or the angular acceptance boundaries of the two detectors. The number and topology of the other tracks radiating in the detectors is also critical to the PID performance. The track reconstruction quality is also relevant, as it affects the determination of the centre of its Cherenkov ring. The choice of the features to be considered is a key point in the application of the method: enough features should be used to describe all the variance in the detector response, while their numbers should be limited to achieve a reasonable training time.
To the purpose of this study, we consider those describing:
\begin{itemize}
\item the particle momentum \ptot,  its longitudinal ($p_z$) and transverse (\pt) components and the pseudorapidity $\eta$;
\item the track position inside the PID detectors: the three coordinates of the track position closest to the beam and the two slopes with respect to the \textit{z} axis;
\item the occupancy in the detector: the number of all reconstructed tracks through the spectrometer ($nTracks$) and the number of energy deposits (hits) in the two RICH ($nRICH1Hits$ and $nRICH2Hits$) and in the \spd detectors ($nSPDHits$);
\item the quality achieved in the reconstruction of the track: the fit \chisq per degree of freedom (\emph{track \chisqndf}) or the number of hits in the tracking detectors consistent with the track (\emph{track ndf}).
\end{itemize}
As a first assessment of  the relevance of these features to the description of the PID response, the distribution of the classifier is studied for pions of the \pHe dataset
for different ranges of each feature. After applying the requirement $\arg(\dllkpi + i\dllppi) < -1$ to reject kaons and protons, for each considered feature the dataset is split in subsets of similar population. The Kolmogorov-Smirnov distance (KS) between the \xsPID distributions is computed for each pair of subsets, taking the maximal distance as a measure of the
dependence of the \xsPID variables on the considered feature.
An example is shown in Fig.~\ref{fig:feature_imp}, where the dependence on the pion pseudorapidity (KS = 0.54) is found to be clearly more important than that on the $y$ position of the \pHe collision (KS = 0.01). The most relevant features, with KS larger than 0.20, are listed in Tab.~\ref{tab_feat:Max_KS}. Some of these features  are known to be redundant or highly correlated with each other. The longitudinal and transverse momentum $p_z$ and $\pt$ will not be considered, as these features are determined from  $\eta$ and \ptot, which exhibit a larger KS. The SPD occupancy is also dropped, as it is highly correlated with the occupancy in the RICH detectors, which are causally connected with the RICH response. On the other hand, variables with low KS value can still be important for explaining differences between the training and validation samples. As an example, the distribution of the longitudinal position of the tracks origin in the training sample, originating from decays, strongly differs from tracks promptly produced at the collision vertex. For this reason, all coordinates of the track position of closest approach to the beam (denoted with ``poca" in the following text and figures) are included, despite a KS value of 0.06, 0.01 and 0.10 for the \textit{x}, \textit{y} and \textit{z} coordinates, respectively.

\begin{table}
\centering
\caption{Values of the maximum Kolmogorov-Smirnov distance for the comparison of the bidimensional PID distributions in bins of the possible features in decreasing order. The higher the indicator, the more relevant is the feature to the PID.}
\label{tab_feat:Max_KS}
\begin{tabular}{cc|cc|cc}
Variable  & Max KS    &   Variable  & Max KS  &   Variable  & Max KS\\
\hline
$\ptot$   & 0.64       &   $p_z$       & 0.64        &   $\eta$    & 0.54   \\
\hline
$p_T$     & 0.51       &   $yz \, slope$     & 0.38     &   $track \, ndf$  & 0.34   \\
\hline
$xz \, slope$   & 0.34    &   $nTracks$     & 0.34      &   $nRICH2Hits$    & 0.33   \\
\hline
$nSPDHits$   & 0.32    &   $nRICH1Hits$    & 0.28    &   $track \, \chisqndf$ & 0.26\\
\hline

\end{tabular}
\end{table}

\subsection{Model training}
\begin{figure} 
\centering 
\includegraphics[width = 0.480000\textwidth]{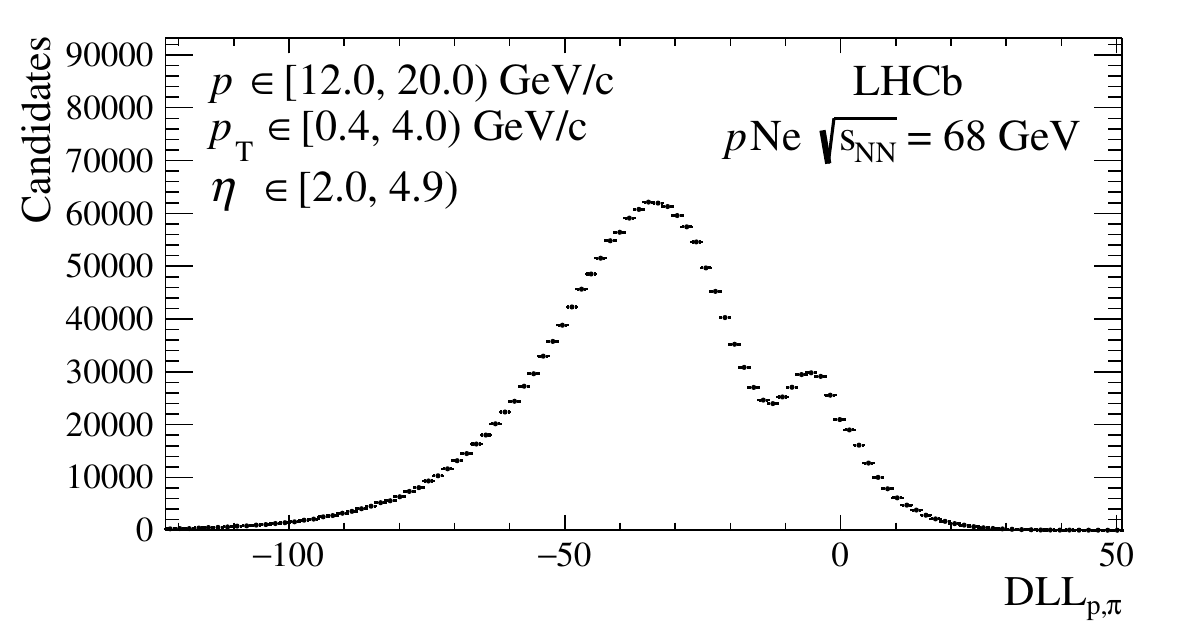} 
\includegraphics[width = 0.480000\textwidth]{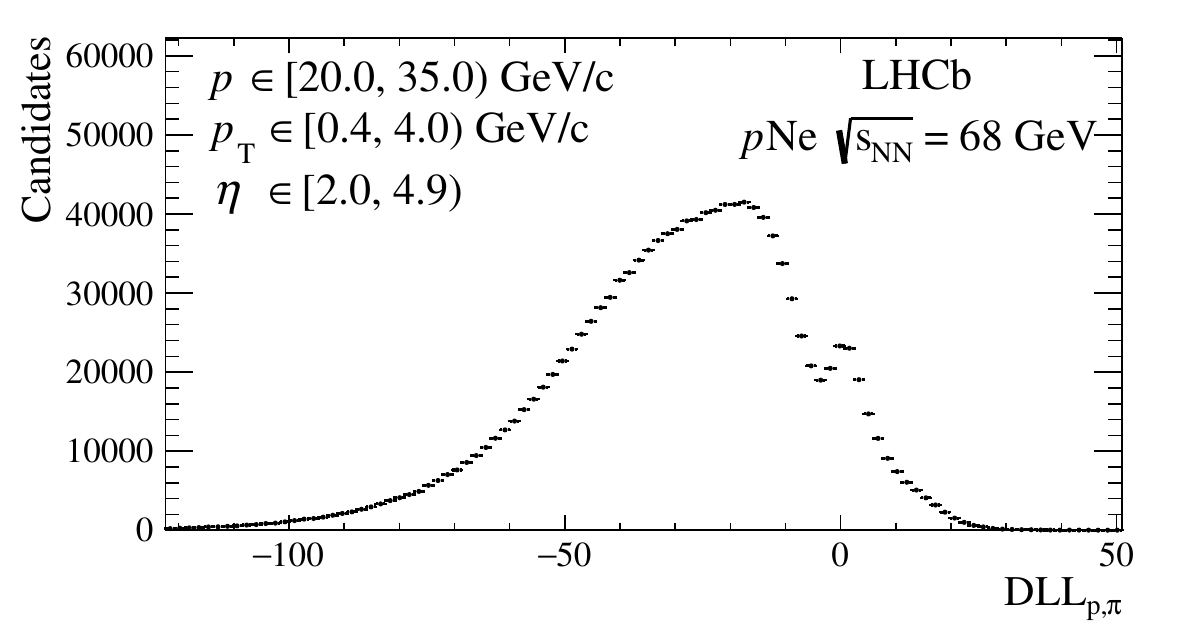} 
\includegraphics[width = 0.480000\textwidth]{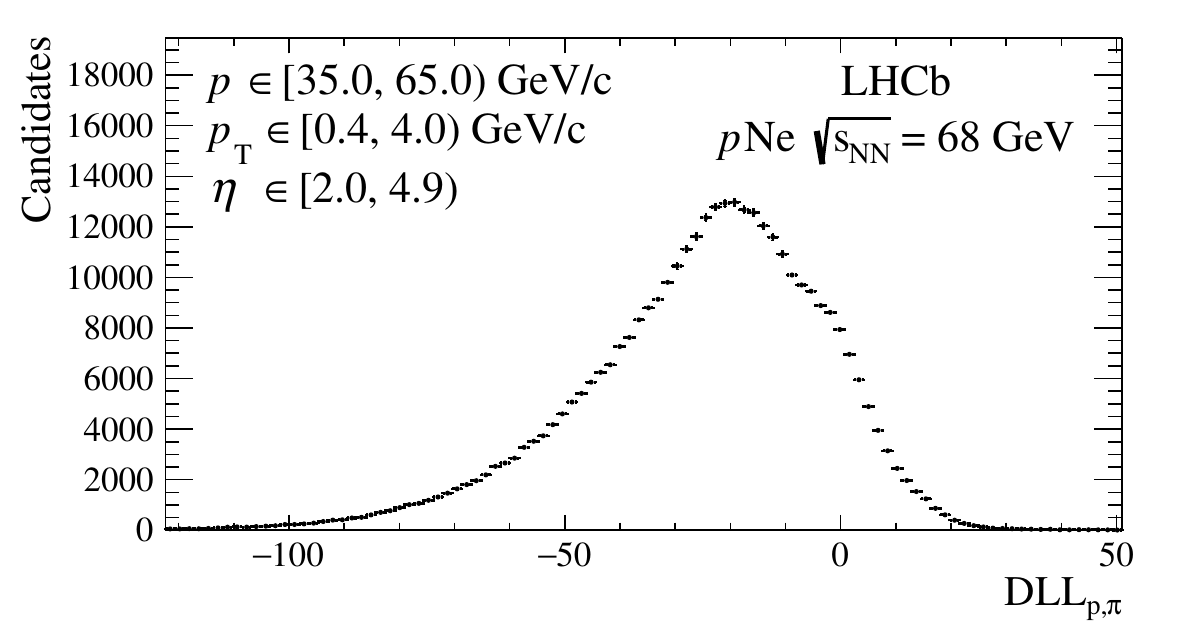} 
\includegraphics[width = 0.480000\textwidth]{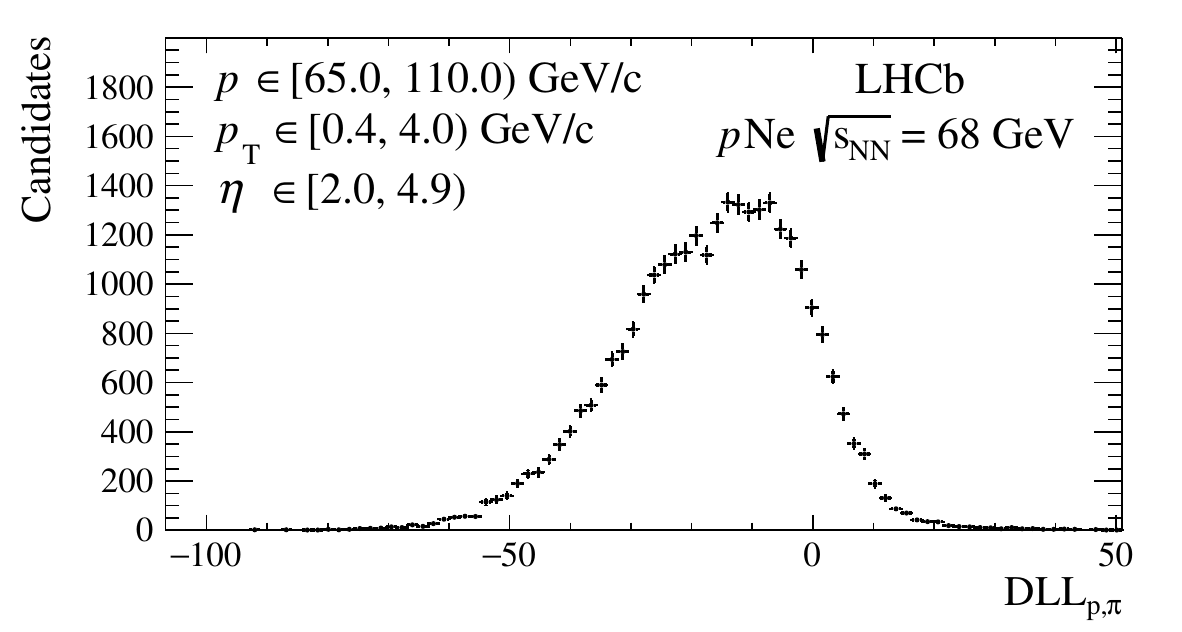} 
\caption{\dllppi distribution for the negative pion in the $\decay{\KS}{\pim\pip}$ calibration channel reconstructed and selected in the \pNe sample in different pion momentum bins.}
\label{fig_ML:KS0_minipeak} 
\end{figure}

The two PID classifiers \dllppi and \dllpk in the training samples show a linear correlation to a good extent.
Therefore, we model their distribution according to Eq.~\ref{eq:model} replacing the Gaussian with a two-dimensional multivariate normal distribution:
\begin{equation}
  \xsPID_p \sim \sum_{j=1}^{N_{g,p}} \alpha_{j,p}(\WV)  \dfrac{\exp(-\frac{1}{2}
(\xsPID_p-\underline{\mu}_{j,p}(\WV))^T \, \Sigma_{j,p}^{-1}(\WV) \, (\xsPID_p-\underline{\mu}_{j,p}(\WV)) )}
  {2\pi \sqrt{\det(\Sigma_{j,p}(\WV))} }
\end{equation}
where the vectors \xsPID represent the two PID classifiers (\dllppi, \dllpk) and
$\Sigma$ is their covariance matrix
\begin{gather}
\Sigma =
\begin{bmatrix}
\sigma^2_1 & \rho \sigma_1 \sigma_2\\
 \rho \sigma_1 \sigma_2 & \sigma^2_2
\end{bmatrix}.
\end{gather}
For each component in the GMM, the free parameters are thus the weight $\alpha$,
the central values $\underline{\mu}$, the standard deviations $(\sigma_1,\sigma_2)$ and the correlation coefficient $\rho$. The model is trained separately for each of the three particle species following the procedure outlined in Section~\ref{sec:method} and with the hyperparameters listed in Tab.~\ref{tab_ML:Parameters}.\\
The choice of the input parameters reflects a higher complexity for the $\decay{\KS}{ \pim \pip}$ and the $\decay{\phiz}{\Km \Kp}$ calibration channels. Indeed, as a consequence of the many threshold effects involved in the process, the distributions of these classifiers may result significantly different from a multivariate normal distribution. For example, in Fig.~\ref{fig_ML:KS0_minipeak} the
distribution of the \dllppi classifier computed on the calibration pions is shown. Comparing the distributions in four different intervals of the pion momentum, a second minor peak, arising at lower momentum, can be clearly seen. For the \phiz line, where the weight for the signal hypothesis obtained with the \sPlot technique is applied, only a fraction of events gives a large variation of the loss function, resulting in a training more prone to statistical fluctuations. To compensate for this effect, a higher batch size is chosen.\\
To differentiate the components in the GMM, the parameters are firstly randomly initialized from uniform distributions in the ranges:
\begin{table}
\caption{Values of the input parameters for the model training of the three calibration channels.}
\label{tab_ML:Parameters}
\centering
\begin{tabular}{cccc}
Input parameter     &  $\decay{\KS}{\pim \pip}$  & $\decay{\Lbar}{\antiproton\pip}$  & $\decay{\phiz}{\Km \Kp}$ \\
\hline
Number of Gaussians    &   64                  &  20                    & 64               \\
\hline
Number of NN nodes     &   128                 &  128                   & 128              \\
\hline
Starting learning rate & $10^{-3}$             & $10^{-4}$              & $5\cdot 10^{-6}$         \\
\hline
Batch size             &  10000 events         & 10000 events           & 20000 events      \\
\hline
\end{tabular}
\end{table}
\begin{itemize}
	\item $\left[
            \langle \underline{\xPID} \rangle - \frac{1}{2}\sqrt{\langle \underline{\xPID}^2\rangle - \langle \underline{\xPID}\rangle^2},
            \langle \underline{\xPID} \rangle + \frac{1}{2}\sqrt{\langle \underline{\xPID}^2\rangle - \langle \underline{\xPID}\rangle^2 }
          \right]$ for the mean values $\underline{\mu}$,
          being $\langle \cdot \rangle$ the average. For the $\phiz$ calibration channel, the background-subtracted distributions are considered;
        \item $[0, 2\pi]$ for the correlation angle;
	\item $\left[
            \frac{1}{10}\sqrt{\langle \underline{\xPID}^2\rangle - \langle \underline{\xPID}\rangle^2} ,
            \sqrt{\langle \underline{\xPID}^2\rangle - \langle \underline{\xPID}\rangle^2 }
          \right]$ for the widths $\sigma_{1,2}$;
	\item $[0, 1]$ for the component weights $\alpha_{j}$.
\end{itemize}

\begin{figure} 
\centering 
\includegraphics[width = 0.480000\textwidth]{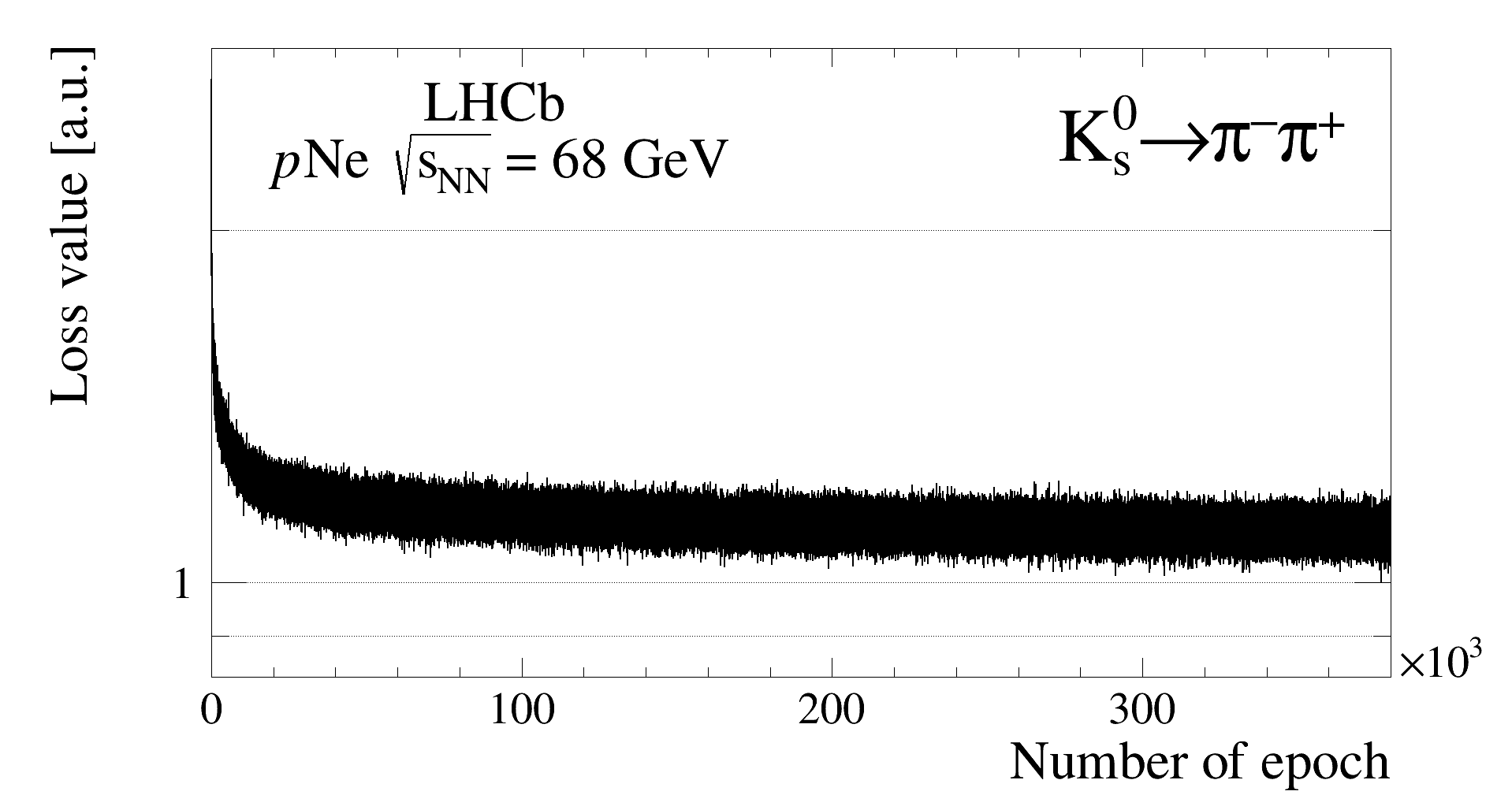} 
\includegraphics[width = 0.480000\textwidth]{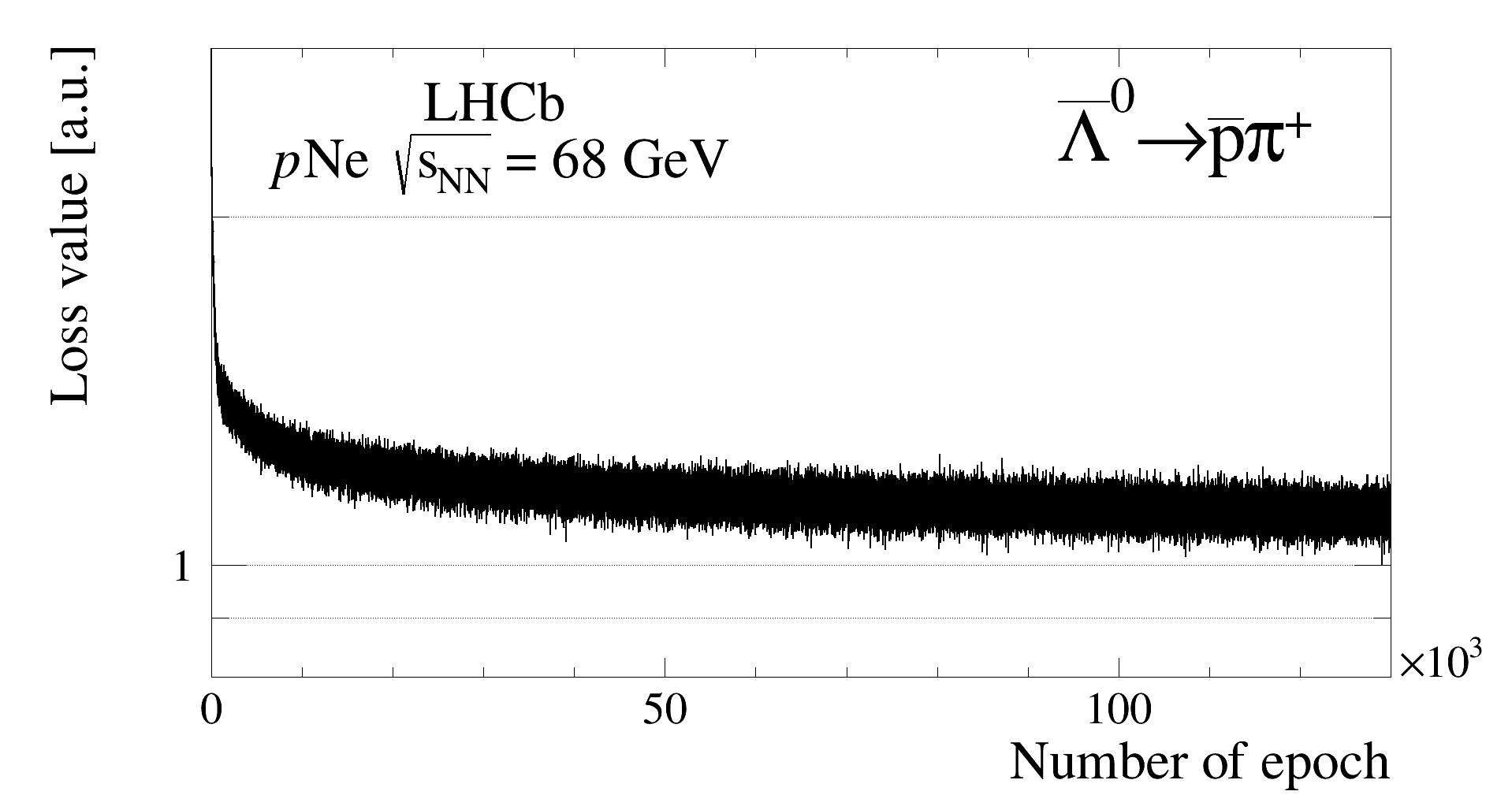} 
\includegraphics[width = 0.480000\textwidth]{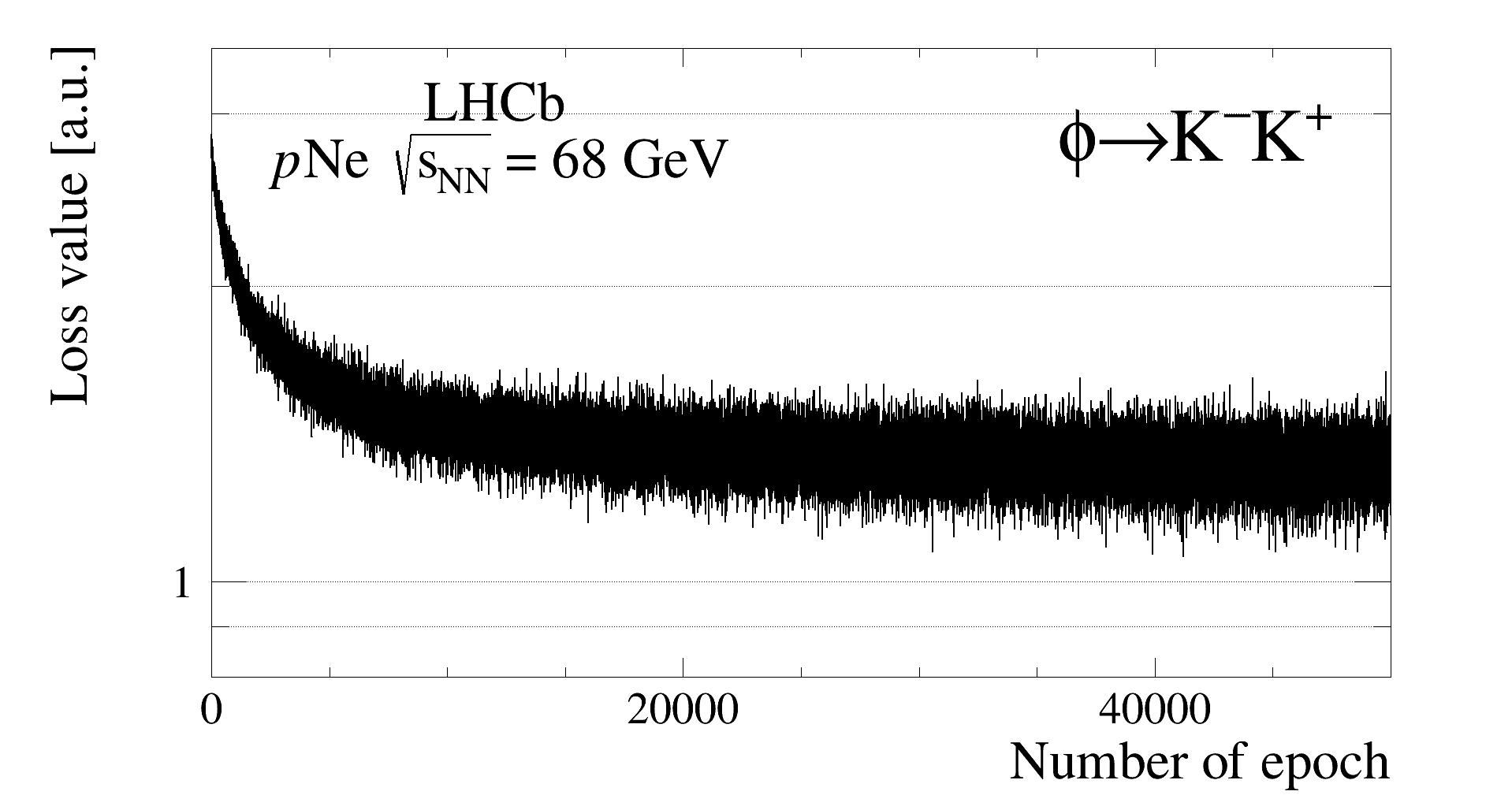} 
\caption{Evolution of the loss function with the number of epochs for the (top left) pion, (top right) proton and (bottom) kaon calibration channels.}
\label{fig_ML:loss_functions} 
\end{figure} 

\begin{figure} 
\centering 
\includegraphics[width = 0.480000\textwidth]{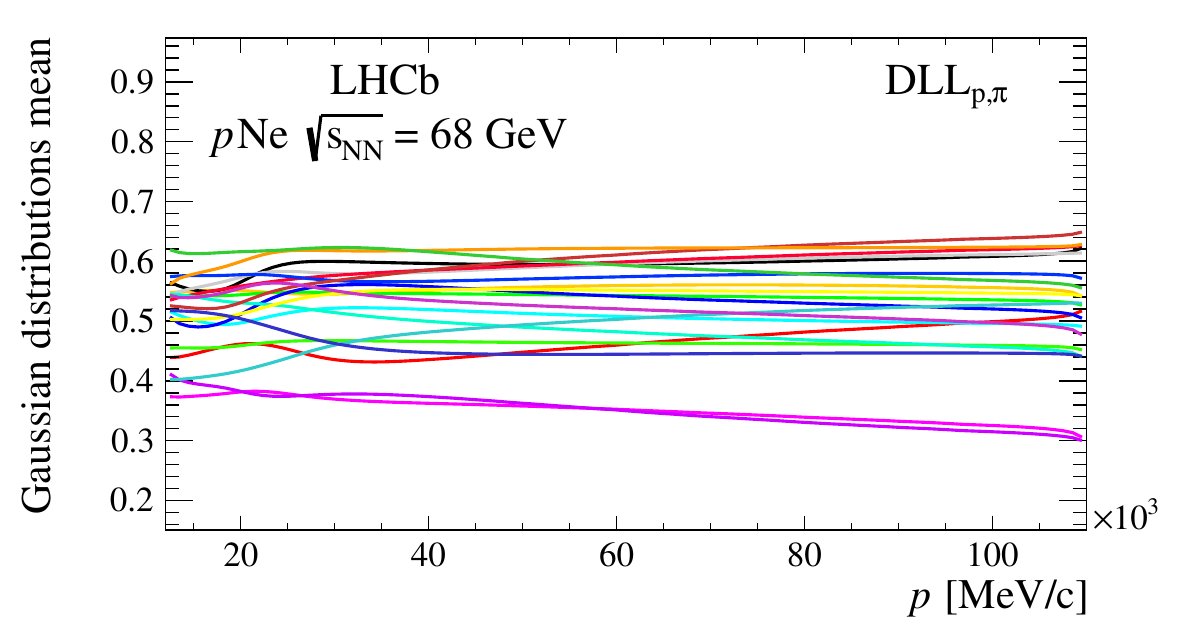} 
\includegraphics[width = 0.480000\textwidth]{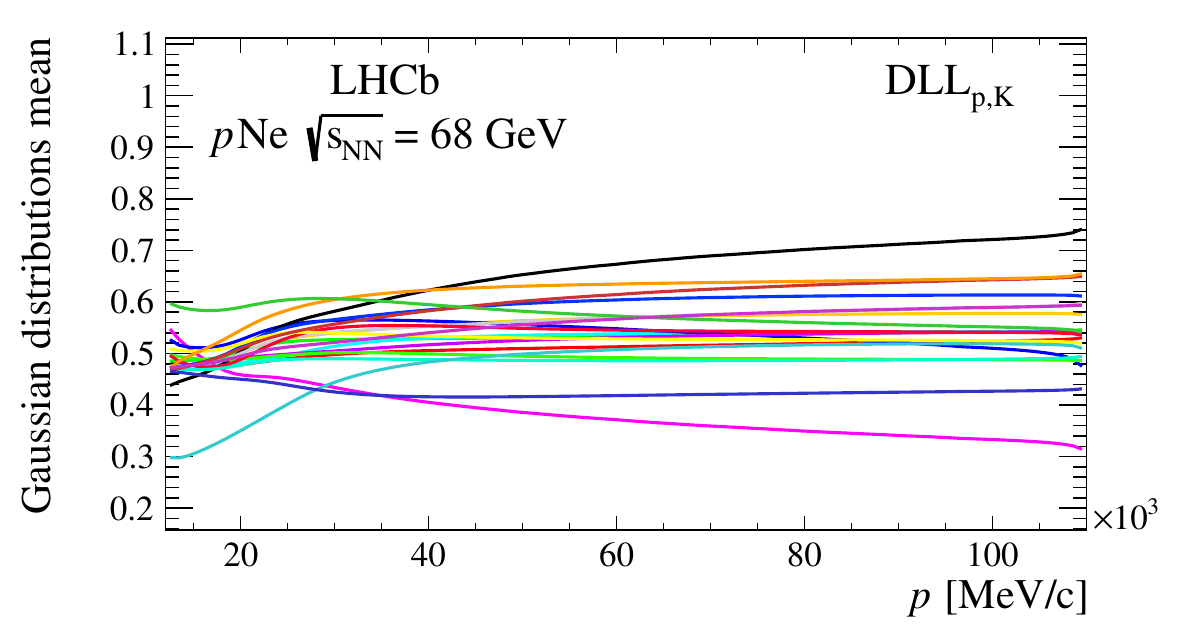} 
\includegraphics[width = 0.480000\textwidth]{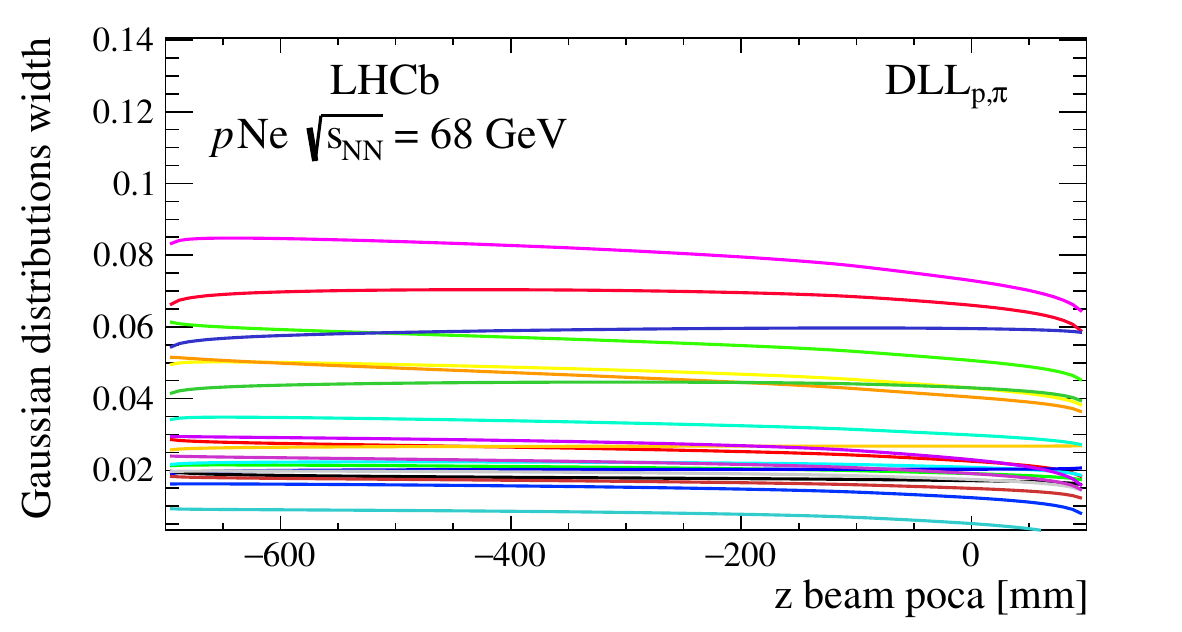} 
\includegraphics[width = 0.480000\textwidth]{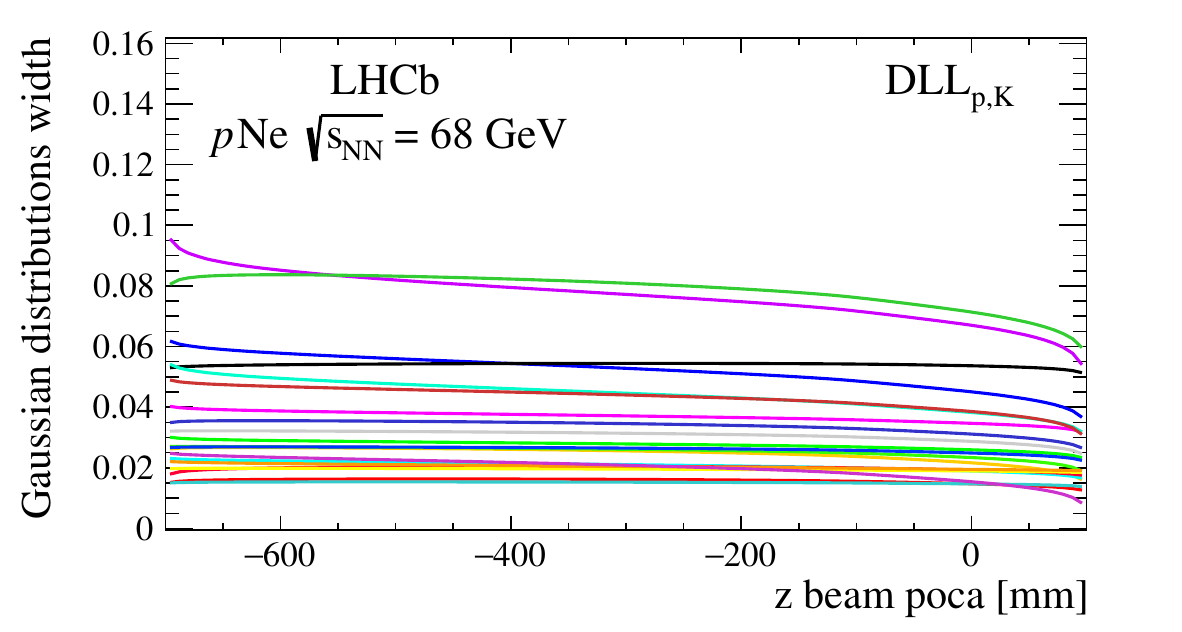} 
\includegraphics[width = 0.480000\textwidth]{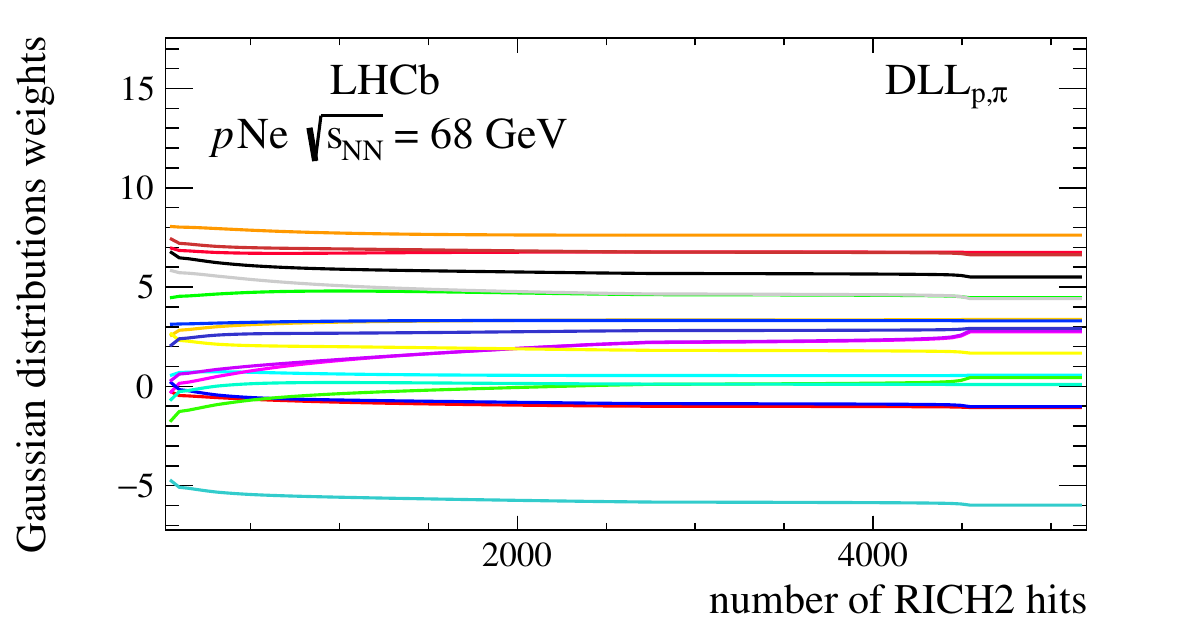} 
\includegraphics[width = 0.480000\textwidth]{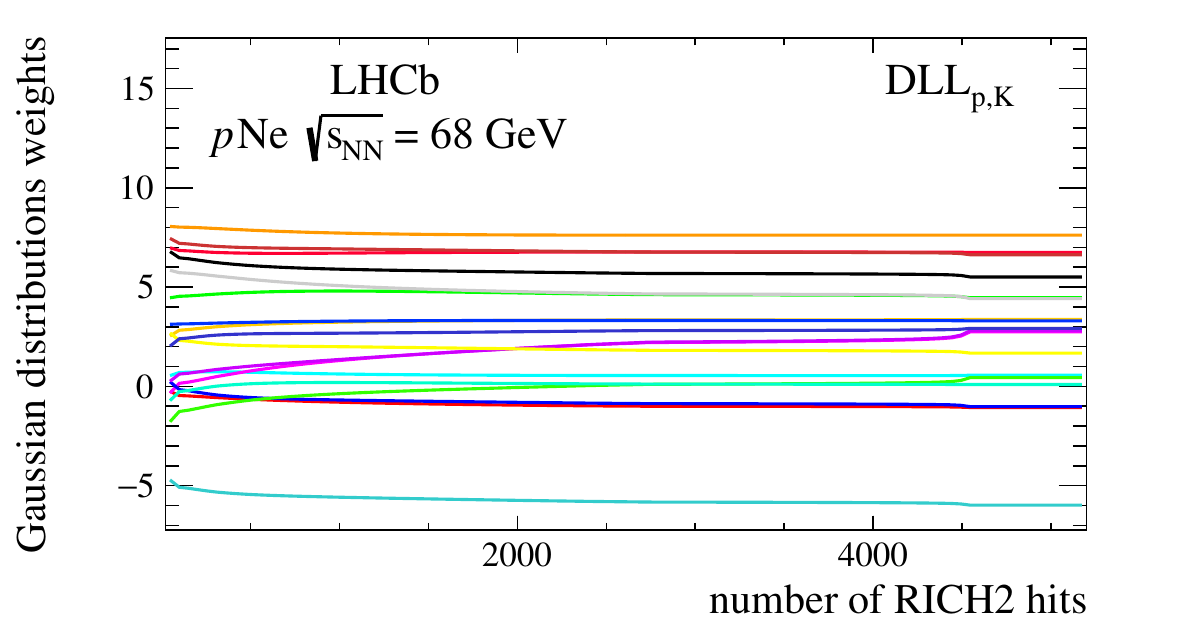} 
\caption{Check against overtraining reported as an example for the $\decay{\Lbar}{\antiproton \pip} $  calibration channel from the evolution of (top) the Gaussian mean values, (middle) widths and (bottom) weights for (left) the \dllppi and (right) the \dllpk target variables as a function of the particle momentum, \textit{z} coordinate of the track position of closest approach to the beam and number of hits in the RICH2 subdetector. Different colours represent the different Gaussian components considered in the model.} 
\label{fig_ML:Overtraining} 
\end{figure}

The training is performed using a mini-batches gradient descent optimized with the RMSProp algorithm. Gaussian weights are left free to fluctuate towards negative values in order to enhance the stability of the training procedure. The evolution of the loss function with the number of epochs is monitored and Fig.~\ref{fig_ML:loss_functions} shows the resulting curves for the three calibration channels, all presenting a first steep decrease followed by a slow one and finally gentle oscillations near the minimum. The trained models present $O(10^5)$ parameters, depending on the chosen complexity for the NN and the GMM. For the \KS calibration line and the parameters listed in Tab.~\ref{tab_ML:Parameters}, this corresponds to a measured training time of approximately 10 hours on a NVIDIA K80 GPU.\\
The dependence of the Gaussian parameters as a function of the features is verified at the end of each training procedure to check for the presence of possible overtraining effects, which would manifest as their rapid
oscillations to adapt to statistical fluctuations in the training sample.
Fig.~\ref{fig_ML:Overtraining} shows an example of this for the  $\decay{\Lbar}{\antiproton \pip}$ calibration channel, where the expected smooth and monotonic behaviour can be observed. The generalization from the training to the application samples described in Sec.~\ref{sec:appl} definitively excludes overtraining effects.

\subsection{Validation}
\label{sec:ML_validation}
\begin{figure} 
\centering 
\includegraphics[width = 0.850000\textwidth]{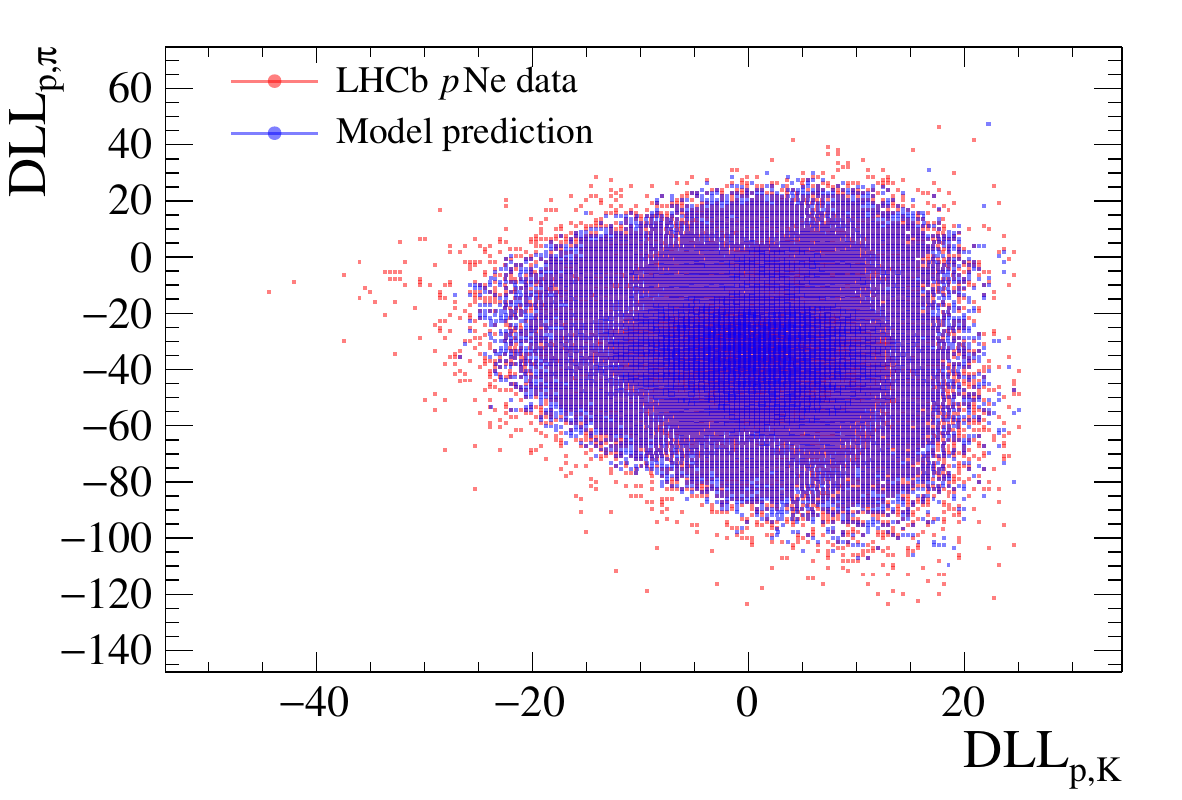} 
\includegraphics[width = 0.480000\textwidth]{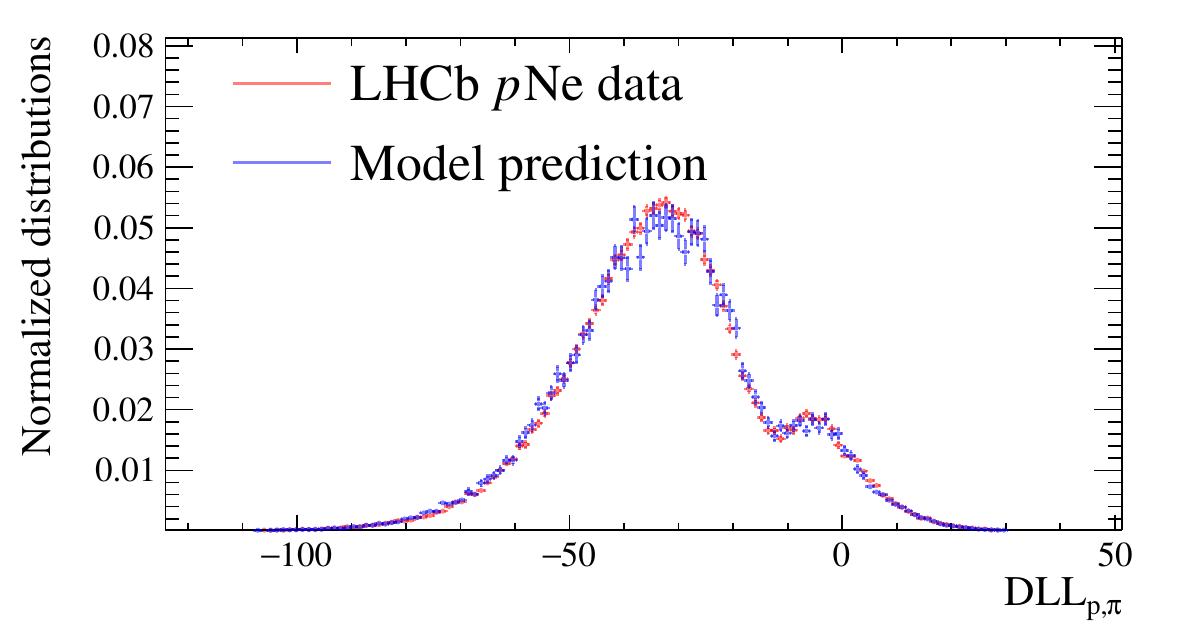} 
\includegraphics[width = 0.480000\textwidth]{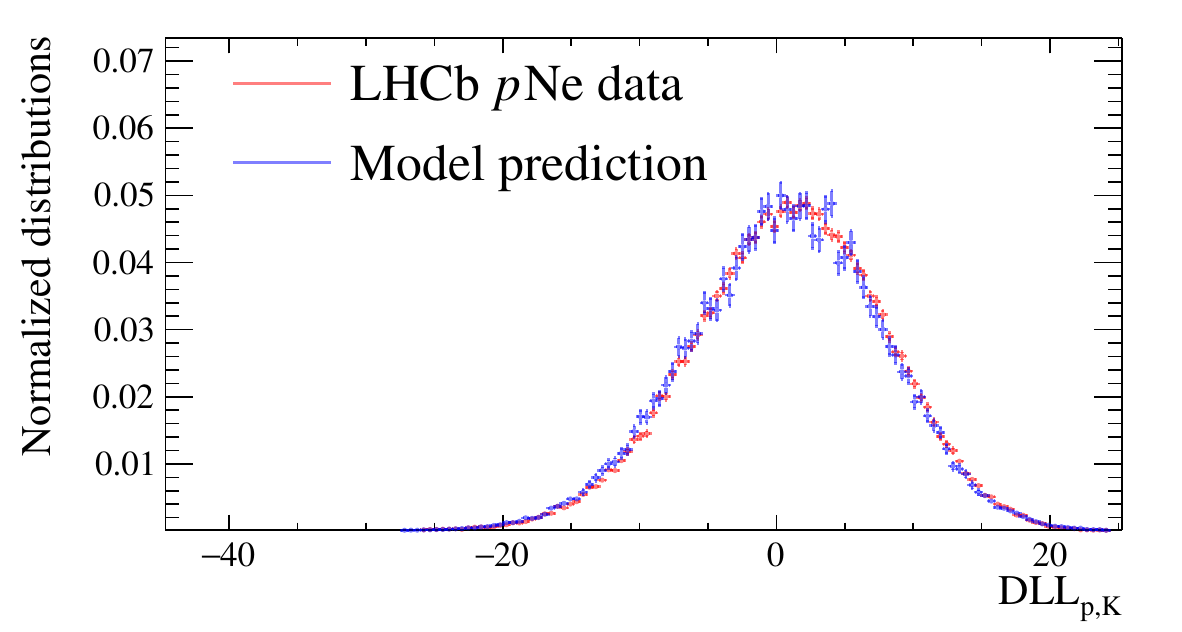} 
\includegraphics[width = 0.480000\textwidth]{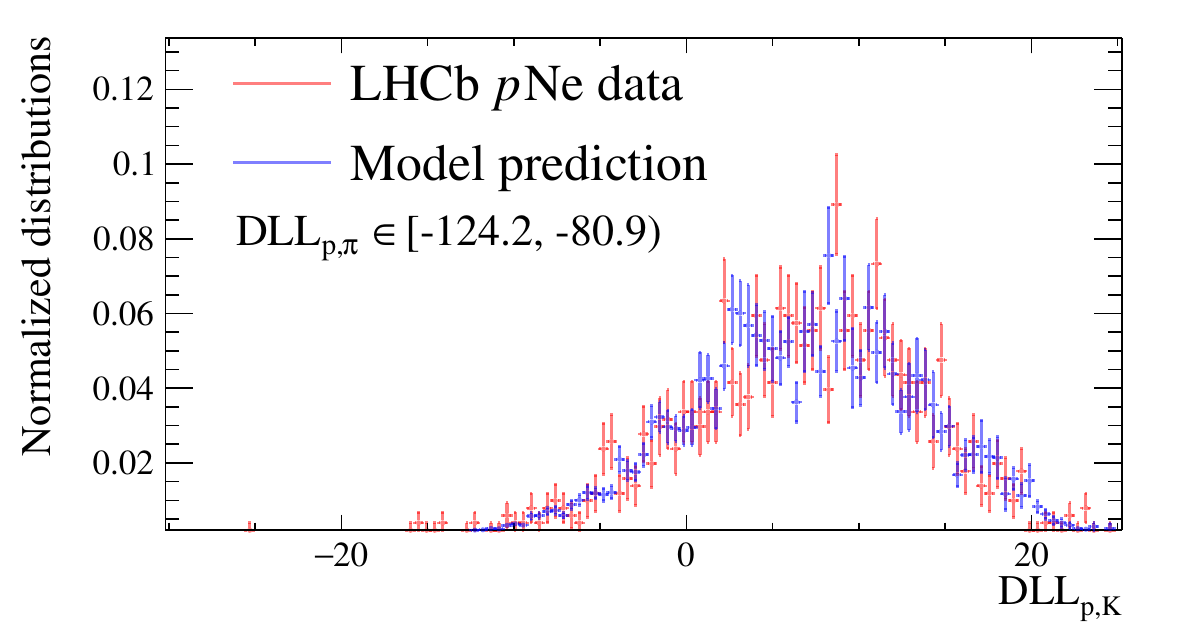} 
\includegraphics[width = 0.480000\textwidth]{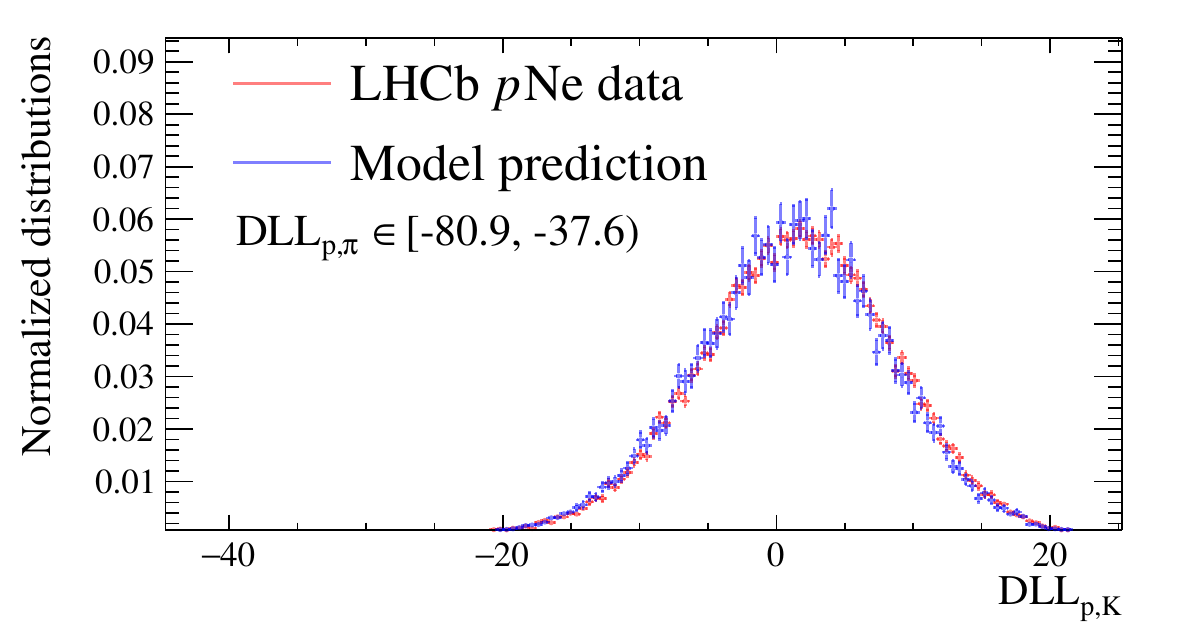} 
\includegraphics[width = 0.480000\textwidth]{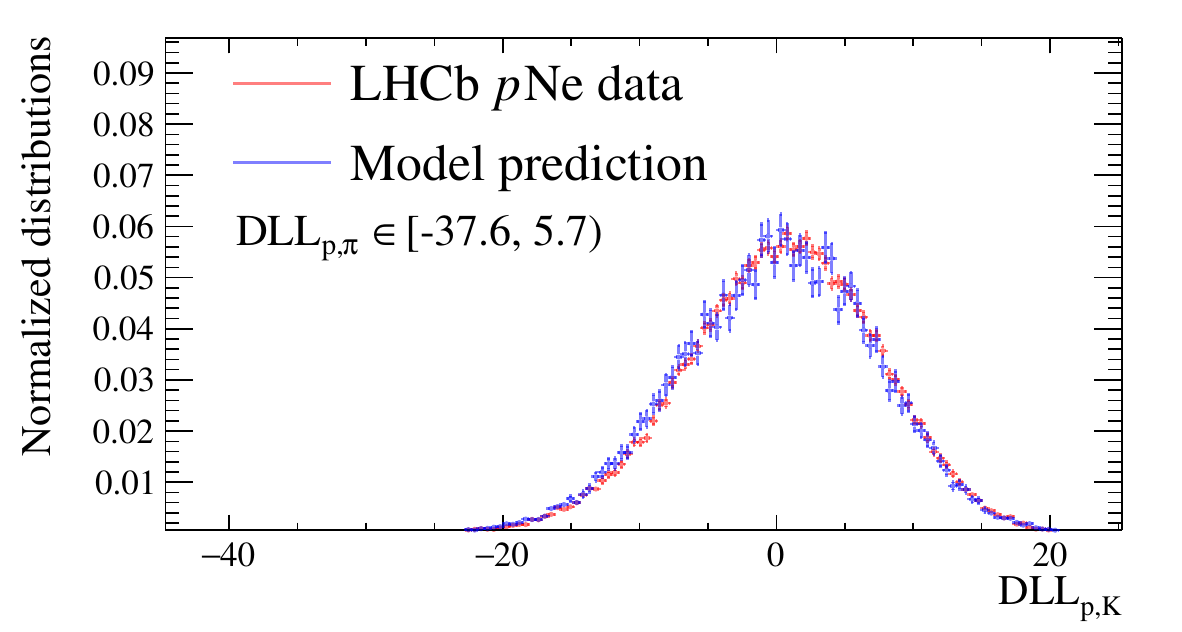} 
\includegraphics[width = 0.480000\textwidth]{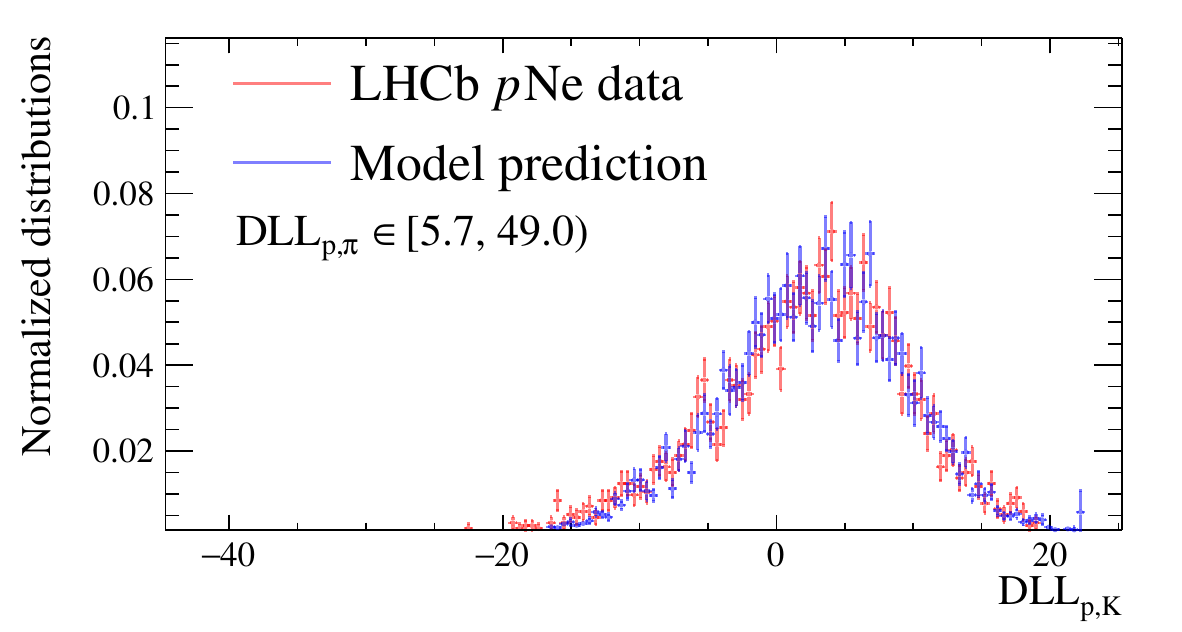} 
\caption{Comparison for the \decay{\KS}{\pim \pip} calibration channel in the $\ptot \in [12.0, 15.5) \mevc$ - $\eta \in [4.1, 4.4)$ bin between the bidimensional \xsPID distributions in (red) the \pNe data and (blue) predicted with the trained model (top plot), its projections onto the two axes (second row) and onto the \dllpk axis in intervals of the \dllppi variable (third and fourth row).}
\label{fig_ML:PID_Validation_KS0} 
\end{figure}

\begin{figure} 
\centering 
\includegraphics[width = 0.850000\textwidth]{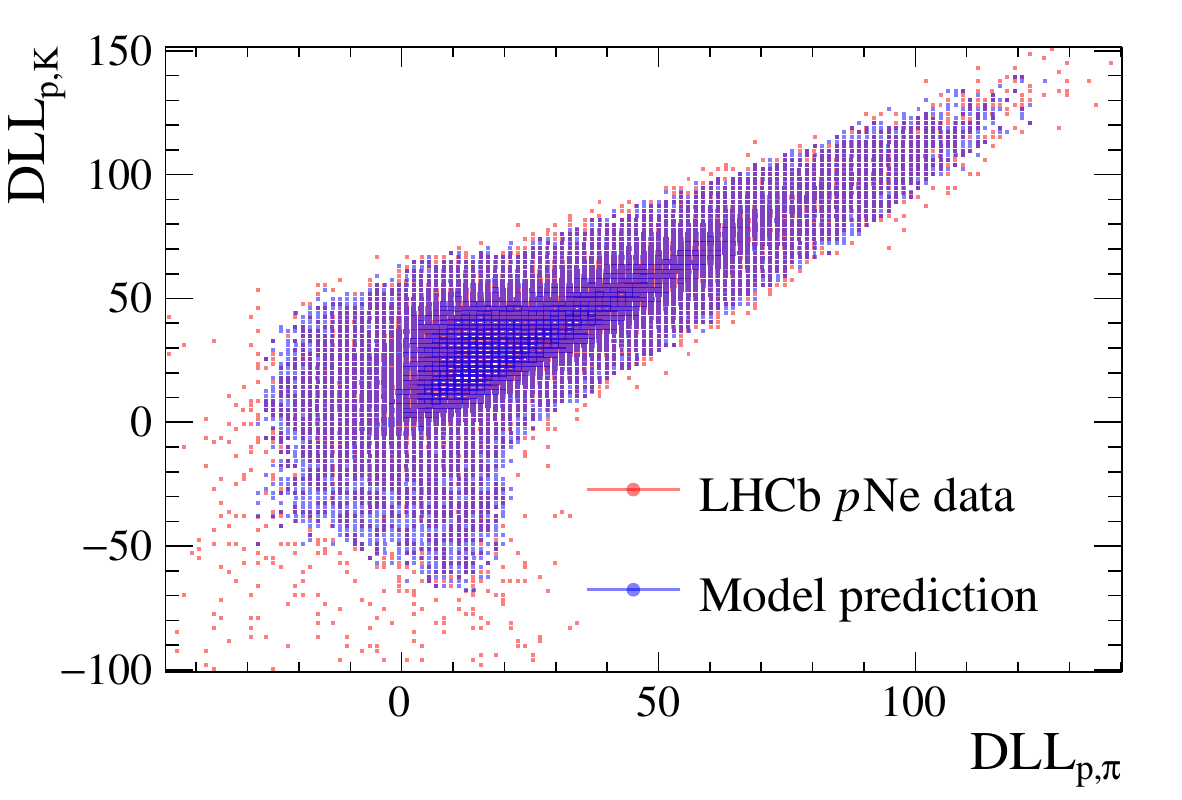} 
\includegraphics[width = 0.480000\textwidth]{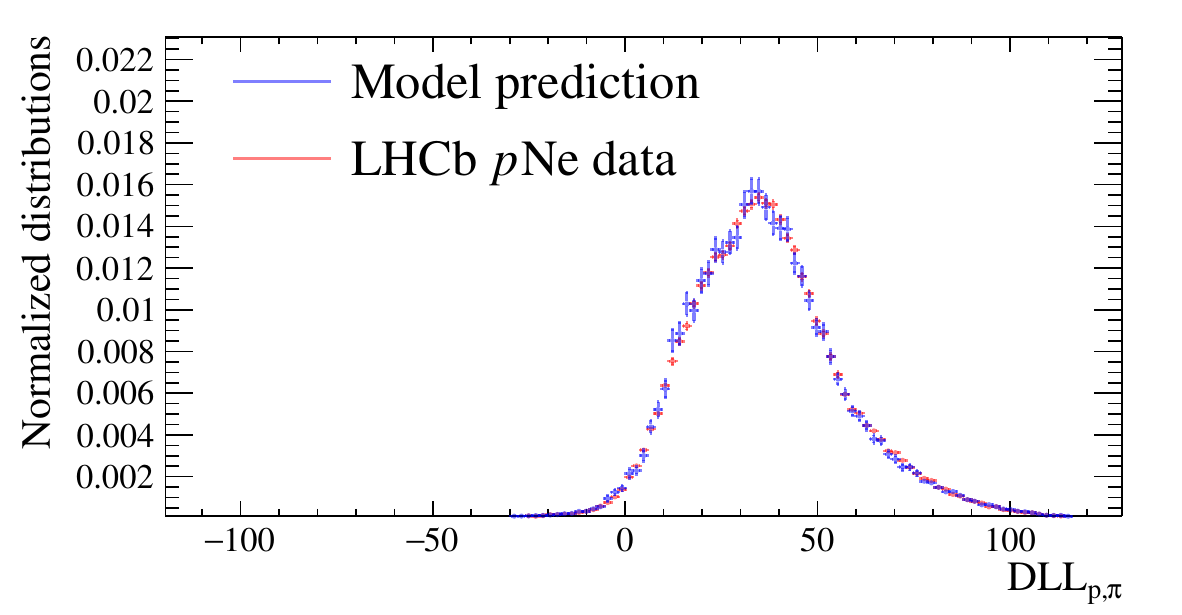} 
\includegraphics[width = 0.480000\textwidth]{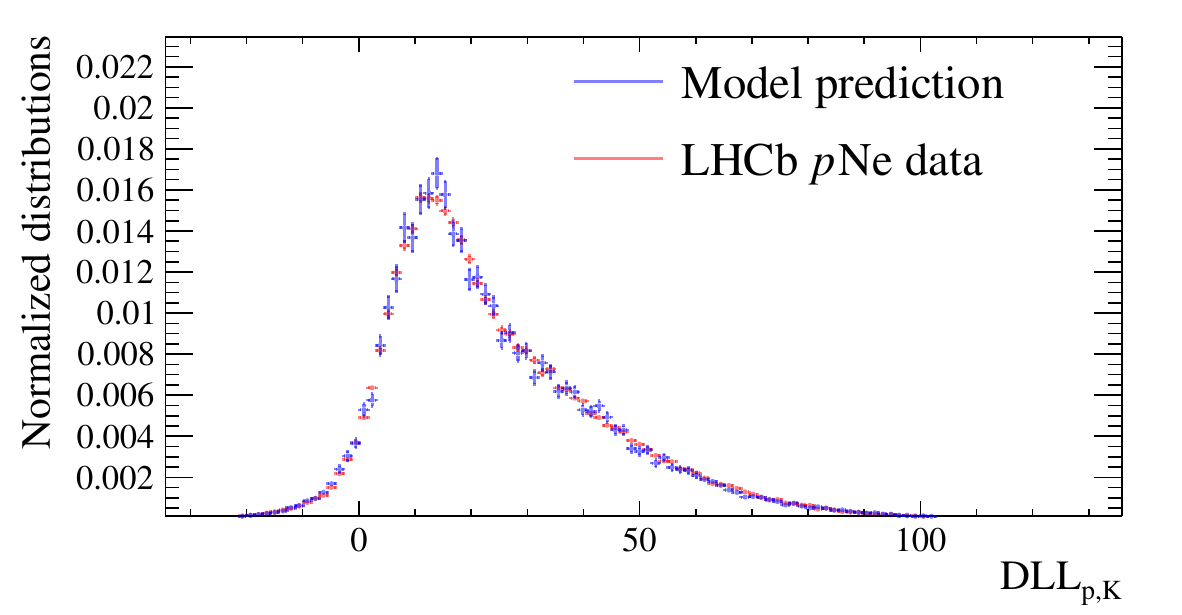} 
\includegraphics[width = 0.480000\textwidth]{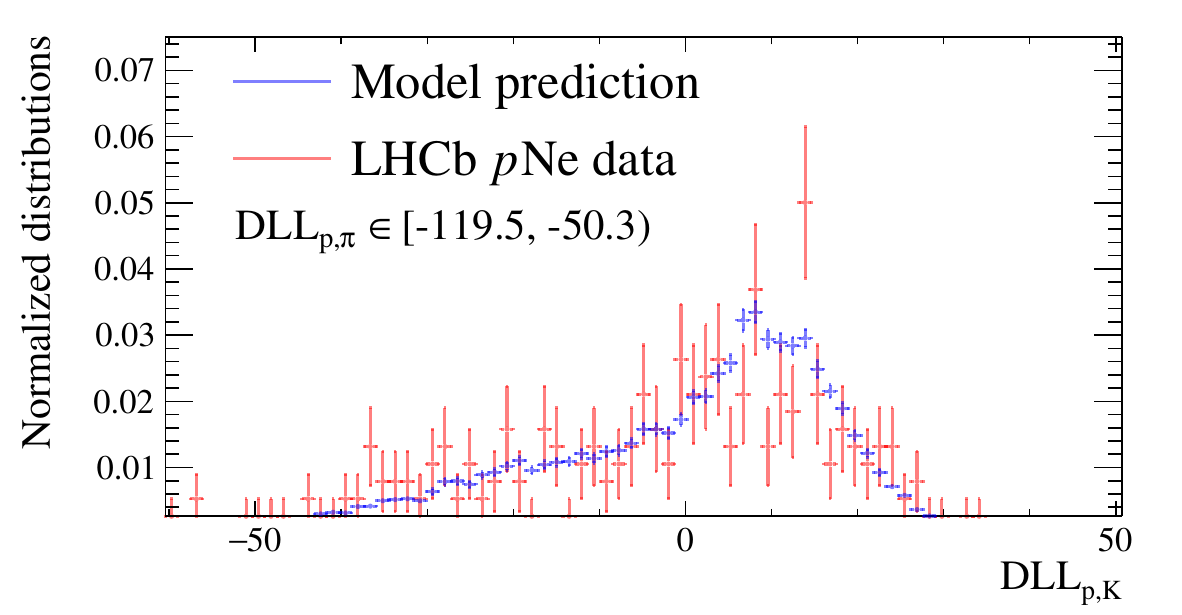} 
\includegraphics[width = 0.480000\textwidth]{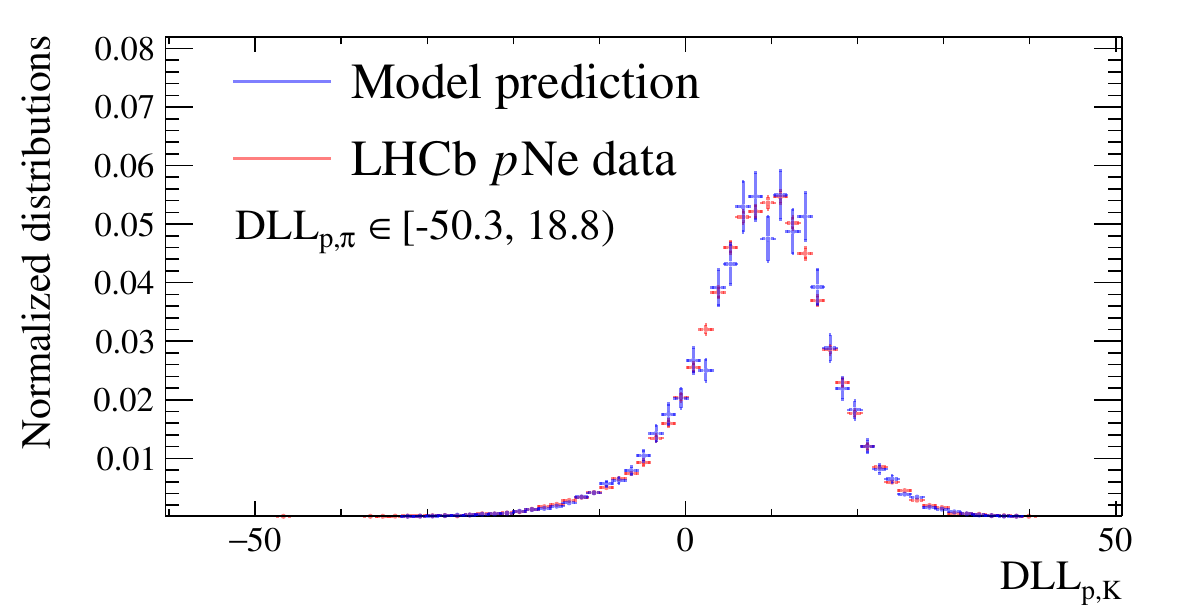} 
\includegraphics[width = 0.480000\textwidth]{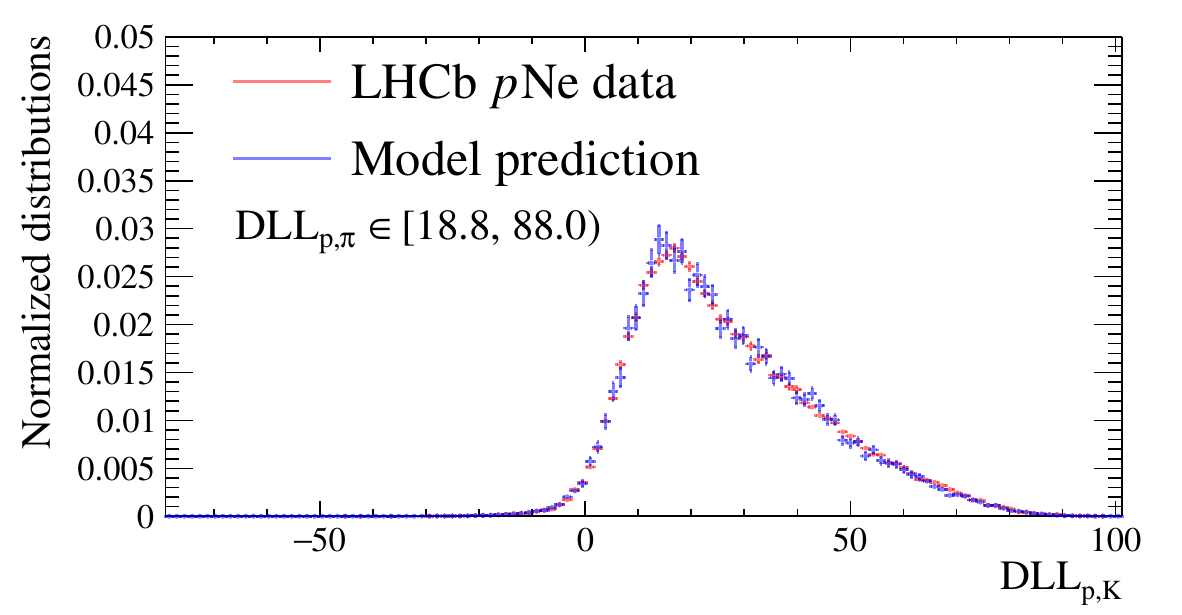} 
\includegraphics[width = 0.480000\textwidth]{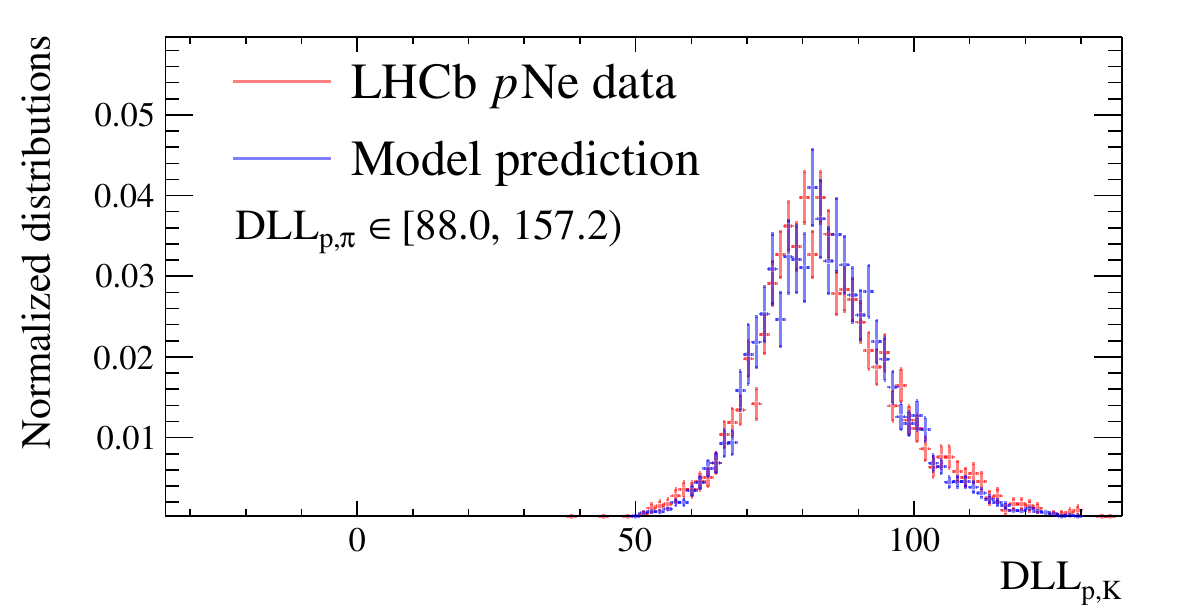} 
\caption{Comparison for the \decay{\Lbar}{\antiproton \pip} calibration channel in the poca$_z \in [-700, -430.8) \mm$ -  $nRICH1Hits \in [21, 688)$ bin between the bidimensional \xsPID distributions in (red) the \pNe data and (blue) predicted with the trained model (top plot), its projections onto the two axes (second row) and onto the \dllpk axis in intervals of the \dllppi variable (third and fourth row).}
\label{fig_ML:PID_Validation_Lambda0} 
\end{figure}

\begin{figure} 
\centering 
\includegraphics[width = 0.850000\textwidth]{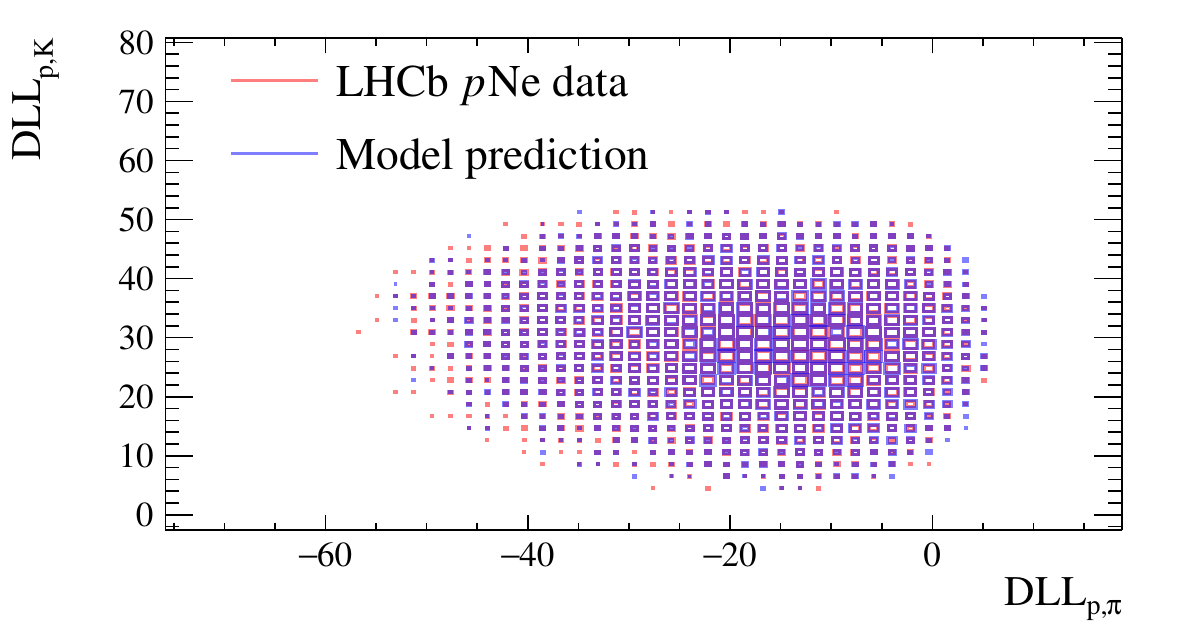} 
\includegraphics[width = 0.480000\textwidth]{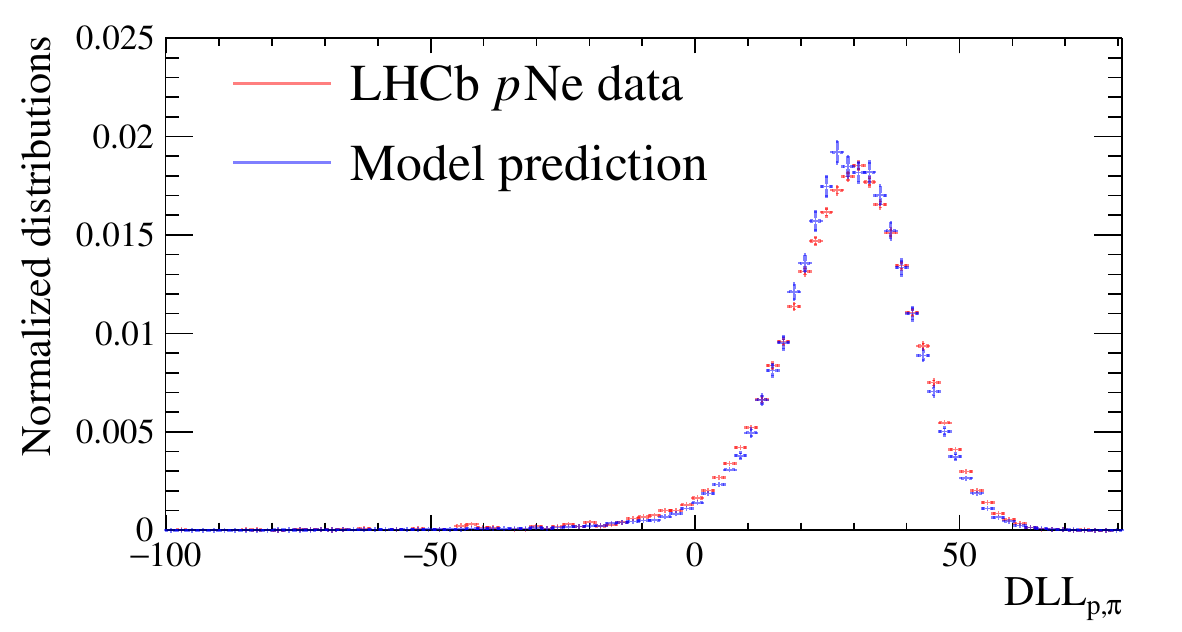} 
\includegraphics[width = 0.480000\textwidth]{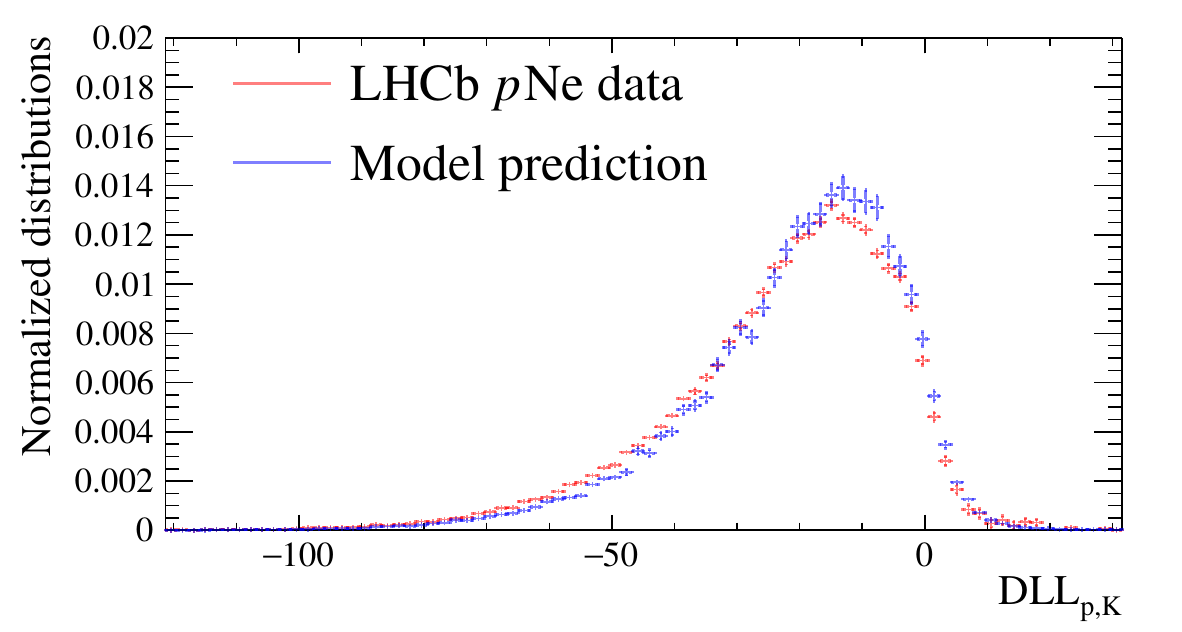} 
\includegraphics[width = 0.480000\textwidth]{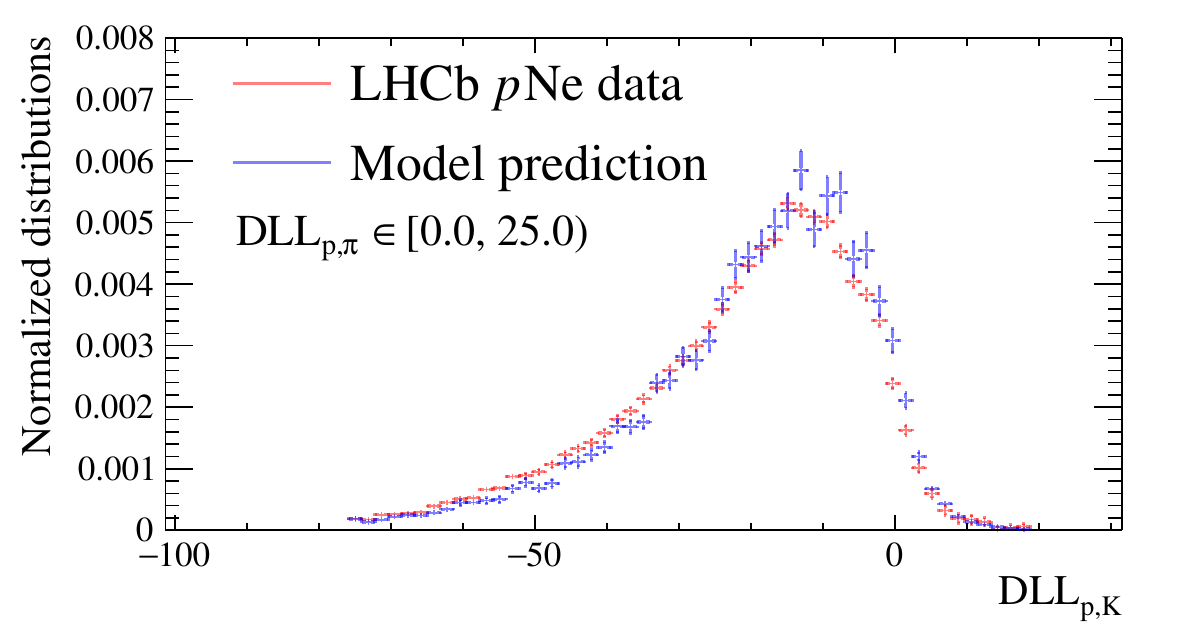}
\includegraphics[width = 0.480000\textwidth]{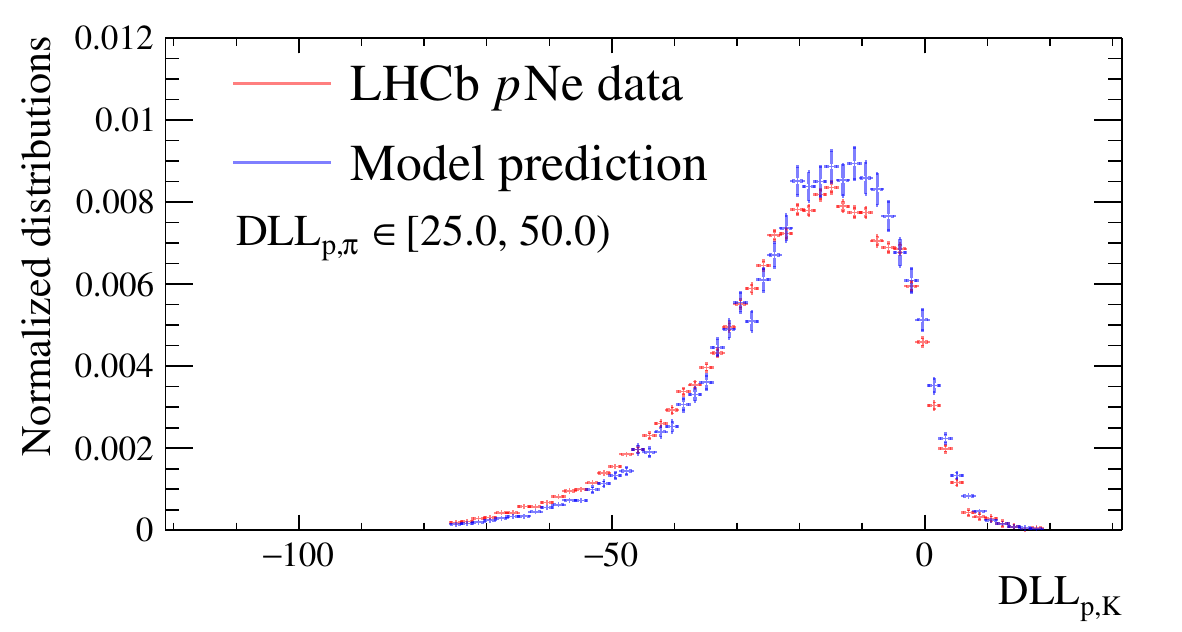}
\caption{Comparison for the \decay{\phiz}{\Km \Kp} calibration channel in the track fit $\chisqndf \in [1.5, 4)$ - $\eta \in [1.8, 3.7)$ bin between the bidimensional \xsPID distributions in (red) the \pNe data and (blue) predicted with the trained model (top plot), its projections onto the two axes (second row) and onto the \dllpk axis in intervals of the \dllppi variable (third row).}
\label{fig_ML:PID_Validation_Phi} 
\end{figure}

\begin{figure}
\centering
\includegraphics[width = 0.480000\textwidth]{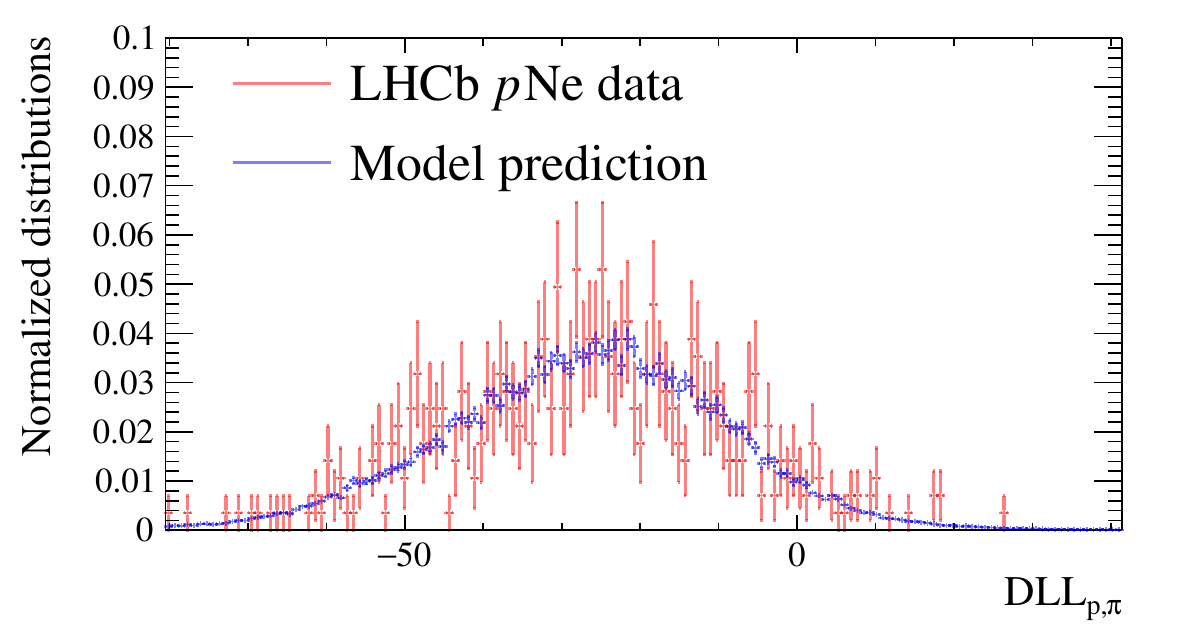}
\includegraphics[width = 0.480000\textwidth]{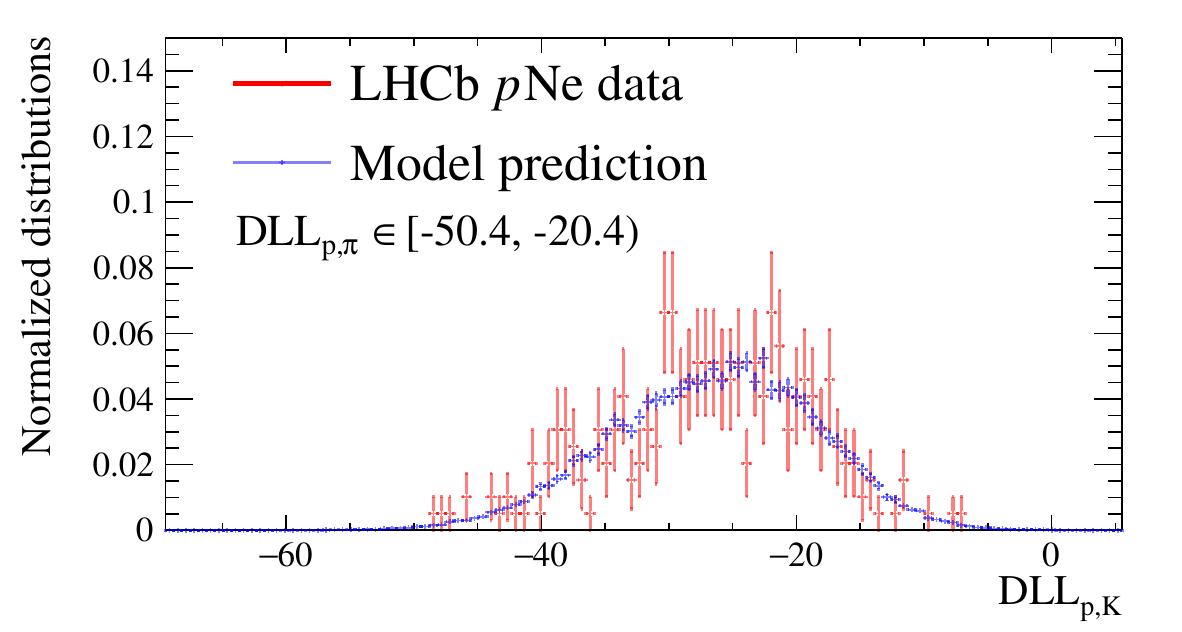}
\caption{Comparison for the \decay{\KS}{\pim \pip} calibration channel in the $\ptot \in [70, 110) \gevc$ - $\eta \in [3,4)$ bin, presenting low statistics. The 2017 \pNe data is shown in red and the model prediction in blue.}
\label{fig_ML:PID_Validation_KS0_lowStat}
\end{figure}

To verify that the model properly learns the non-trivial correlations between the PID classifiers and the considered features, its predictions are compared to the actual distributions for control data in two-dimensional intervals of all possible pairs of features \WV. The interval boundaries for each feature are set to roughly contain the same number of calibration entries. For each interval, a prediction is obtained by randomly choosing calibration entries in the selected interval and generating values of the target variables according to the fitted model for the corresponding values of \WV.
Fig.~\ref{fig_ML:PID_Validation_KS0} shows an example of this validation for the
$\decay{\KS}{\pim \pip}$ calibration channel in intervals of the $\pim$ particle momentum and transverse momentum, Figs.~\ref{fig_ML:PID_Validation_Lambda0} and \ref{fig_ML:PID_Validation_Phi} for the proton and kaon ones considering other feature pairs, respectively. The curves illustrate the non-trivial correlations of the target variables with the features and the excellent agreement between the data and the predictions demonstrates that the correlations are correctly learned by the model.
For the pion calibration channel, the validation is also repeated considering an interval with low \pNe statistics and Fig.~\ref{fig_ML:PID_Validation_KS0_lowStat} shows that, also in this bin,  the trained model is able to generate a smooth template based on the parametric interpolation of the available data.

\subsection{Application}
\label{sec:appl}
\begin{figure} 
\centering 
\includegraphics[width = 0.480000\textwidth]{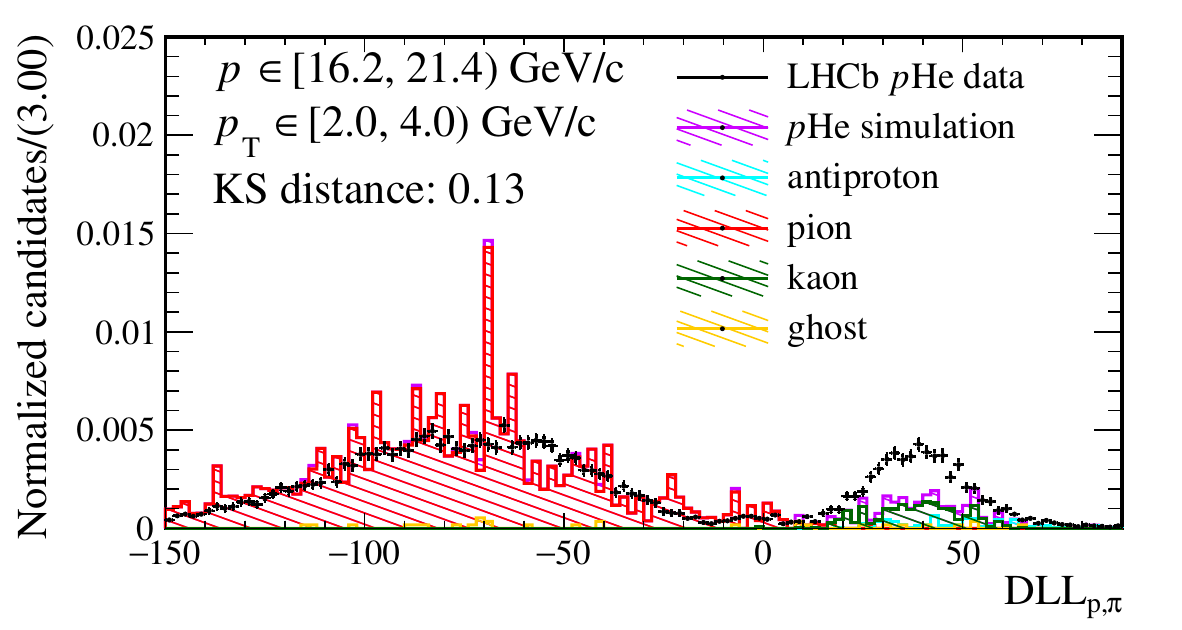} 
\includegraphics[width = 0.480000\textwidth]{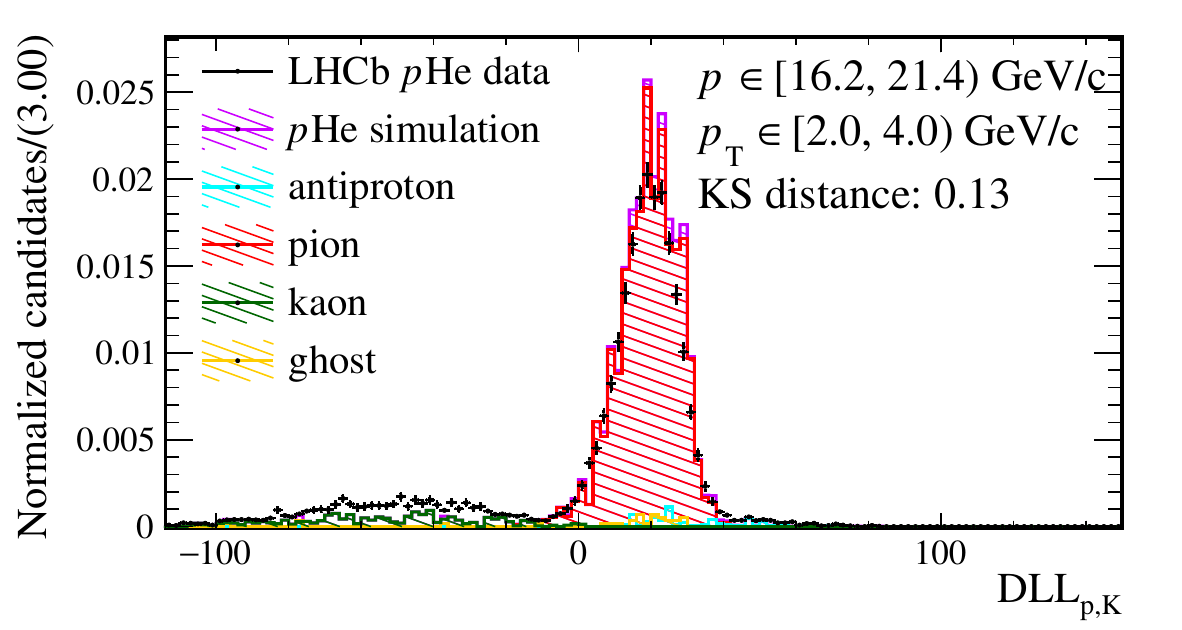} 
\includegraphics[width = 0.480000\textwidth]{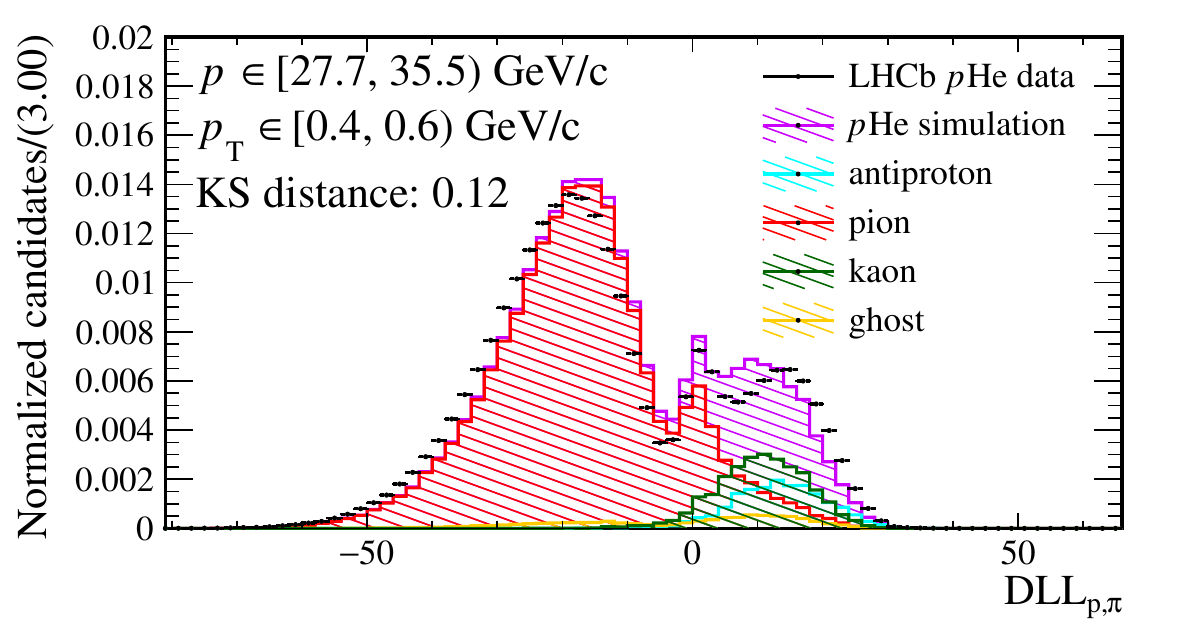} 
\includegraphics[width = 0.480000\textwidth]{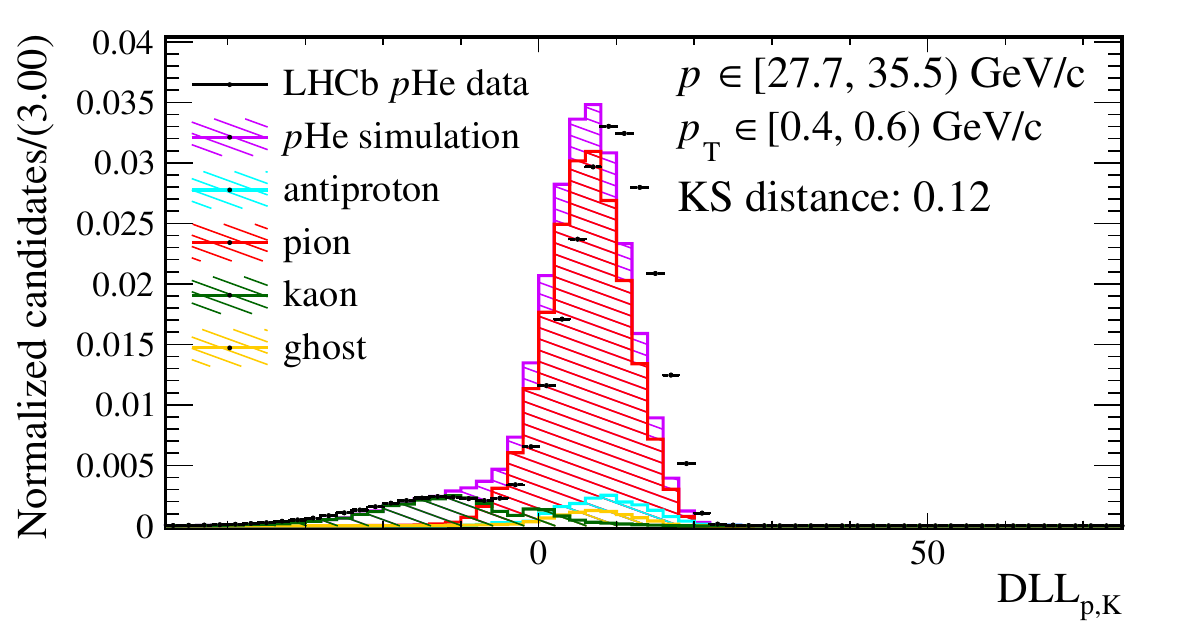}
\includegraphics[width = 0.480000\textwidth]{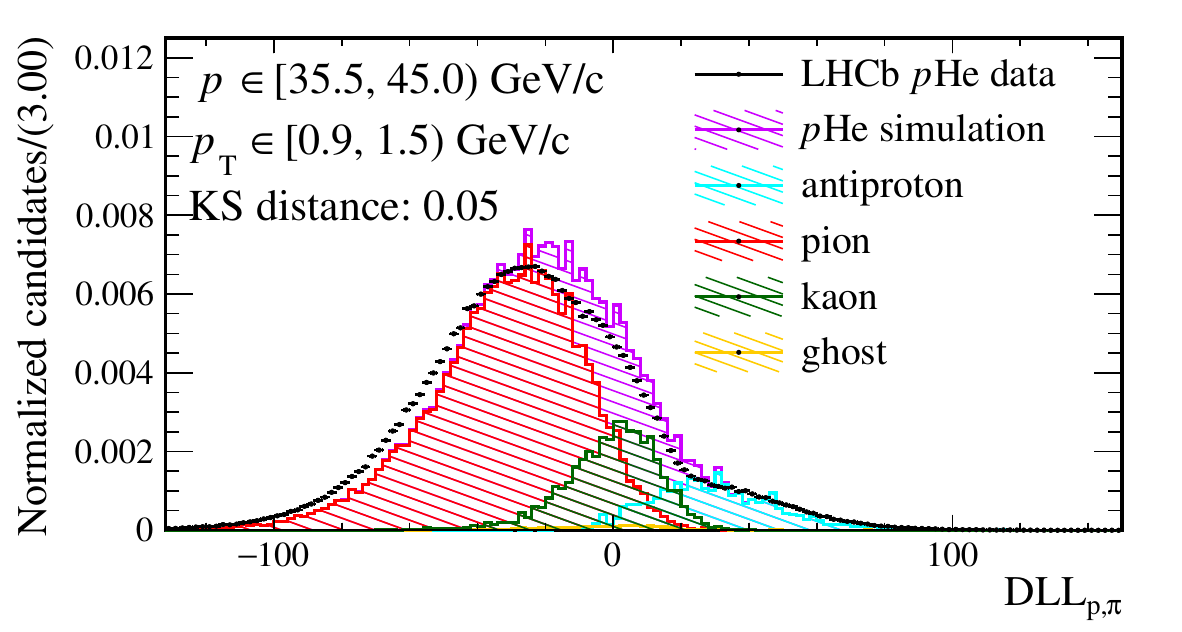}  
\includegraphics[width = 0.480000\textwidth]{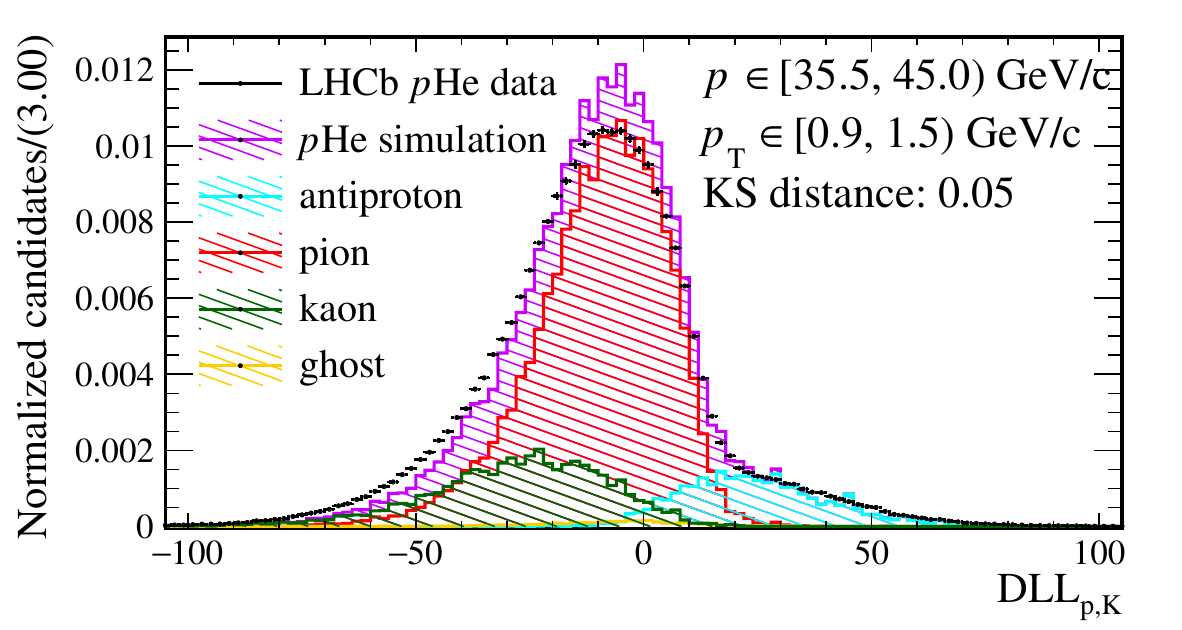} 
\includegraphics[width = 0.480000\textwidth]{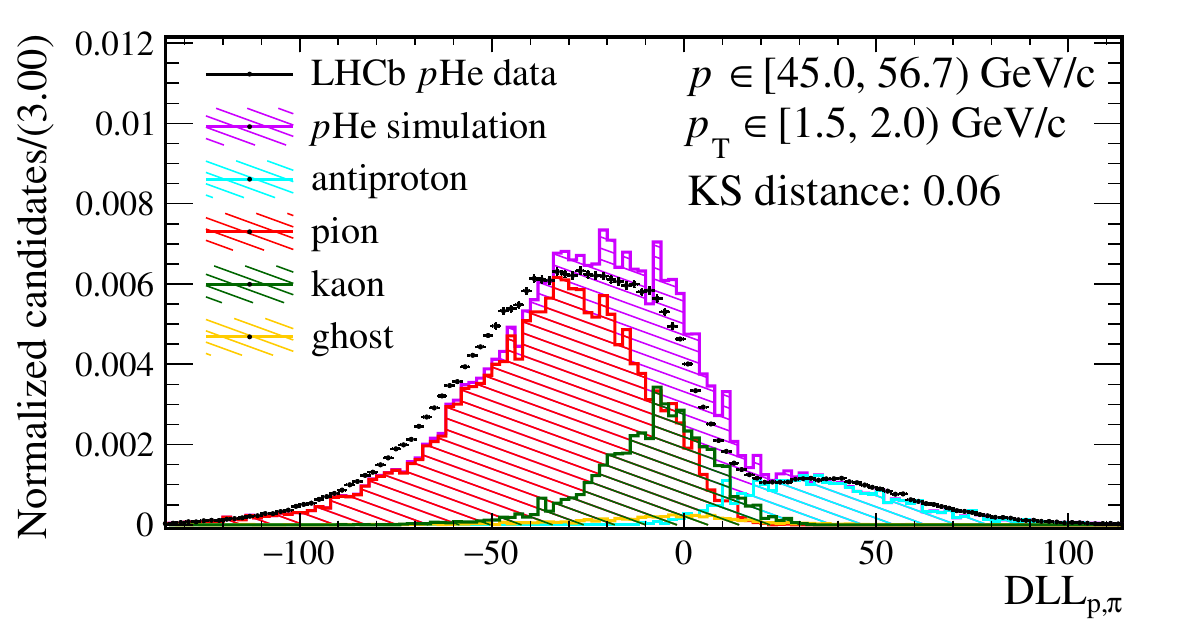} 
\includegraphics[width = 0.480000\textwidth]{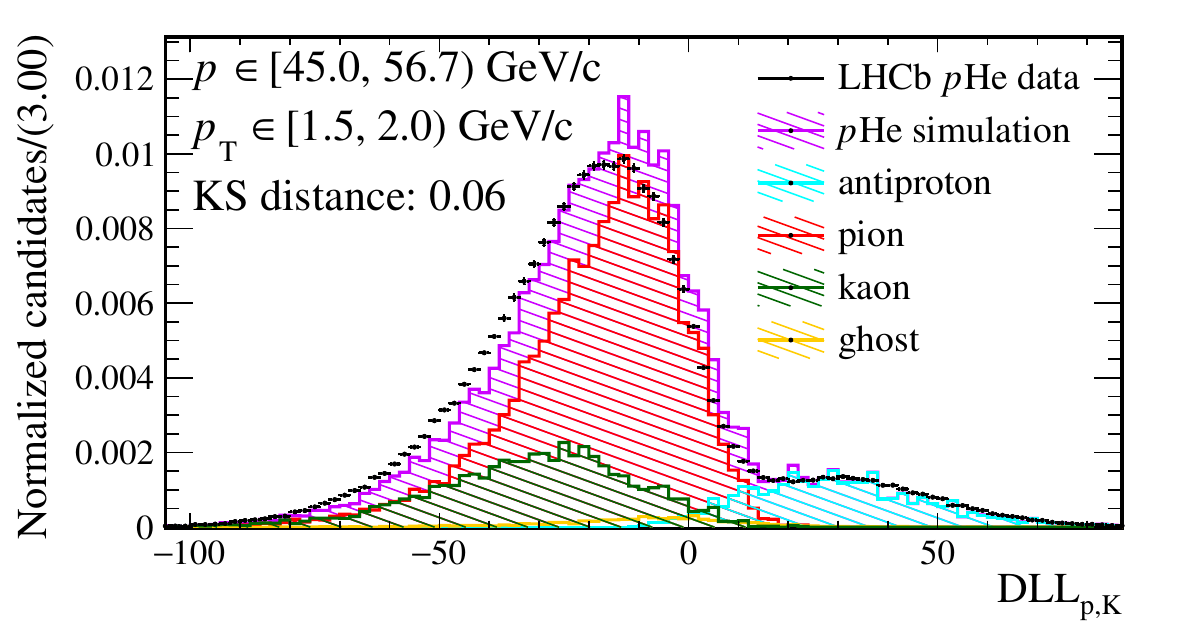} 
\includegraphics[width = 0.480000\textwidth]{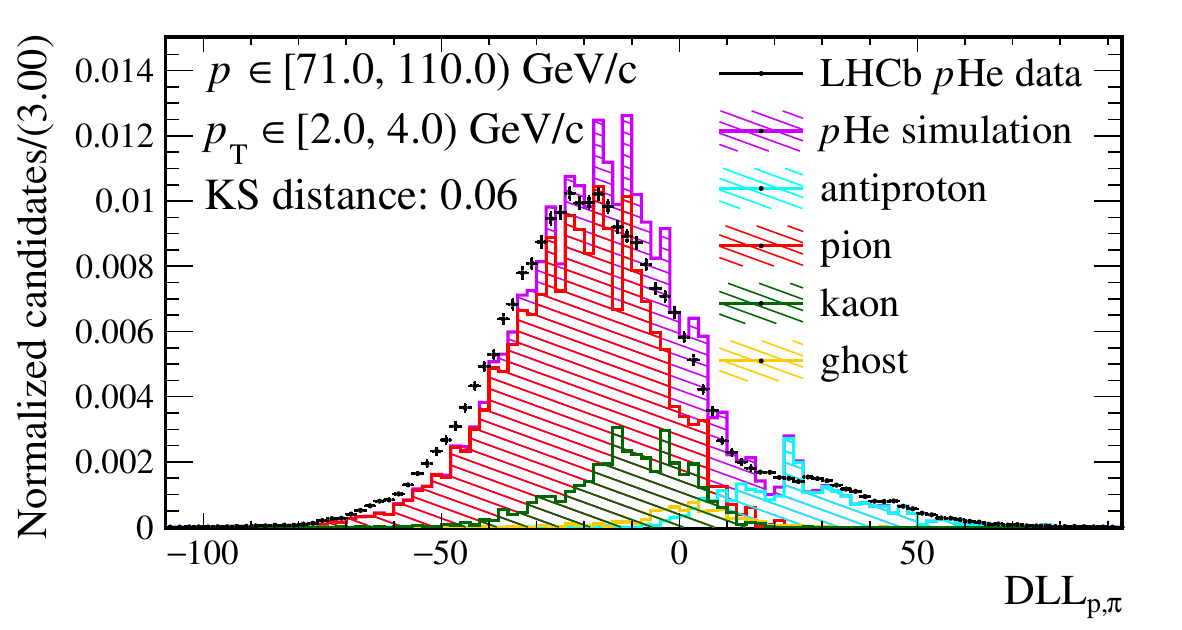} 
\includegraphics[width = 0.480000\textwidth]{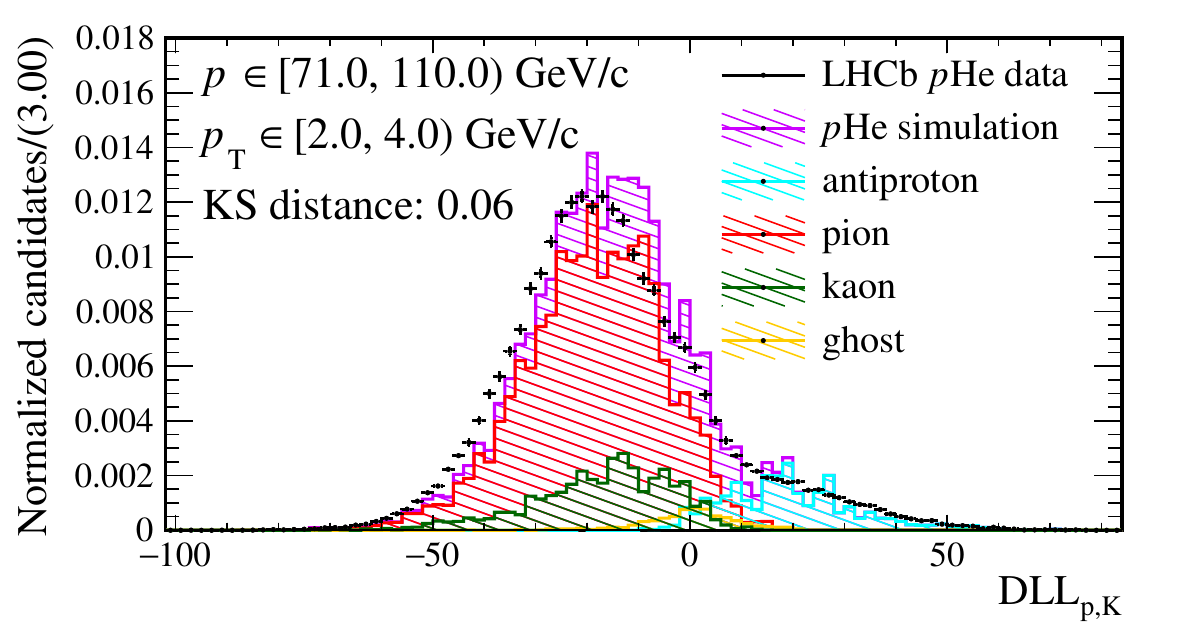} 
\caption{Projections onto the (left) \dllppi and (right) \dllpk axes of the fit to \pHe data employing simulation-based templates in five momentum, transverse momentum intervals.} 
\label{fig_GenpHe:fit_simulated} 
\end{figure}

\begin{figure} 
\centering 
\includegraphics[width = 0.480000\textwidth]{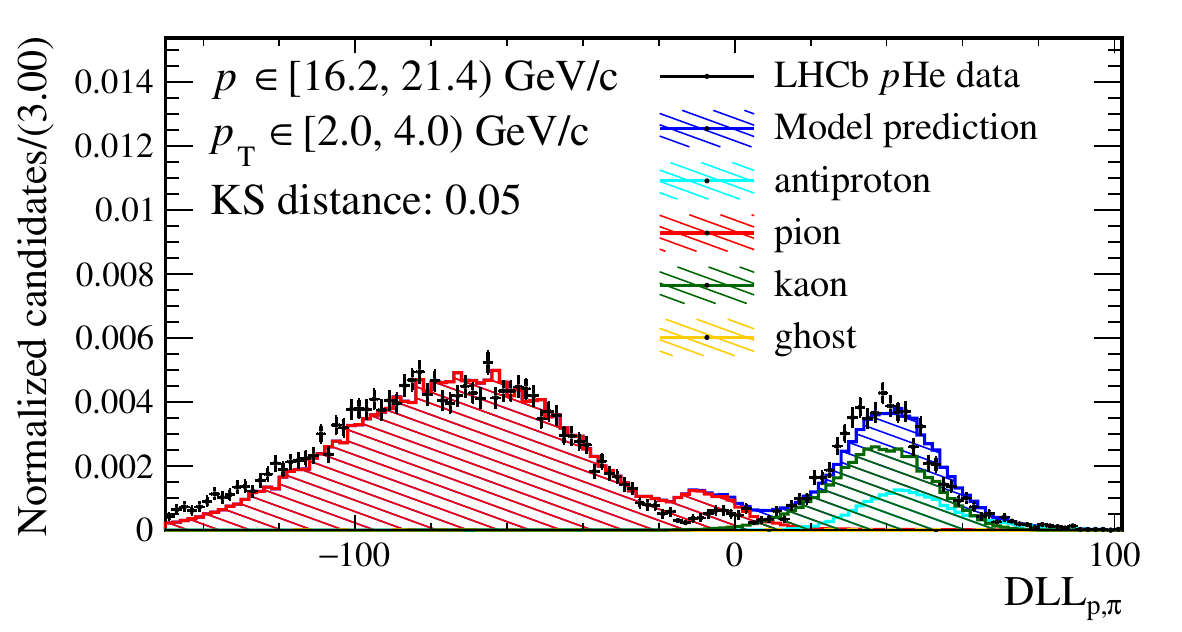} 
\includegraphics[width = 0.480000\textwidth]{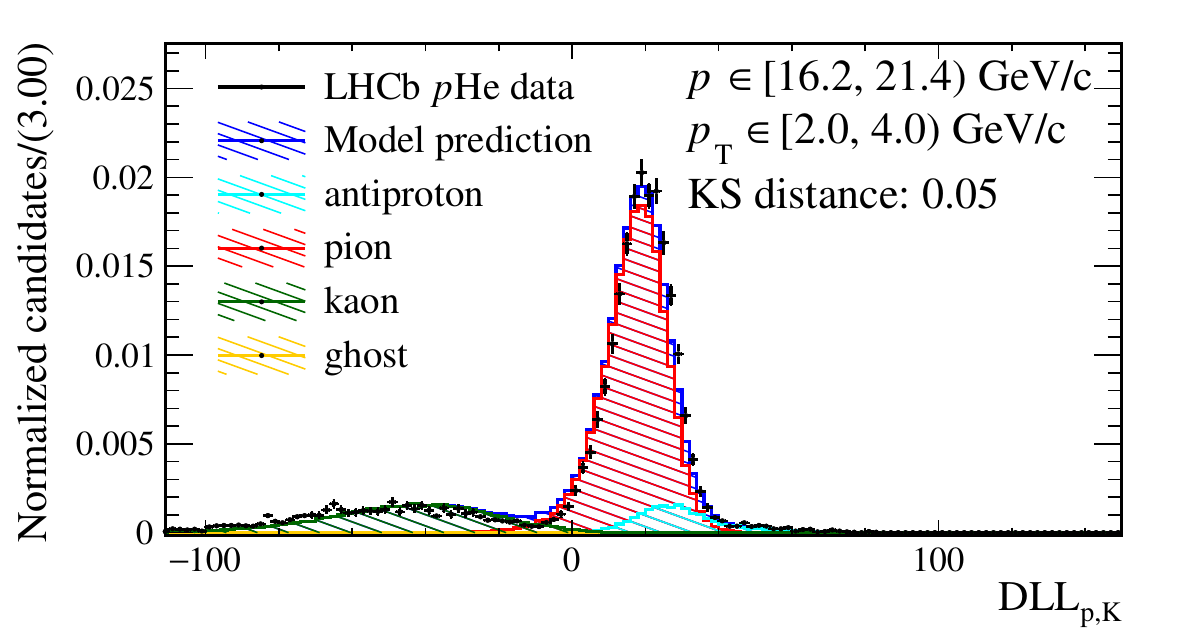} 
\includegraphics[width = 0.480000\textwidth]{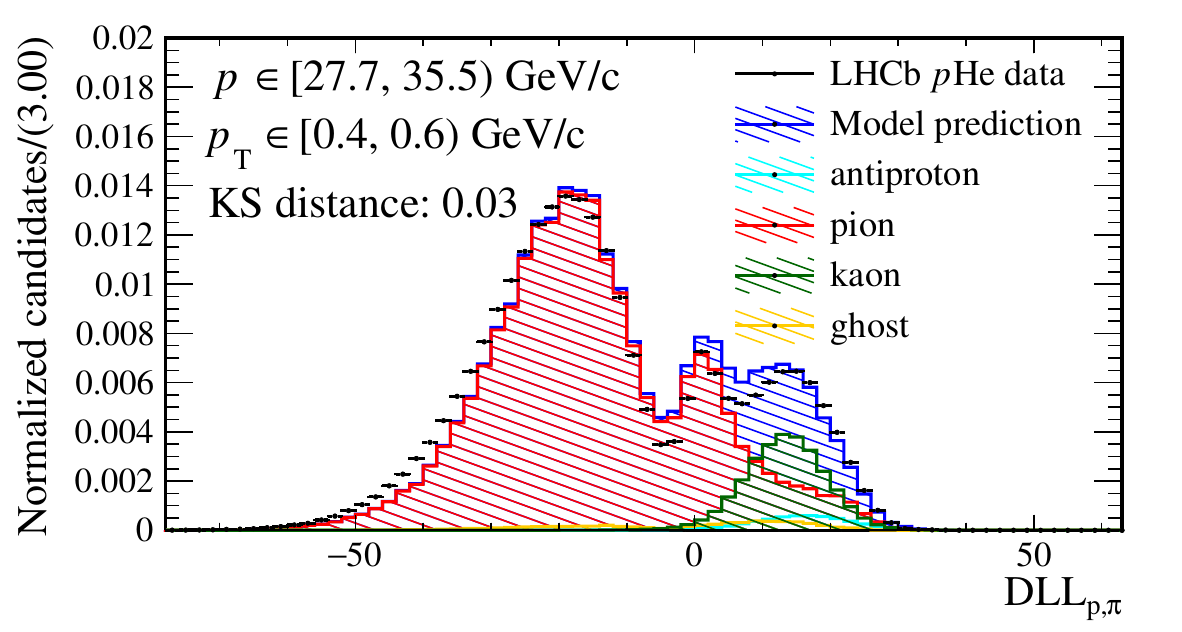} 
\includegraphics[width = 0.480000\textwidth]{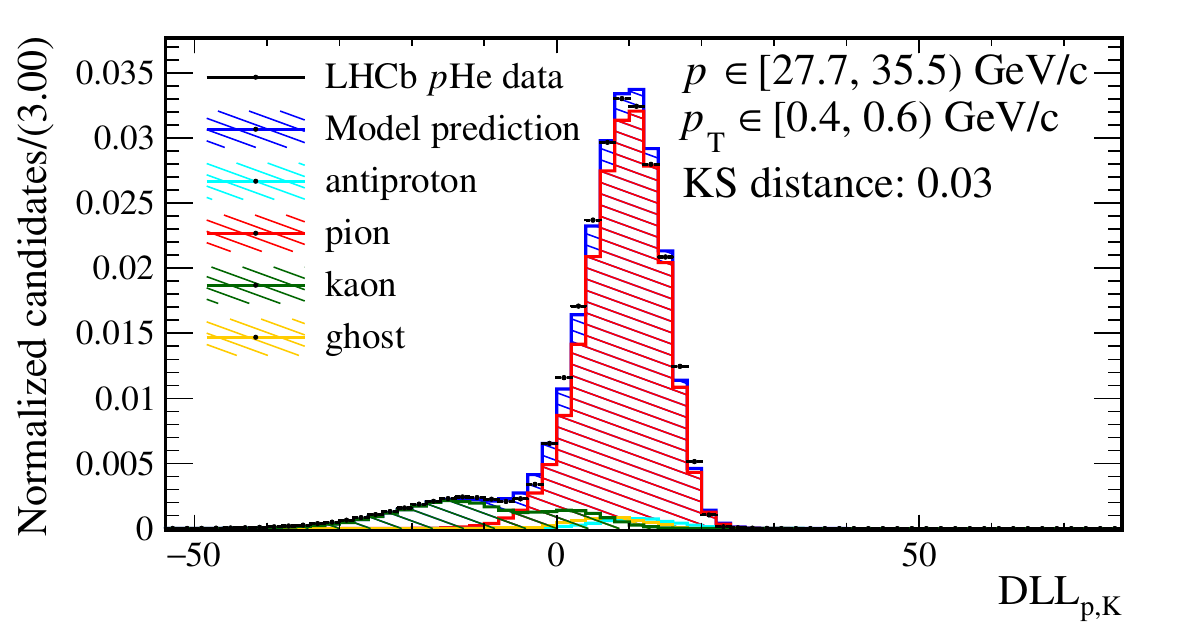} 
\includegraphics[width = 0.480000\textwidth]{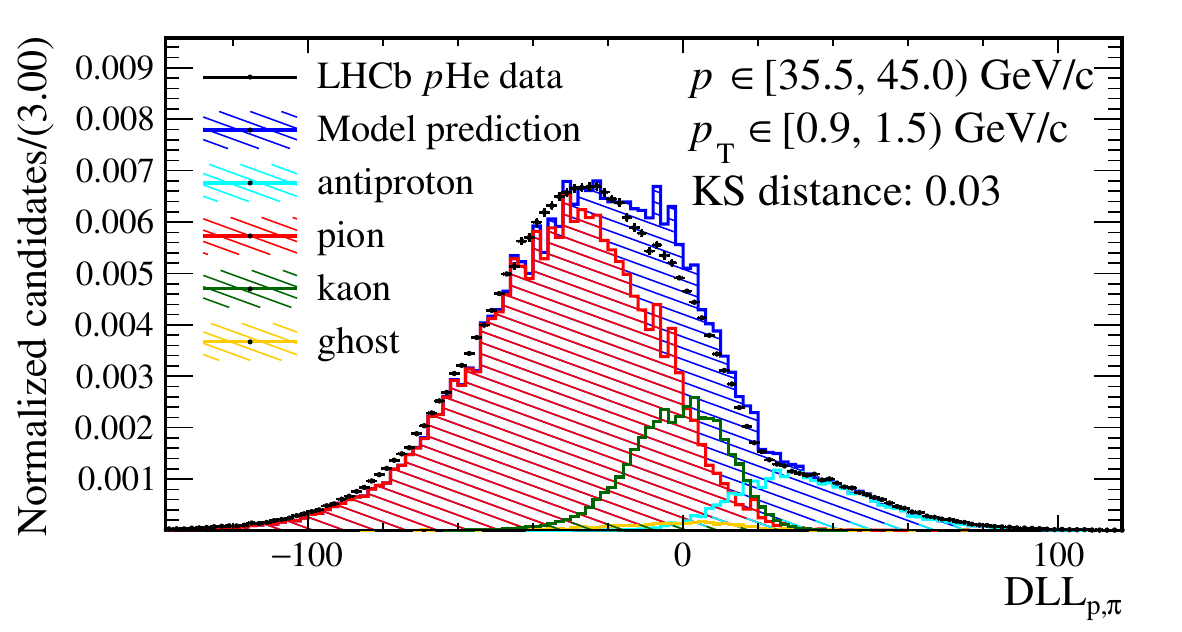} 
\includegraphics[width = 0.480000\textwidth]{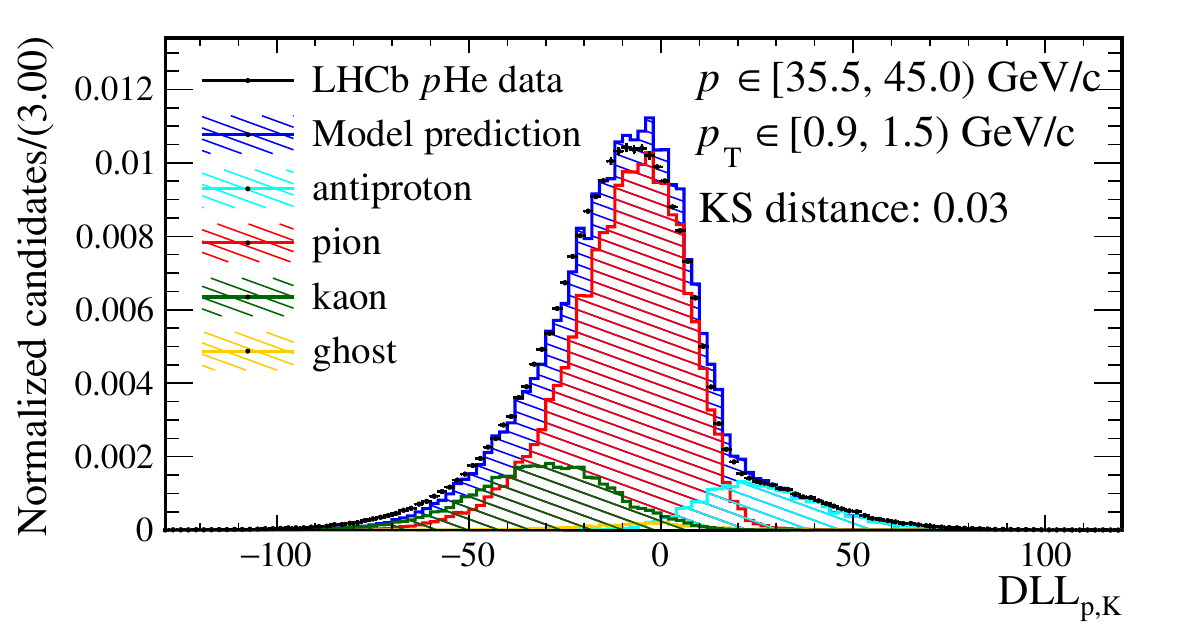} 
\includegraphics[width = 0.480000\textwidth]{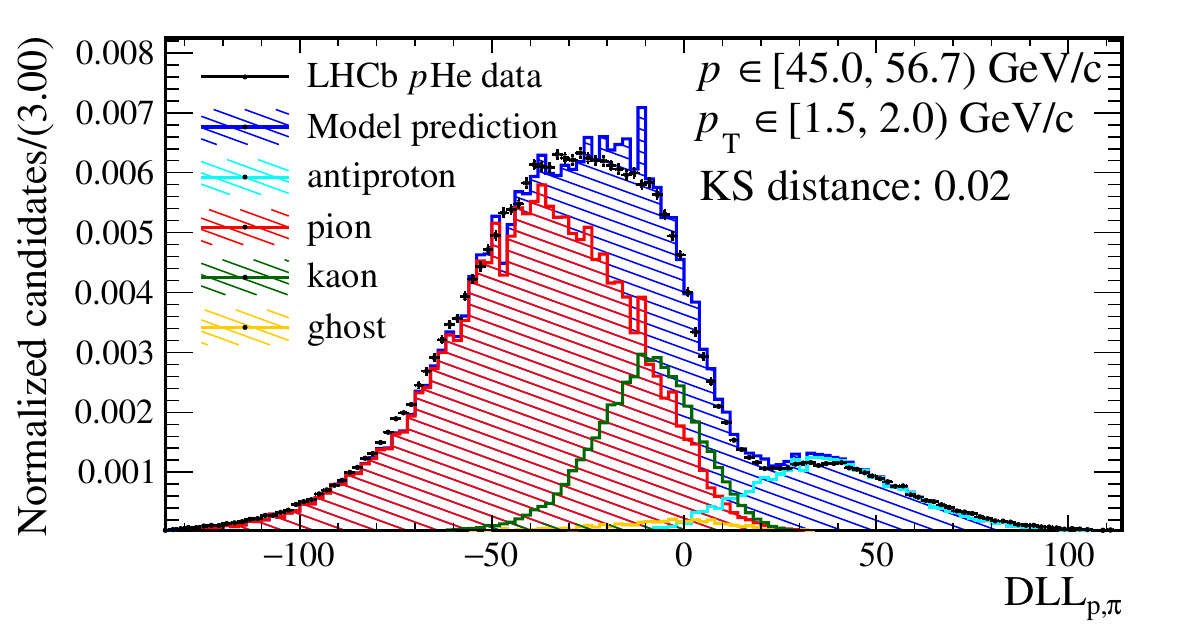} 
\includegraphics[width = 0.480000\textwidth]{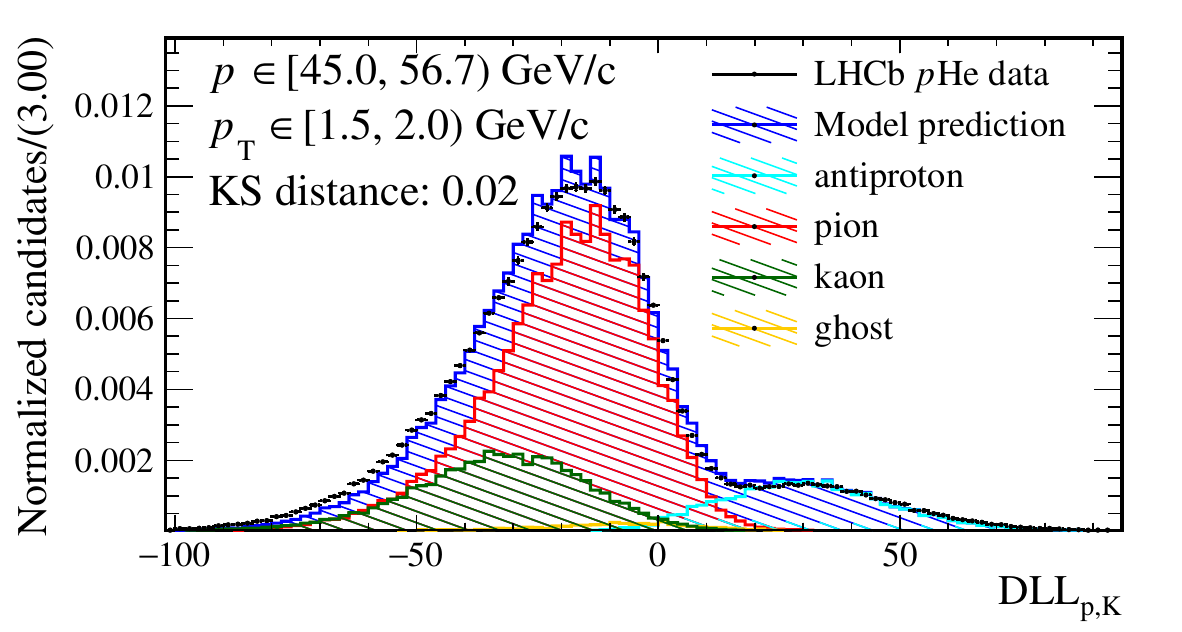} 
\includegraphics[width = 0.480000\textwidth]{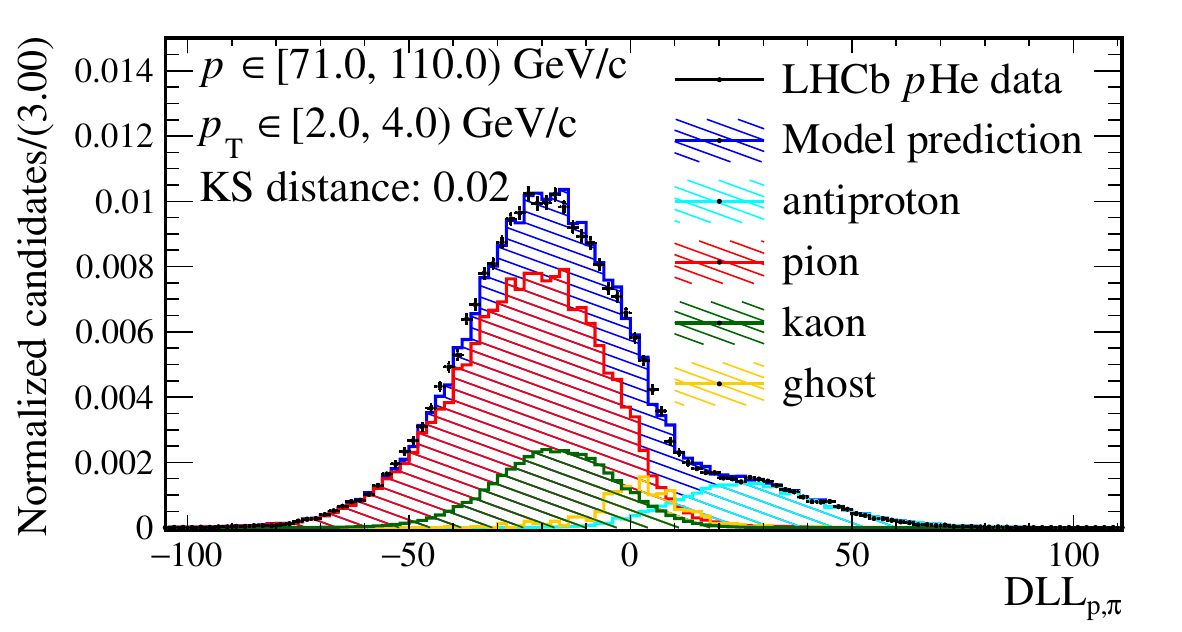} 
\includegraphics[width = 0.480000\textwidth]{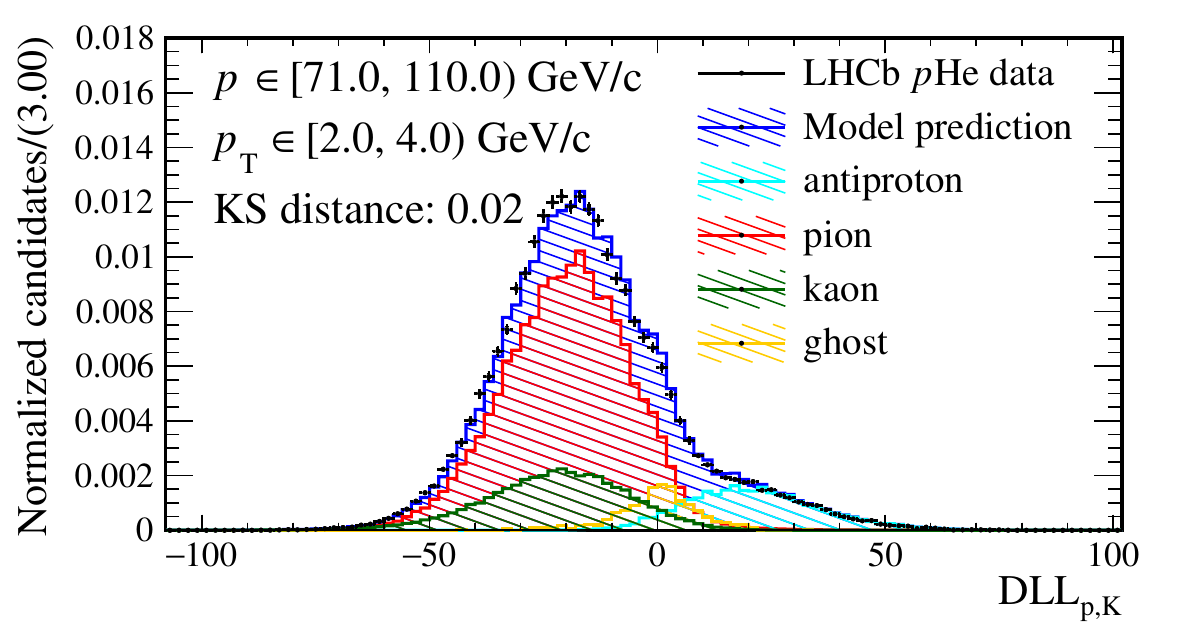} 
\caption{Projections onto the (left) \dllppi and (right) \dllpk axes of the fit to \pHe data employing the data-driven templates in five momentum, transverse momentum intervals.}
\label{fig_GenpHe:fit_TF} 
\end{figure}

\begin{figure} 
\centering 
\includegraphics[width = 0.480000\textwidth]{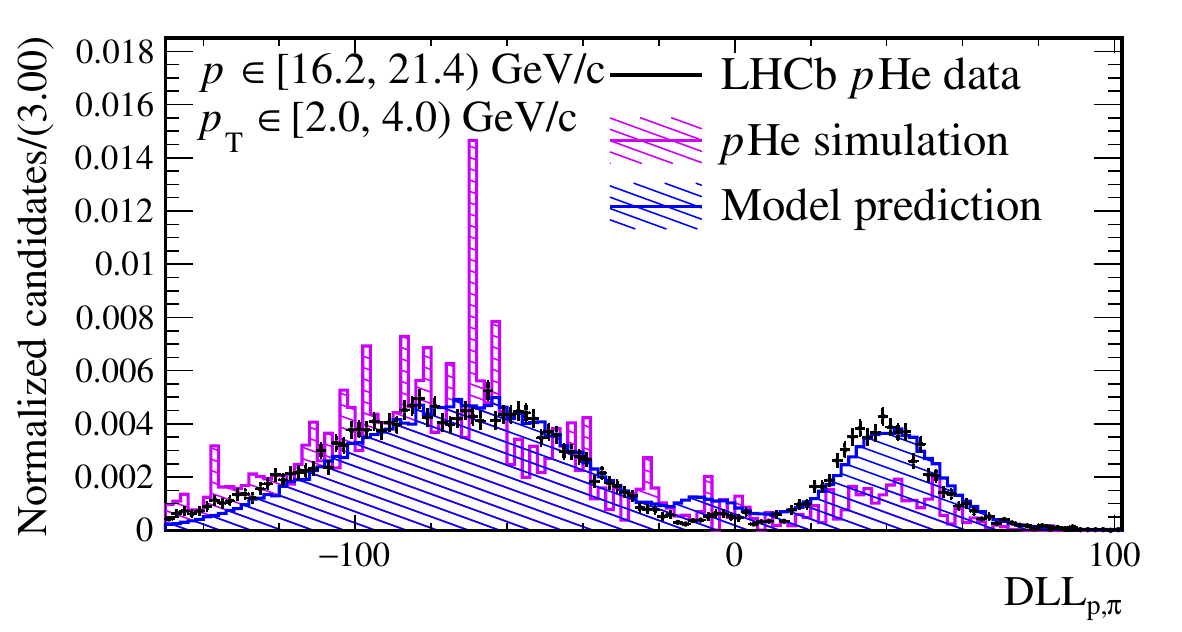} 
\includegraphics[width = 0.480000\textwidth]{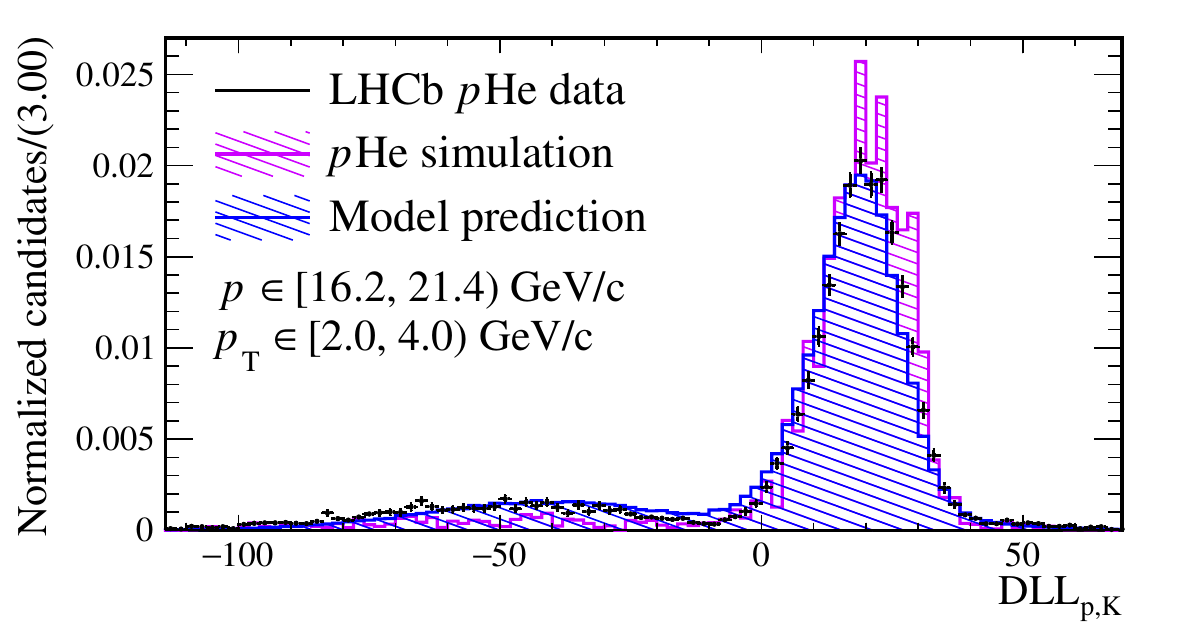} 
\includegraphics[width = 0.480000\textwidth]{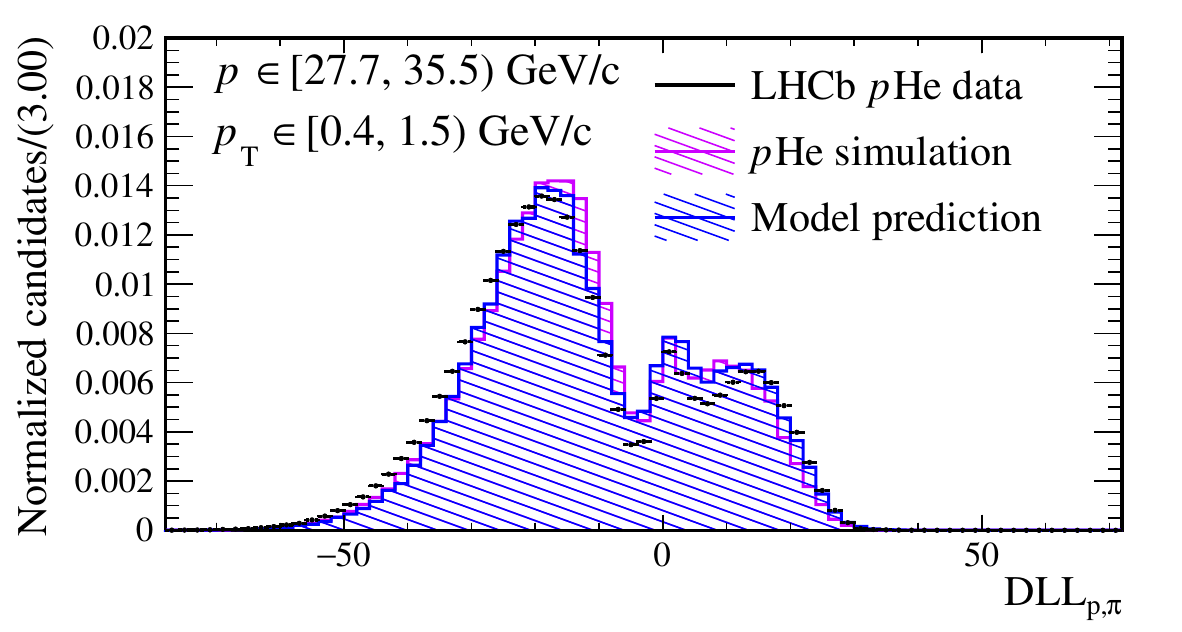} 
\includegraphics[width = 0.480000\textwidth]{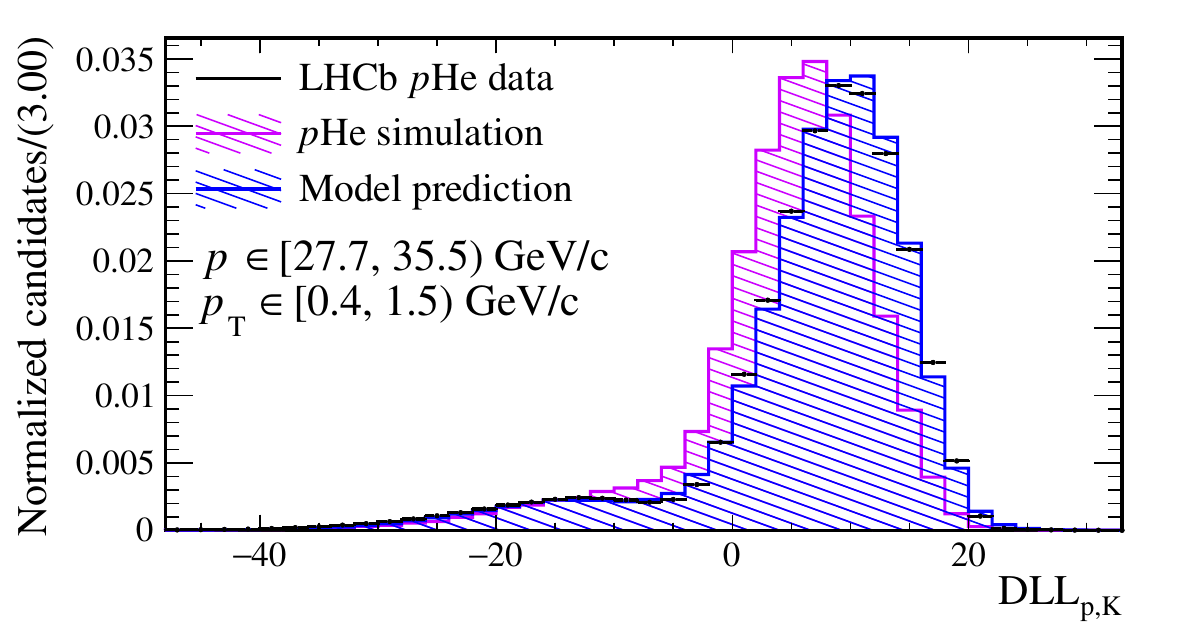} 
\includegraphics[width = 0.480000\textwidth]{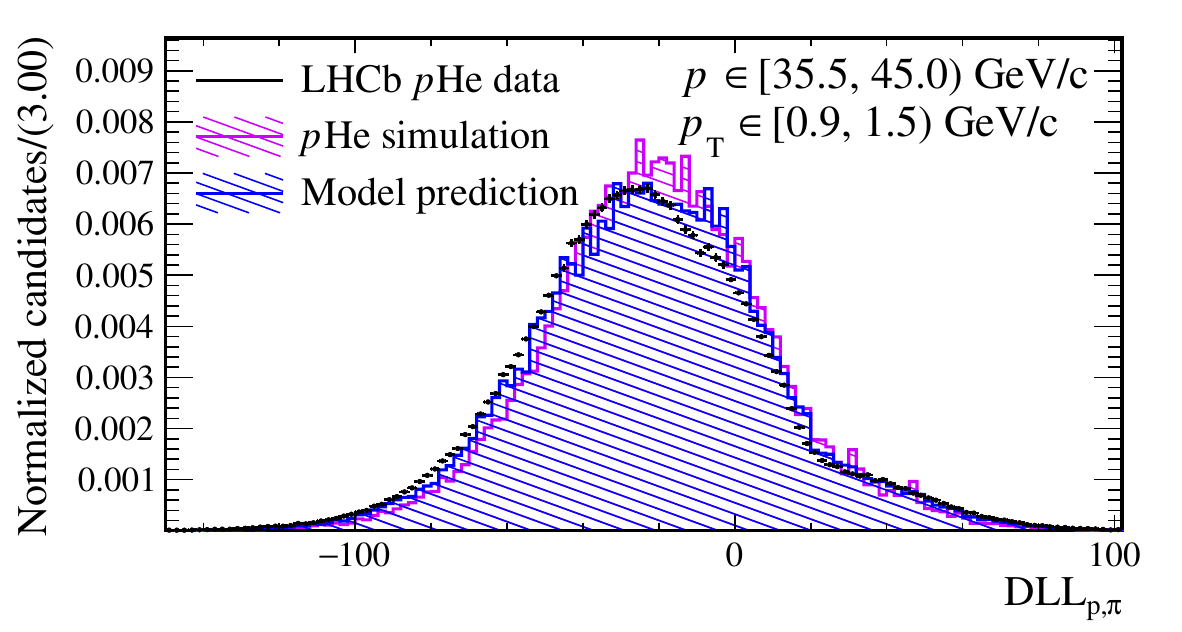} 
\includegraphics[width = 0.480000\textwidth]{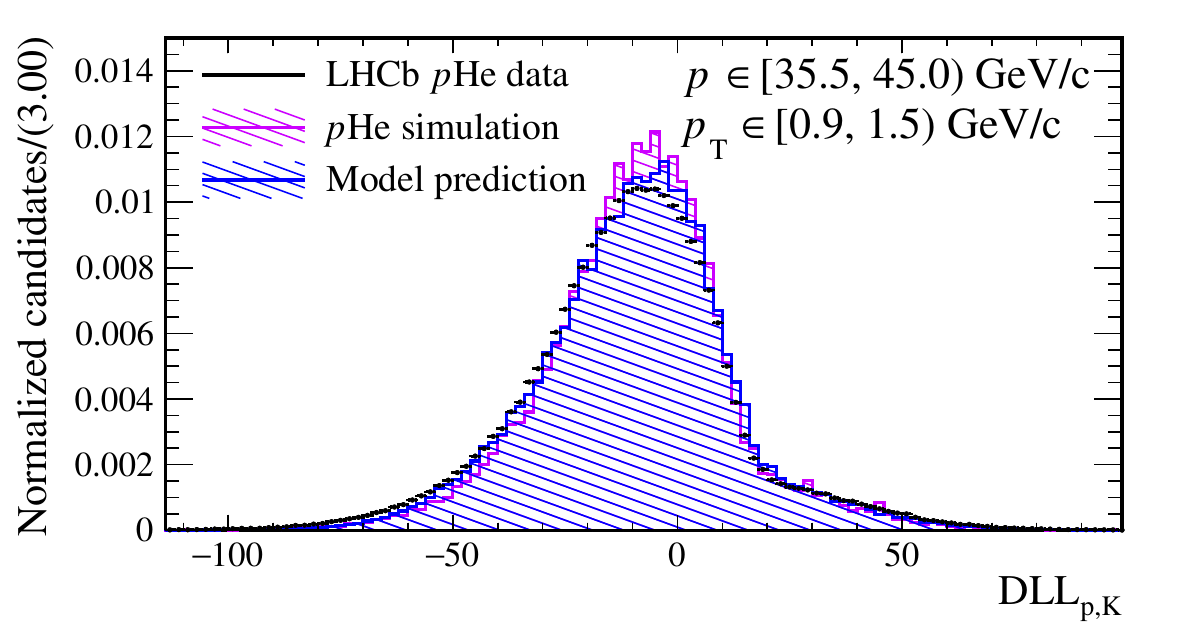} 
\includegraphics[width = 0.480000\textwidth]{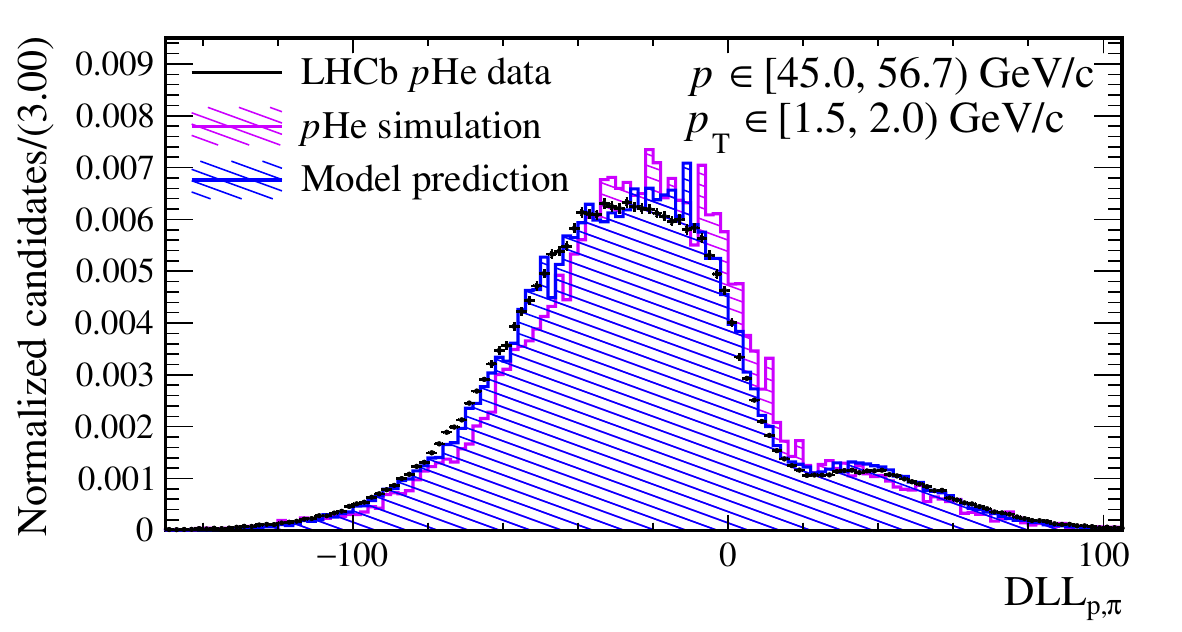} 
\includegraphics[width = 0.480000\textwidth]{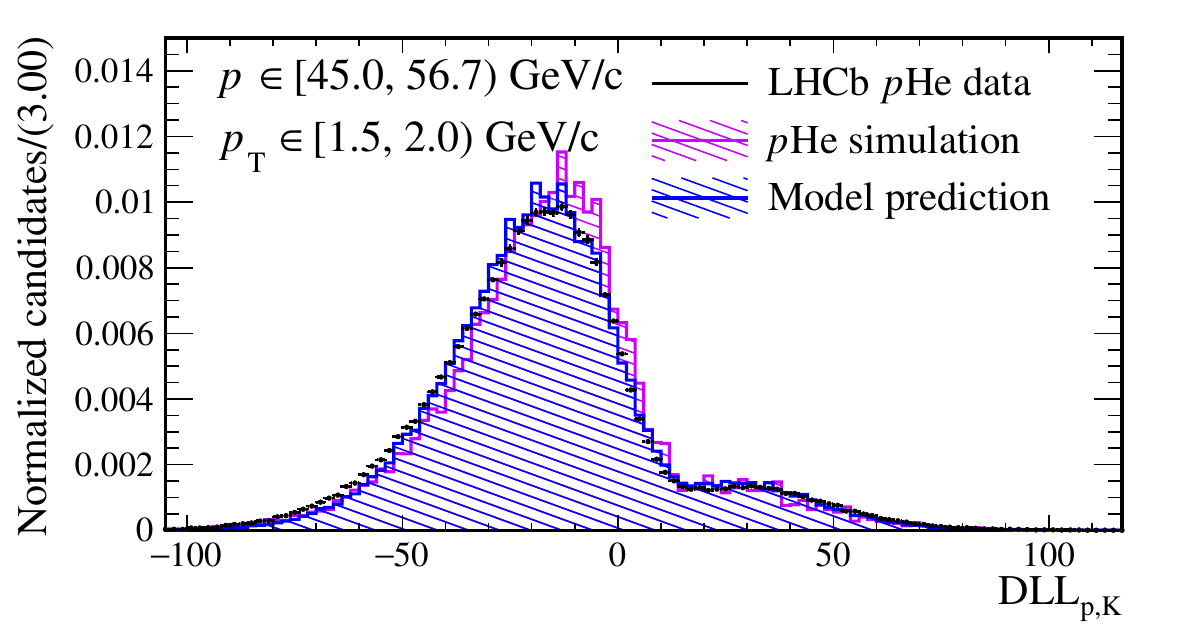} 
\includegraphics[width = 0.480000\textwidth]{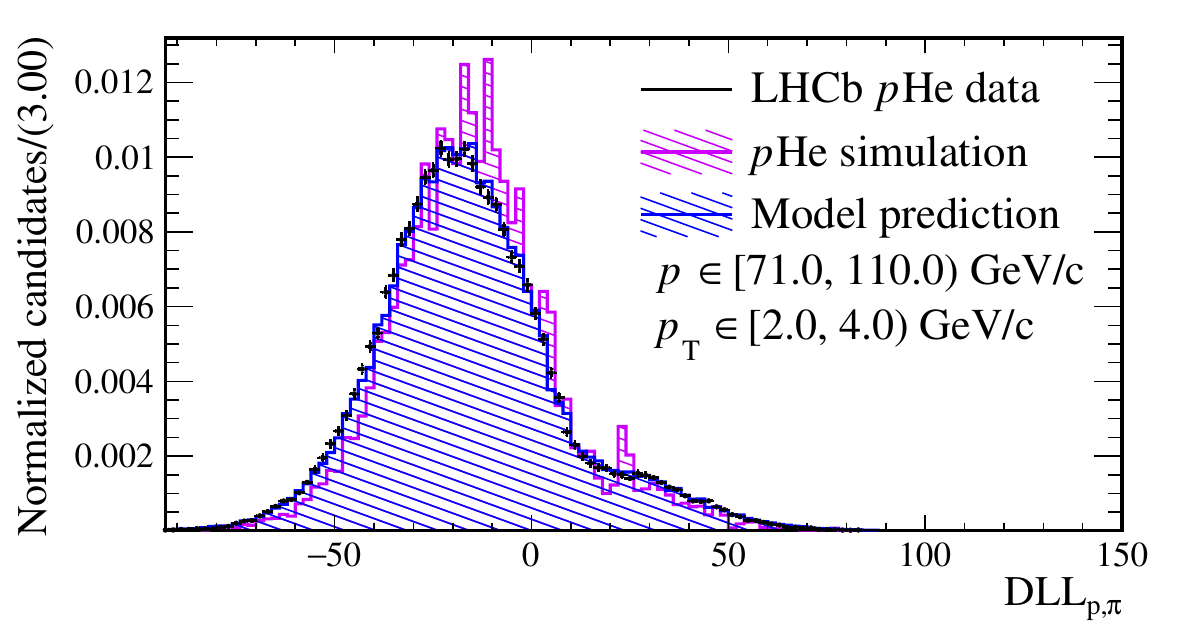} 
\includegraphics[width = 0.480000\textwidth]{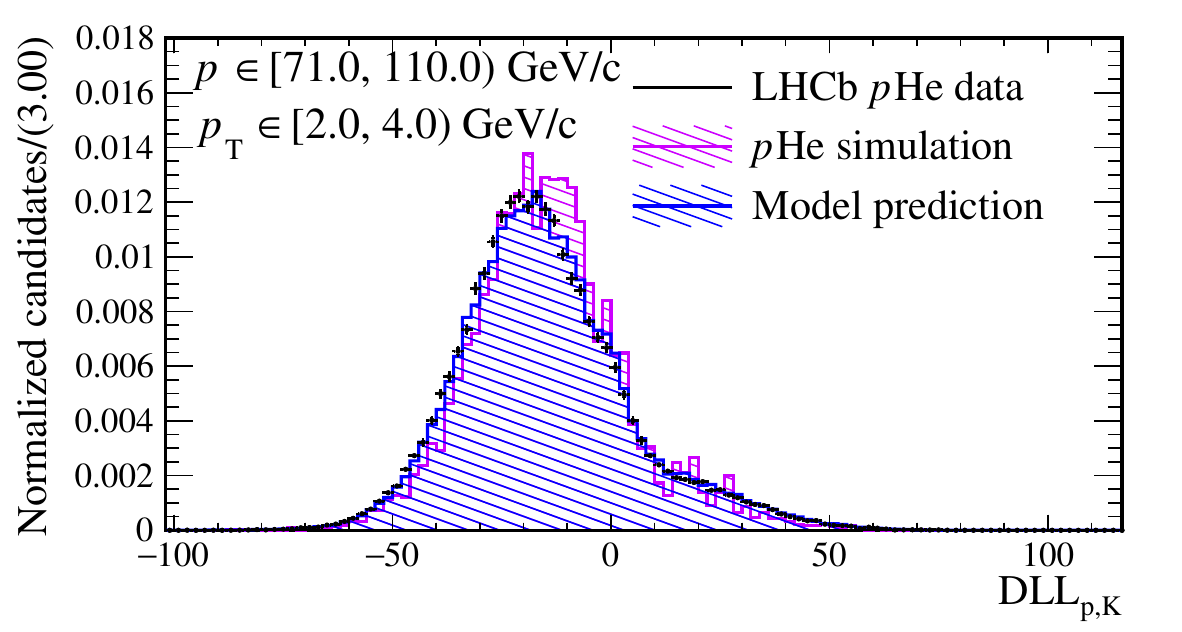} 
\caption{Comparison of the (left) \dllppi and (right) \dllpk distributions in \pHe data modelled with simulation (violet) and data-based (blue) templates in five momentum, transverse momentum intervals.} 
\label{fig_GenpHe:fit_Comparison} 
\end{figure}

\begin{figure} [h]
\centering 
\includegraphics[width = 0.480000\textwidth]{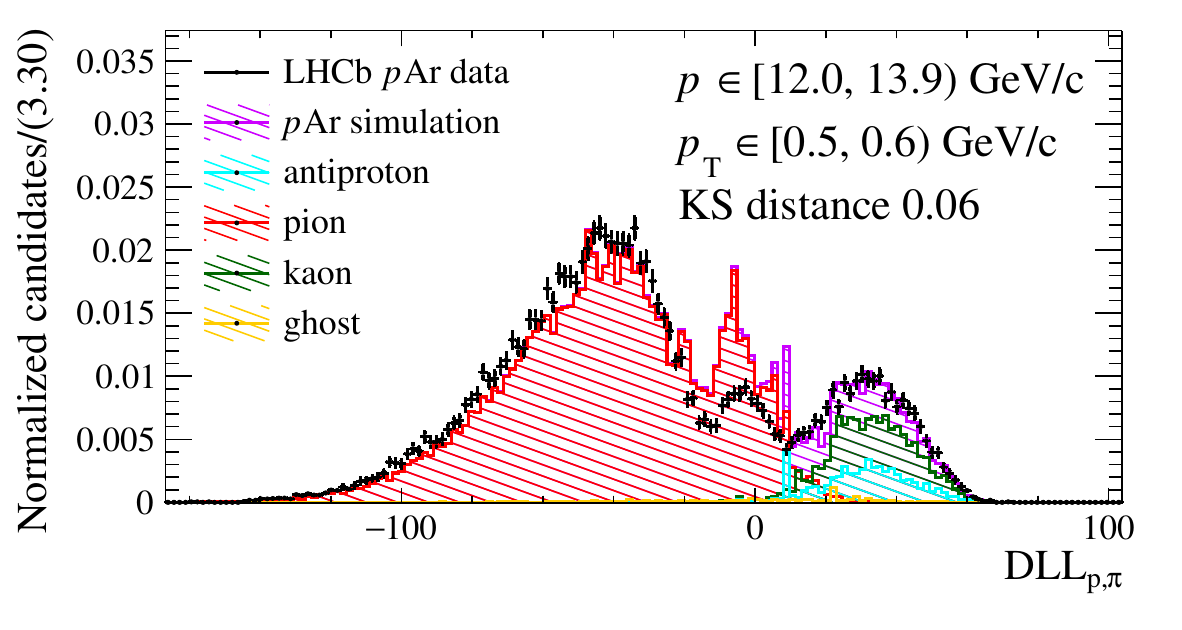} 
\includegraphics[width = 0.480000\textwidth]{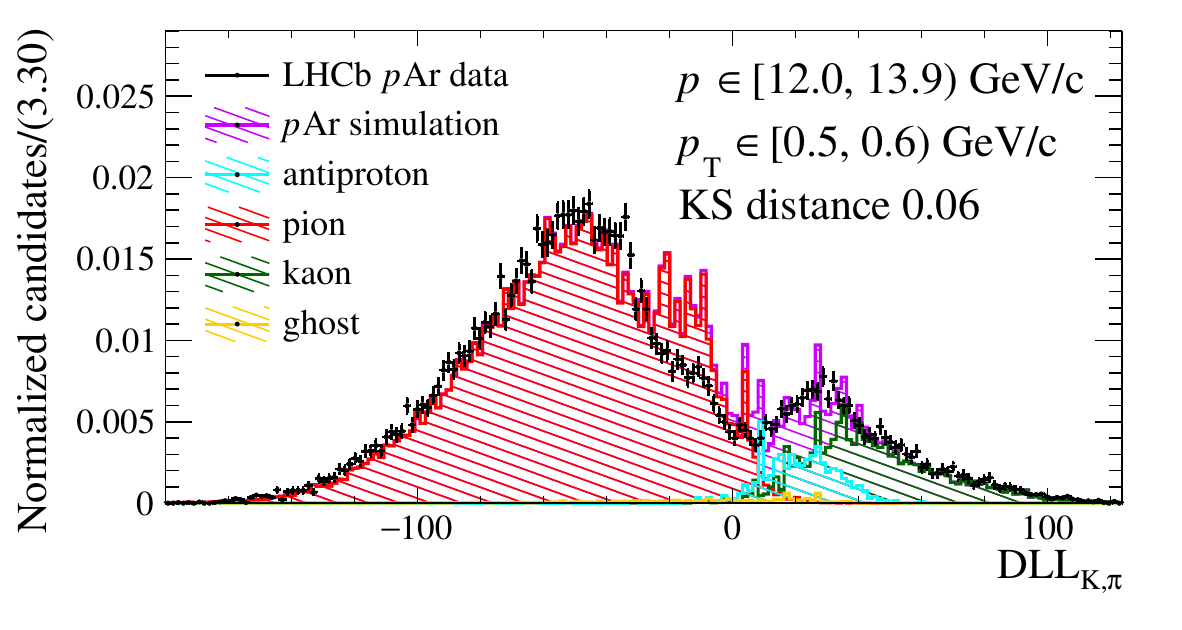} 
\includegraphics[width = 0.480000\textwidth]{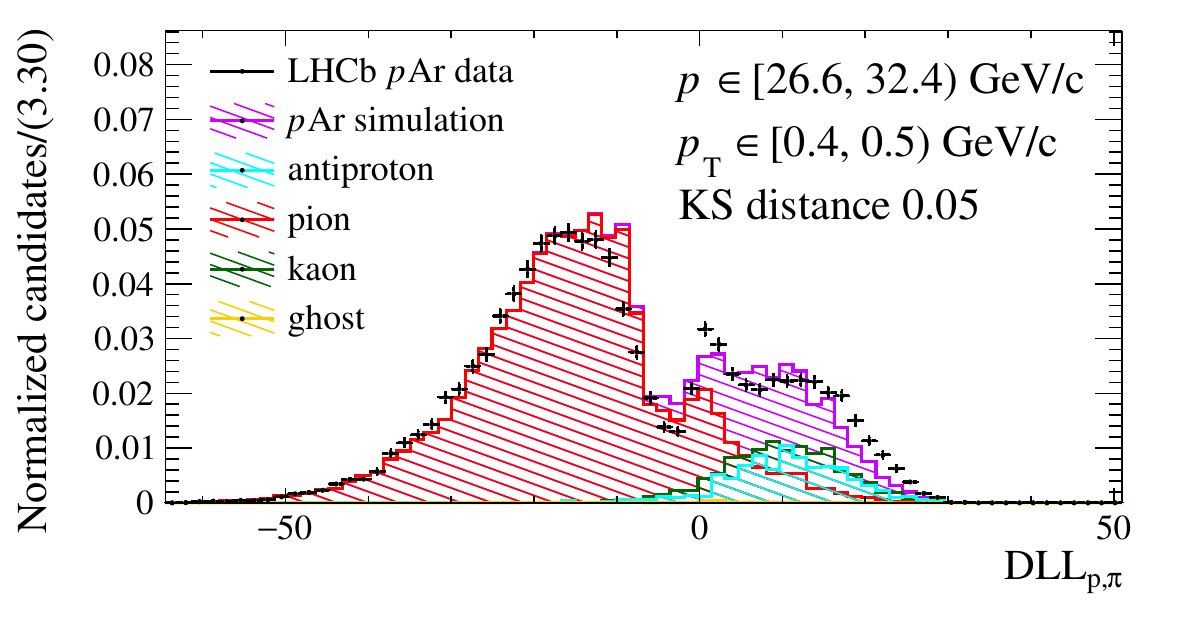} 
\includegraphics[width = 0.480000\textwidth]{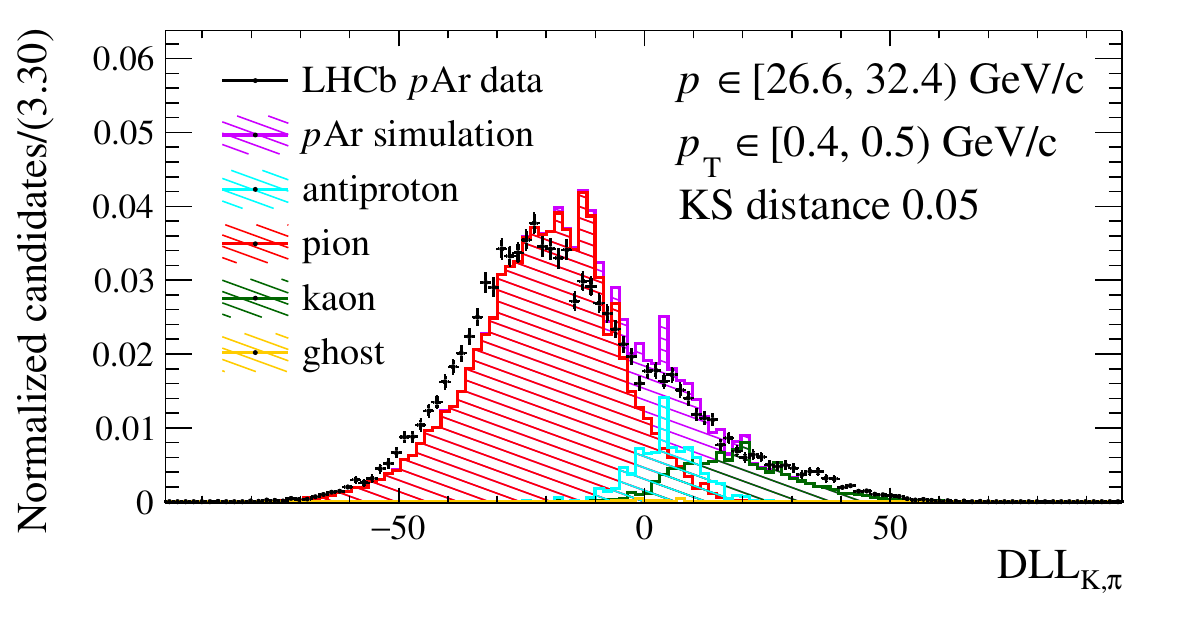} 
\includegraphics[width = 0.480000\textwidth]{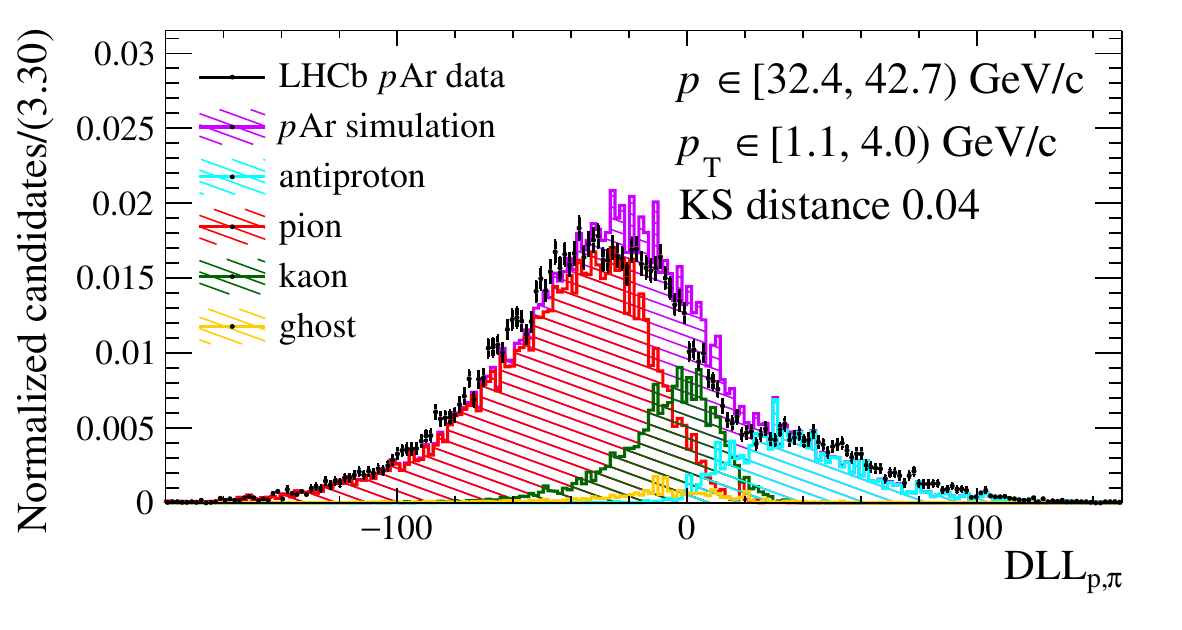} 
\includegraphics[width = 0.480000\textwidth]{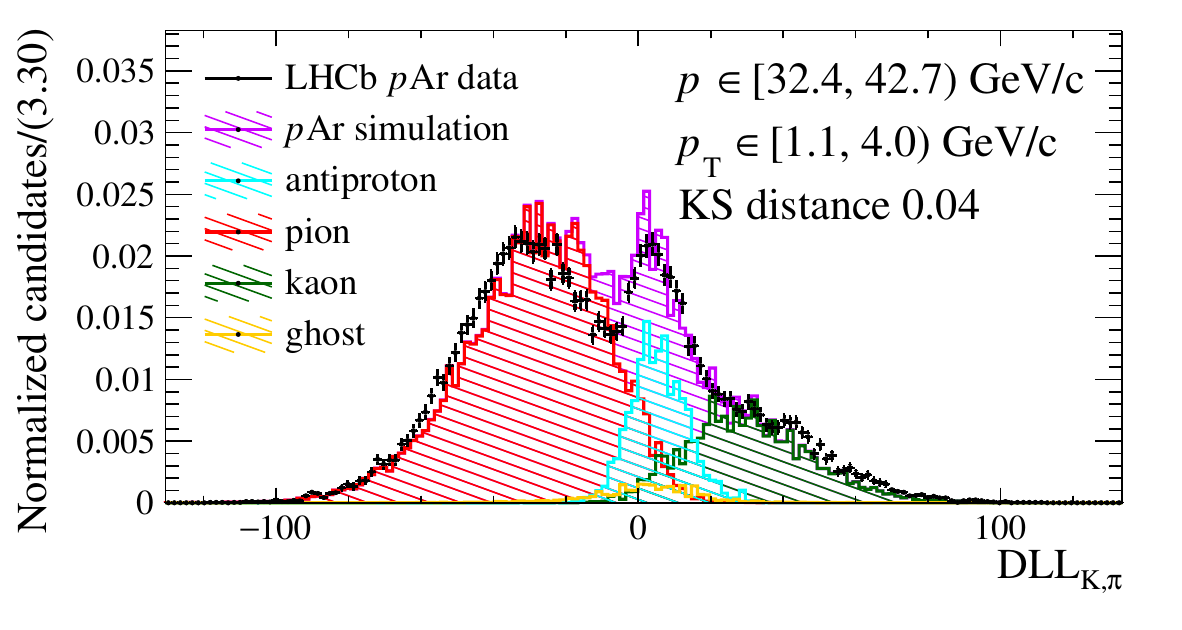} 
\includegraphics[width = 0.480000\textwidth]{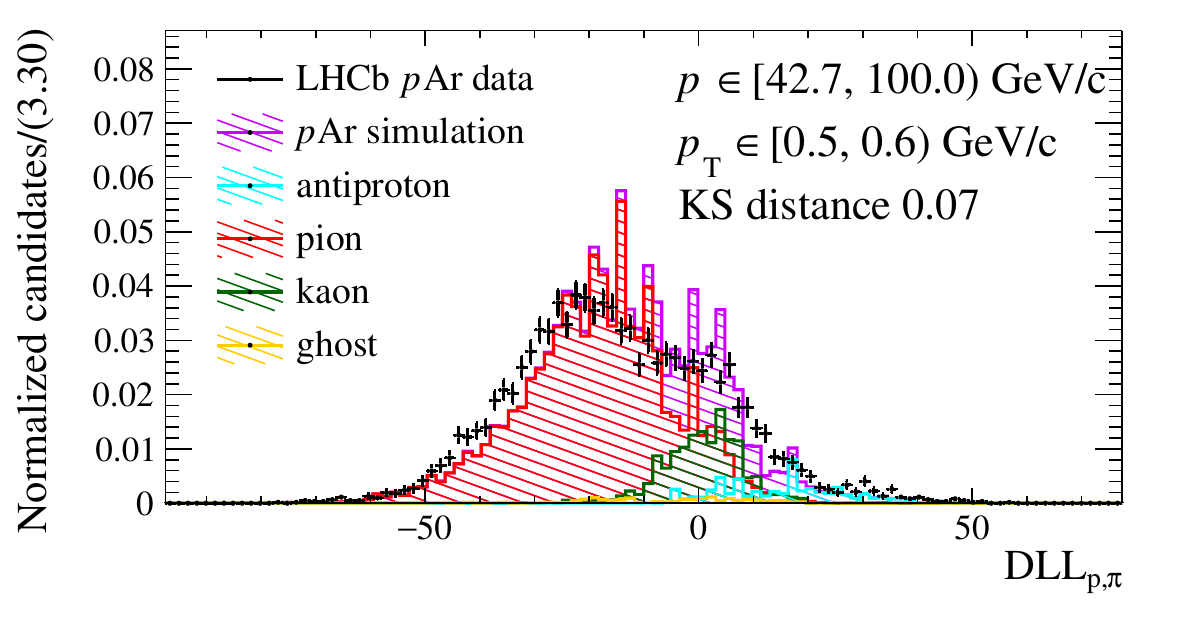} 
\includegraphics[width = 0.480000\textwidth]{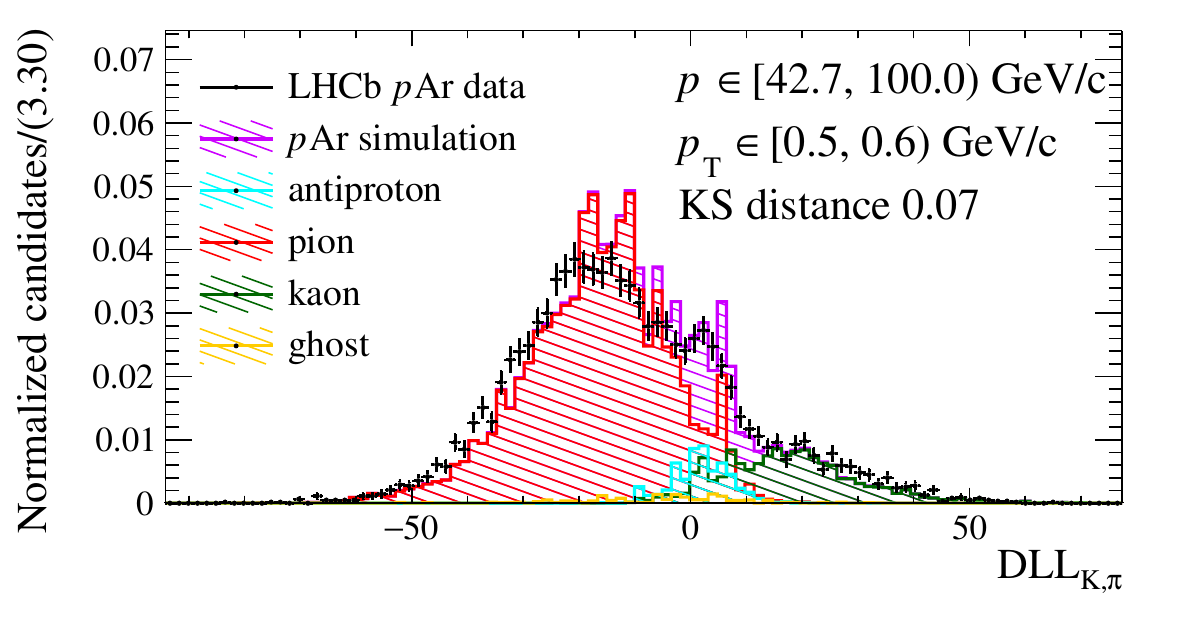} 
\includegraphics[width = 0.480000\textwidth]{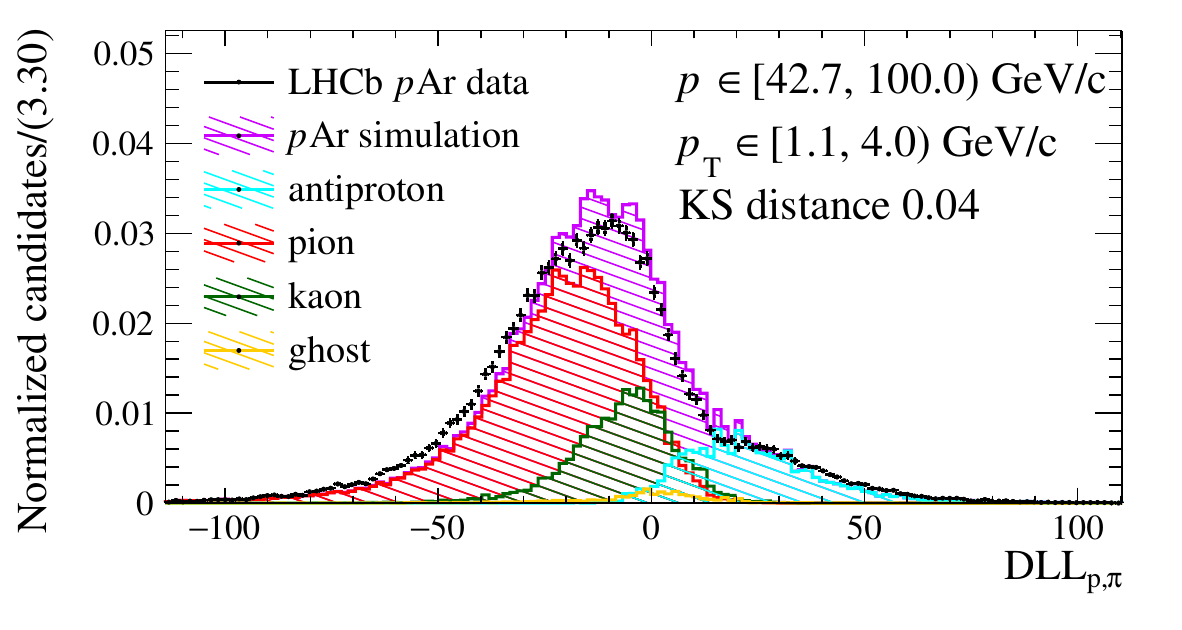} 
\includegraphics[width = 0.480000\textwidth]{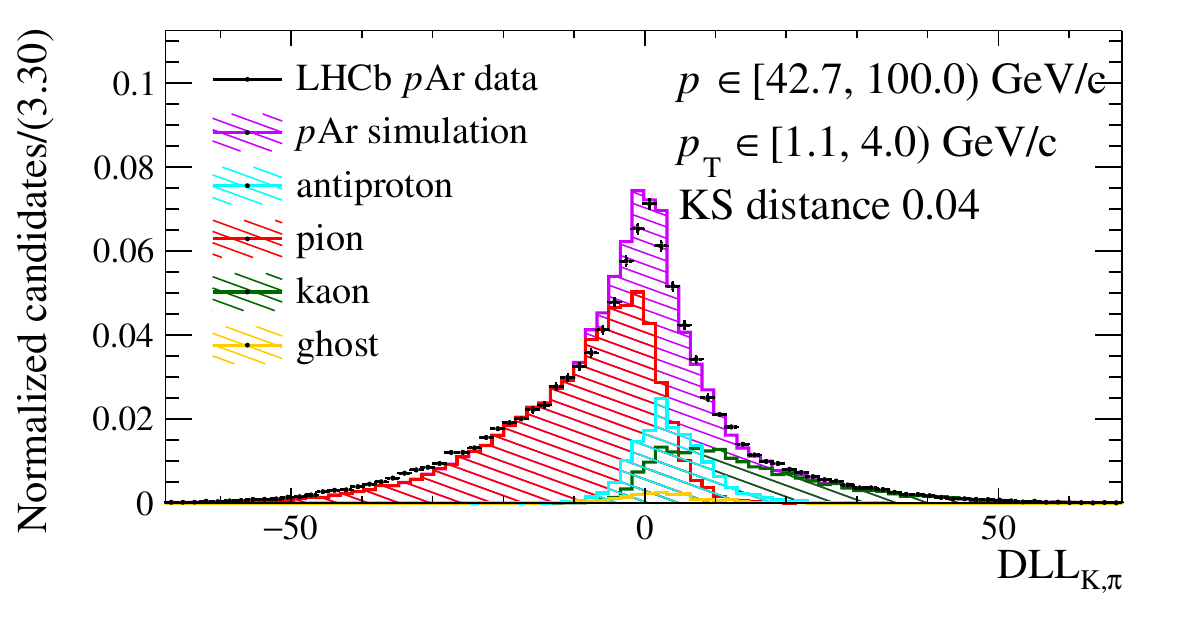} 
\caption{Projections onto the (left) \dllppi and (right) \dllpk axes of the fit to \pAr data employing simulation-based templates for five kinematic bins.} 
\label{fig_ar:fit_simulated_ar} 
\end{figure}

\begin{figure} [h]
\centering 
\includegraphics[width = 0.480000\textwidth]{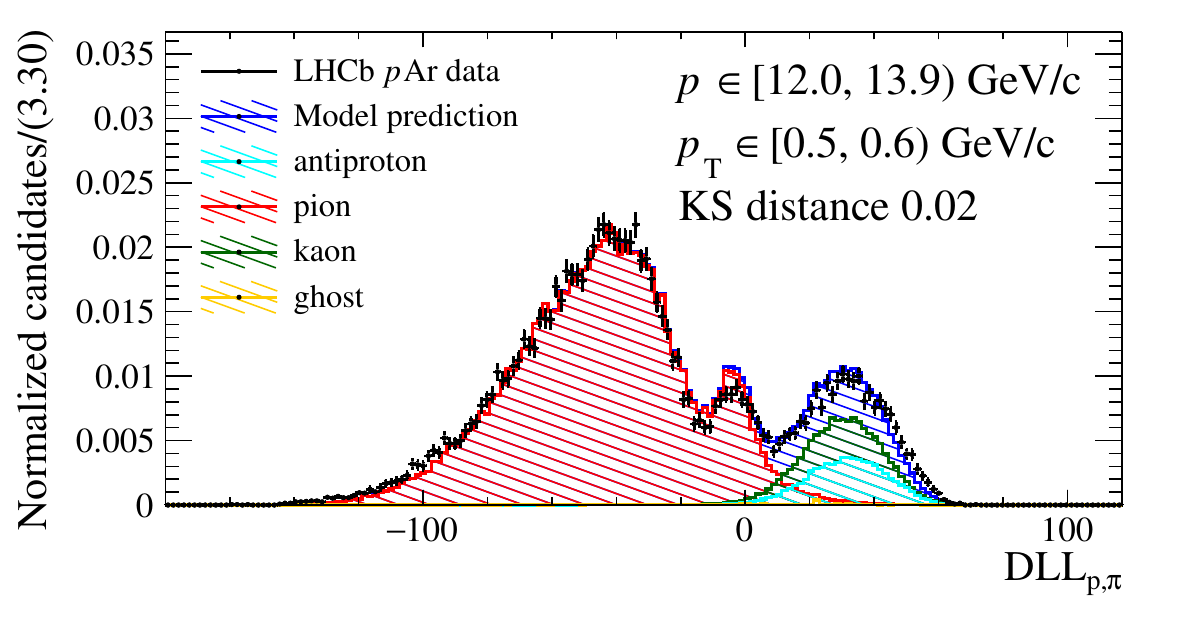} 
\includegraphics[width = 0.480000\textwidth]{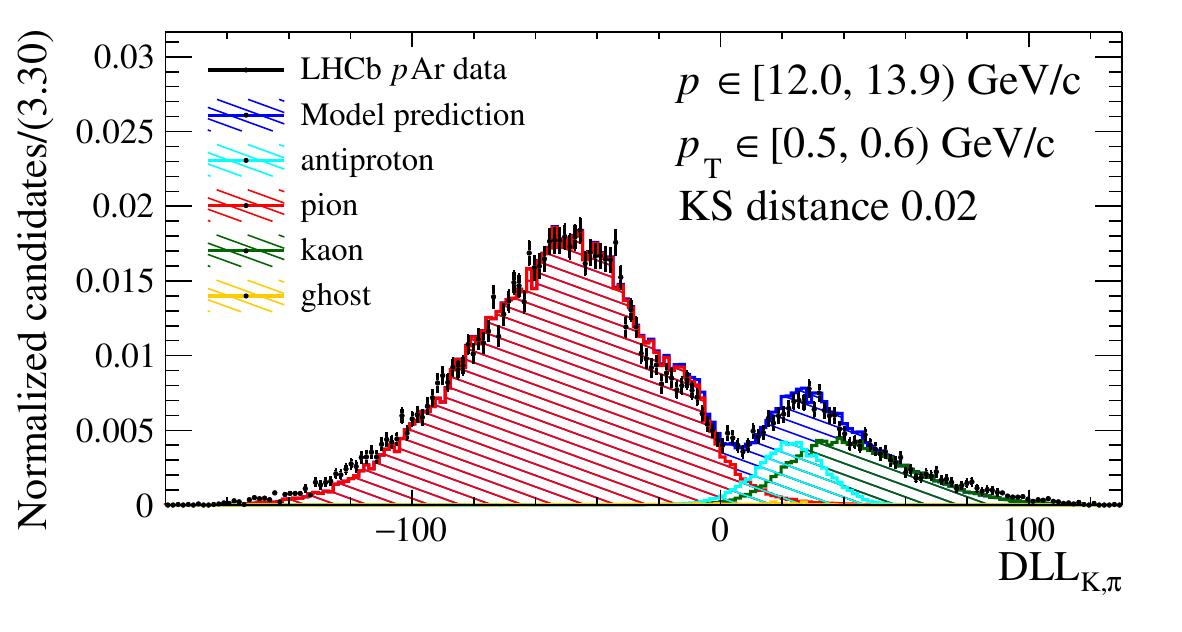} 
\includegraphics[width = 0.480000\textwidth]{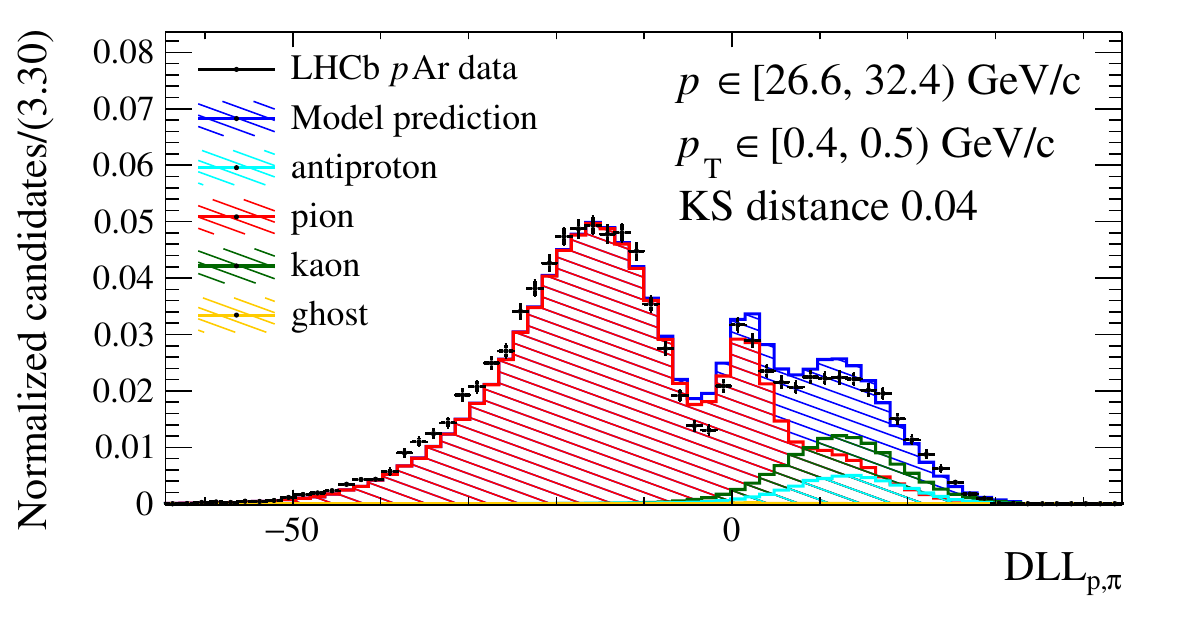} 
\includegraphics[width = 0.480000\textwidth]{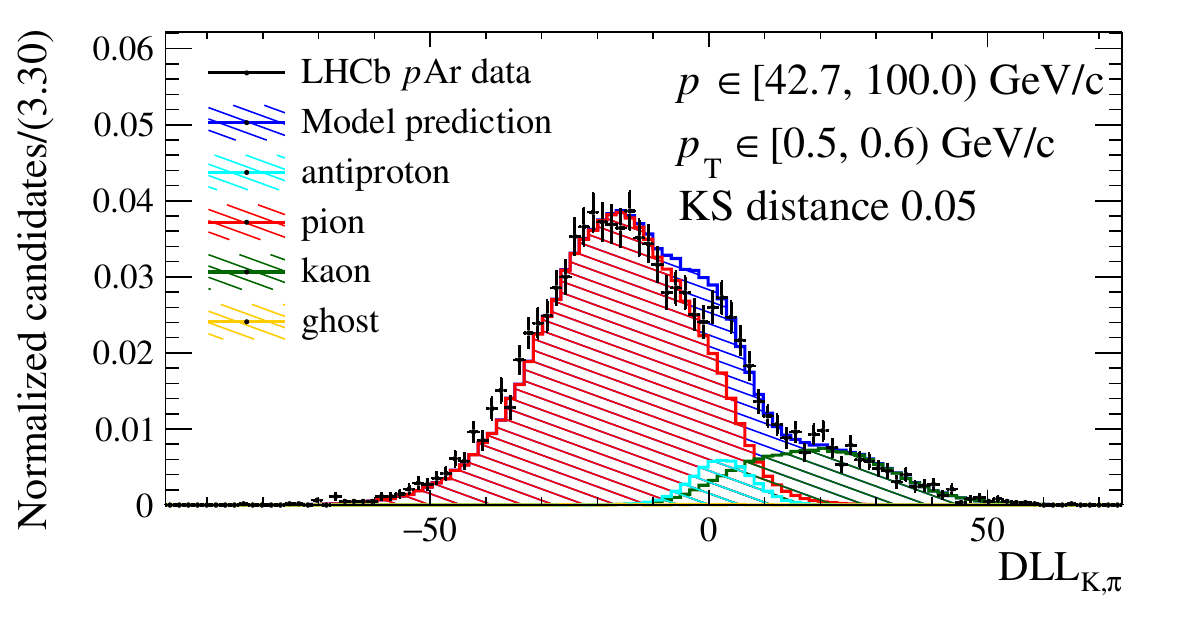} 
\includegraphics[width = 0.480000\textwidth]{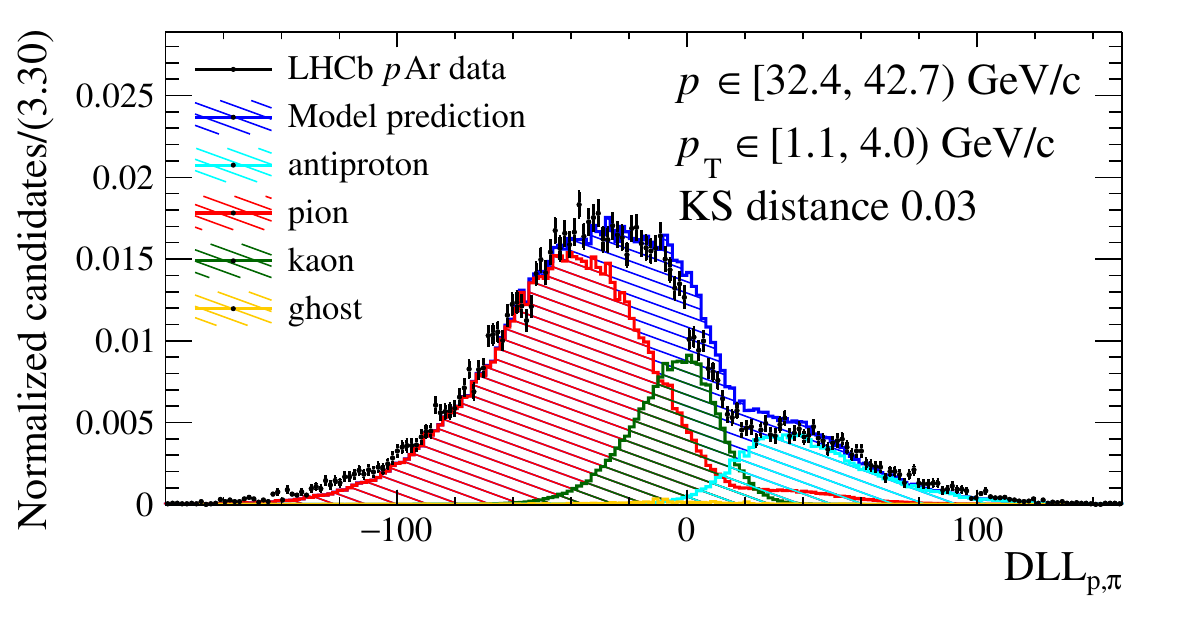} 
\includegraphics[width = 0.480000\textwidth]{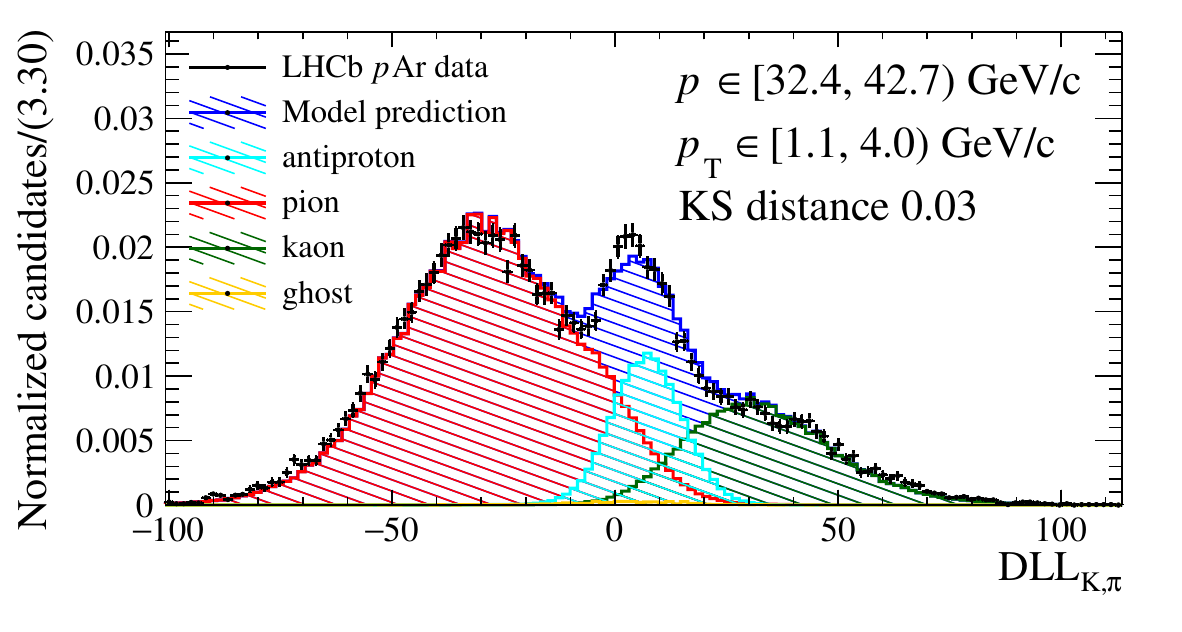} 
\includegraphics[width = 0.480000\textwidth]{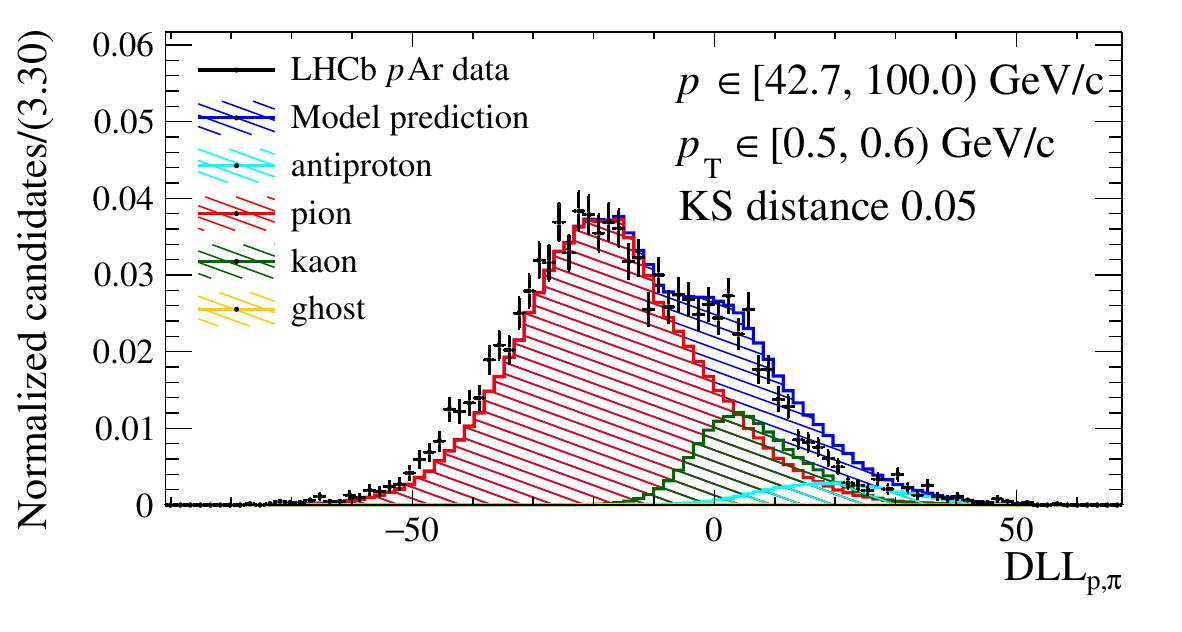} 
\includegraphics[width = 0.480000\textwidth]{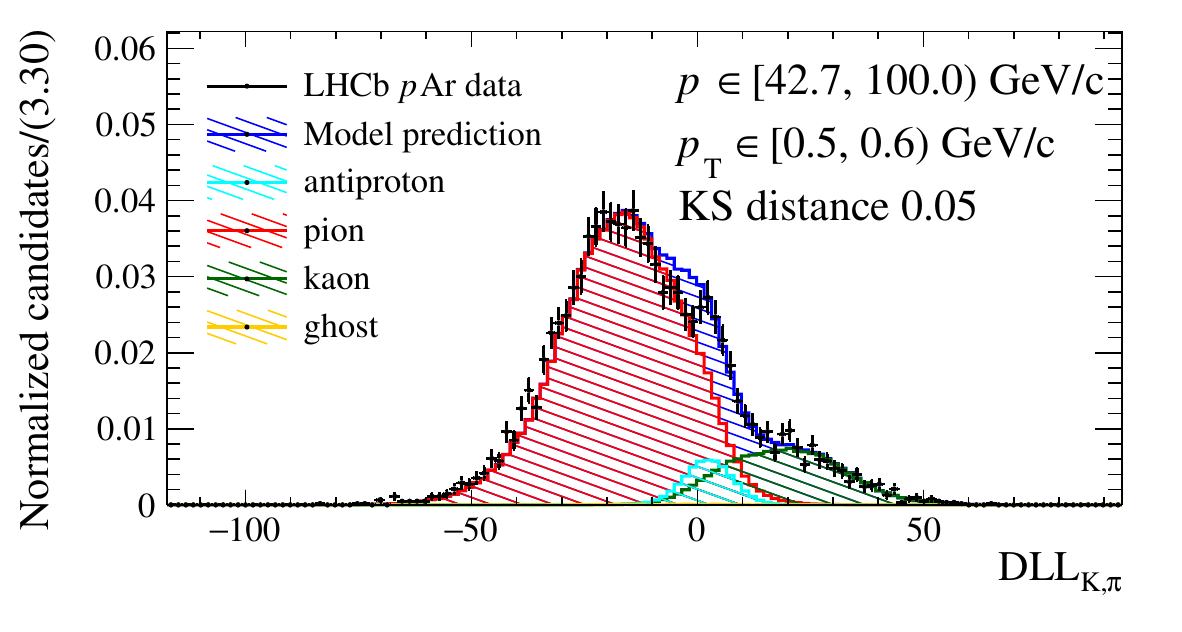} 
\includegraphics[width = 0.480000\textwidth]{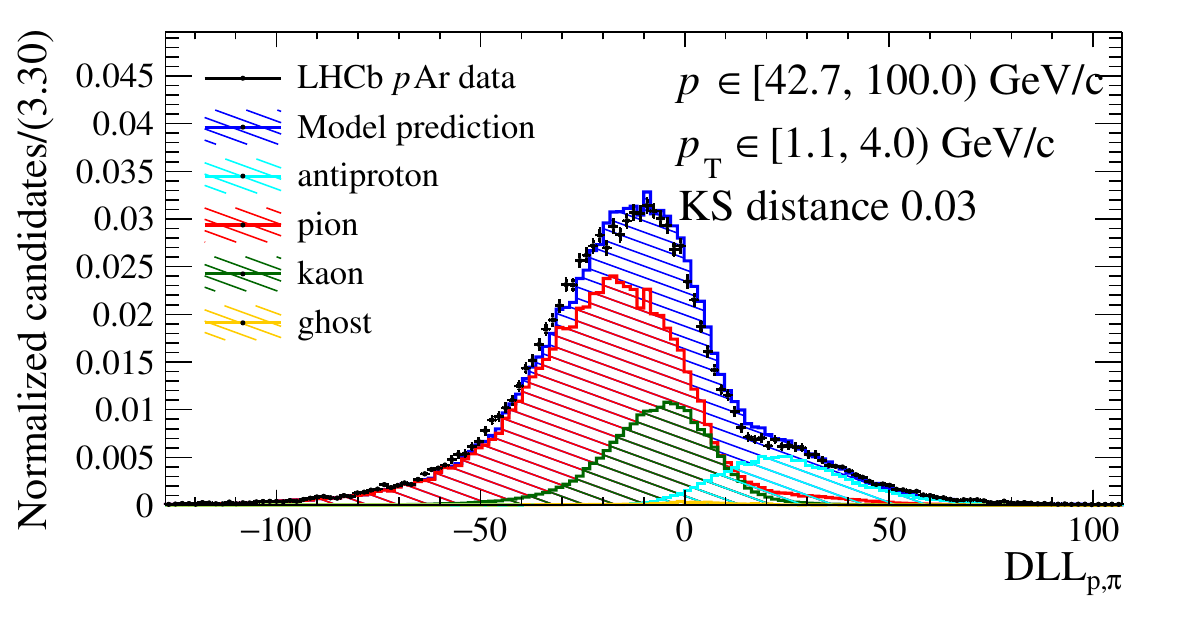} 
\includegraphics[width = 0.480000\textwidth]{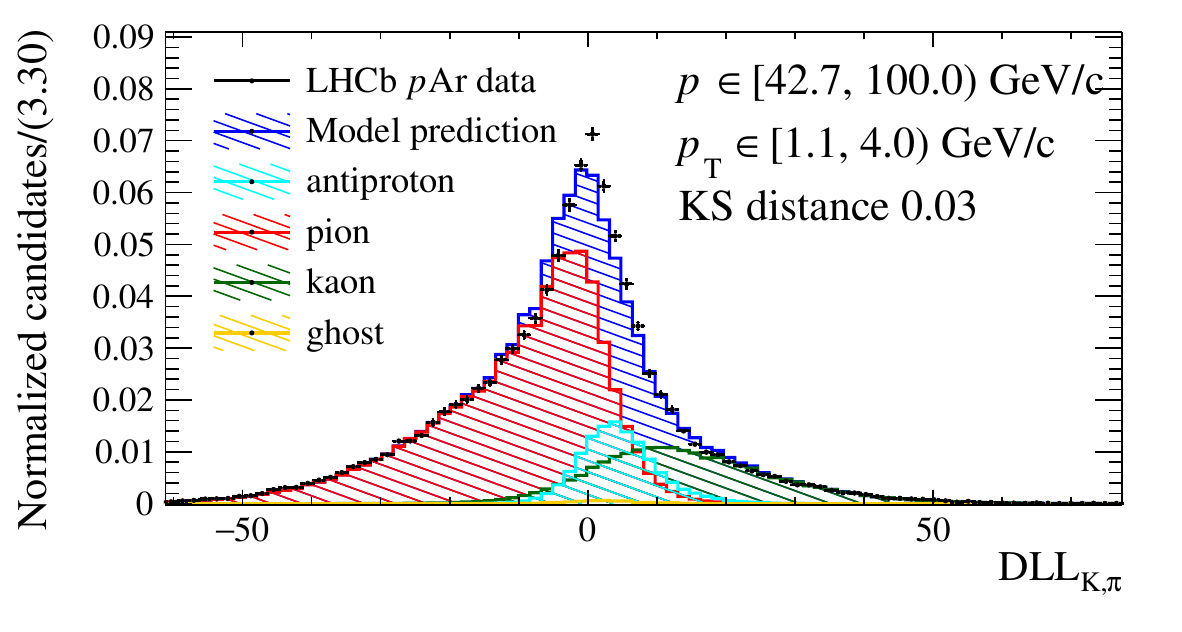} 
\caption{Projections onto the (left) \dllppi and (right) \dllpk axes of the fit to \pAr data employing the generated templates in five momentum, transverse momentum intervals.} 
\label{fig_ar:fit_gmm_ar} 
\end{figure}

\begin{figure} [h]
\centering 
\includegraphics[width = 0.480000\textwidth]{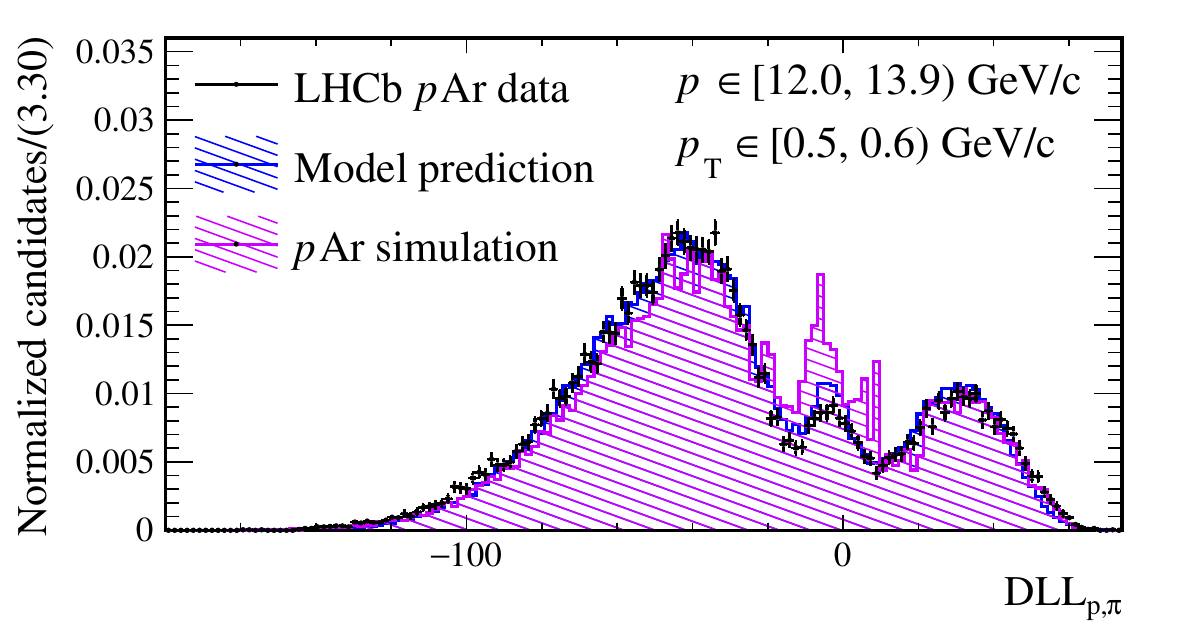} 
\includegraphics[width = 0.480000\textwidth]{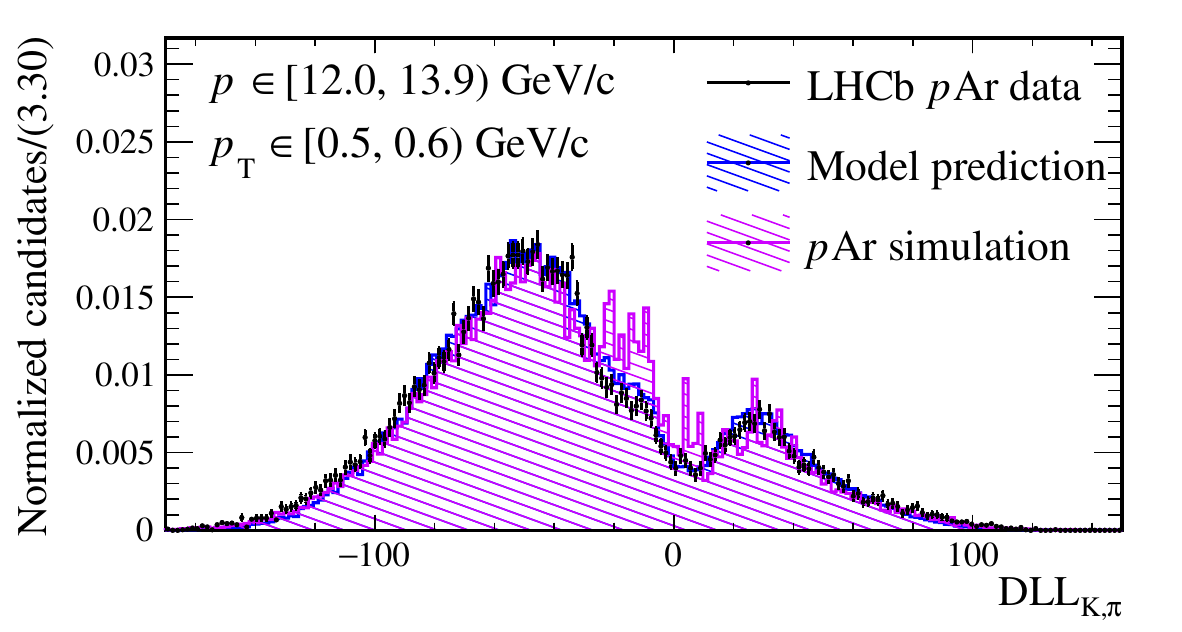} 
\includegraphics[width = 0.480000\textwidth]{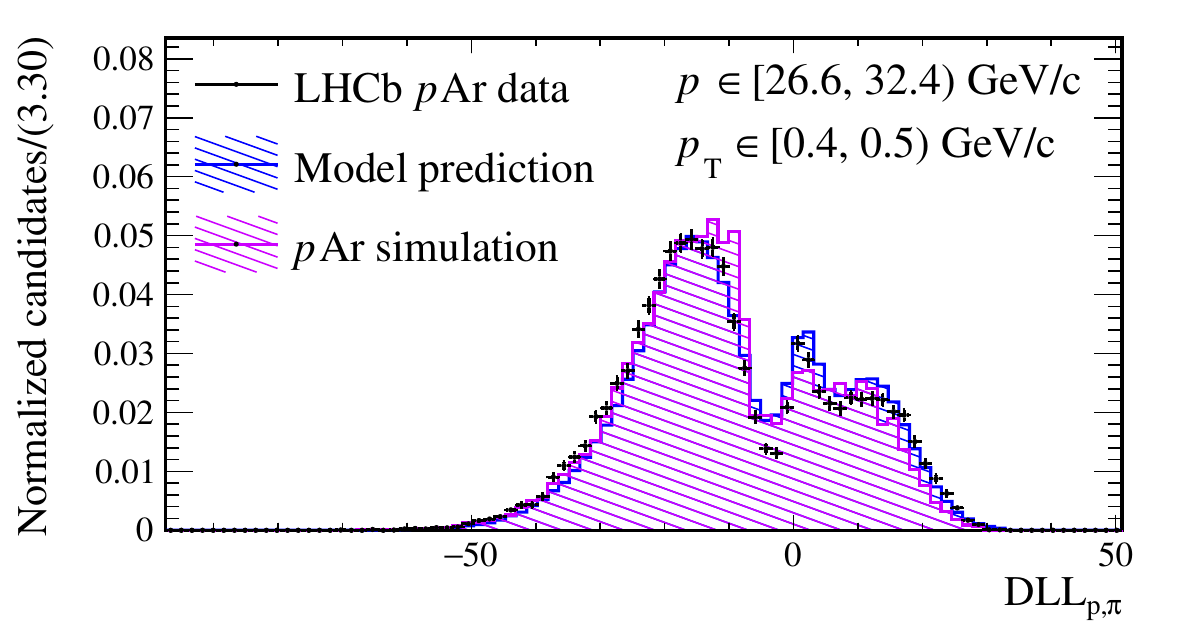} 
\includegraphics[width = 0.480000\textwidth]{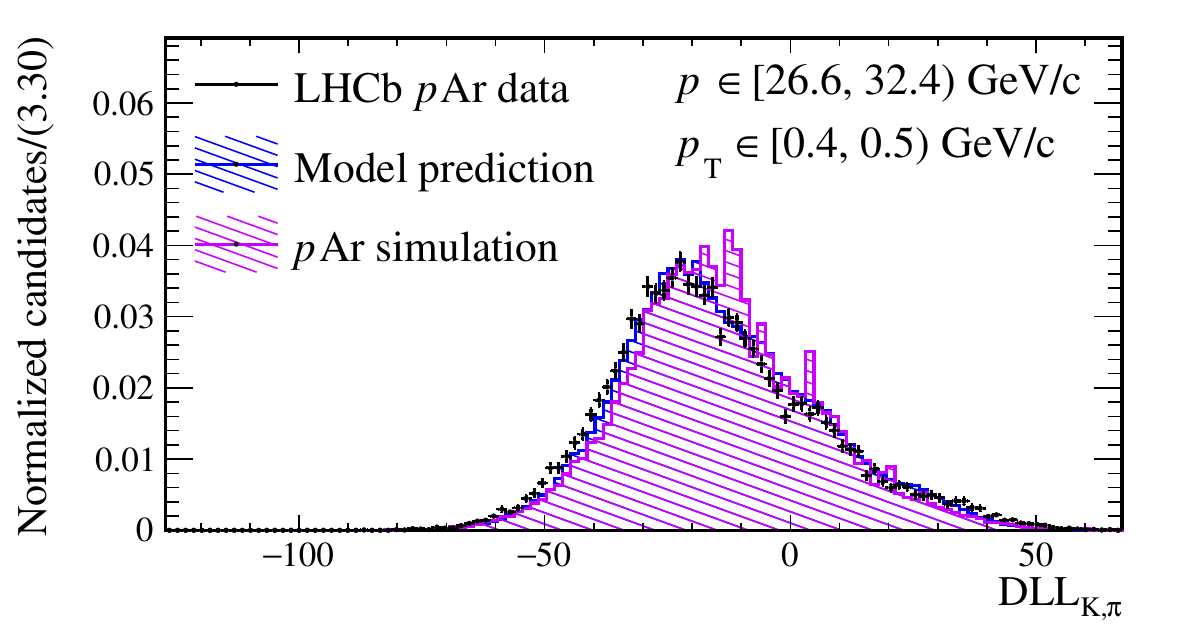} 
\includegraphics[width = 0.480000\textwidth]{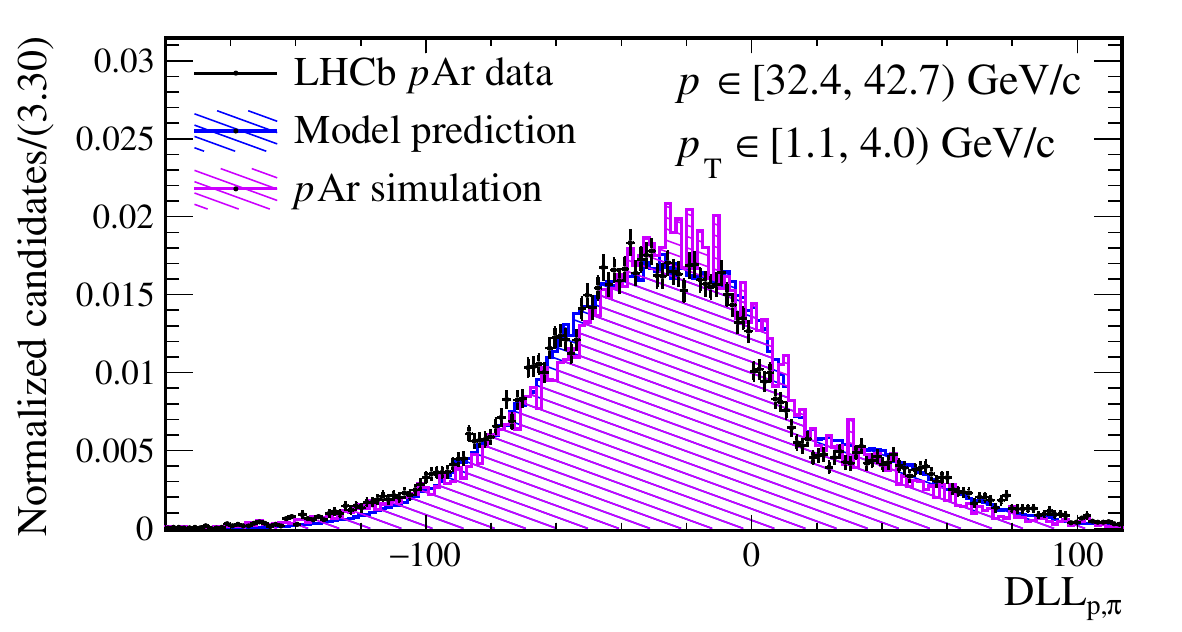} 
\includegraphics[width = 0.480000\textwidth]{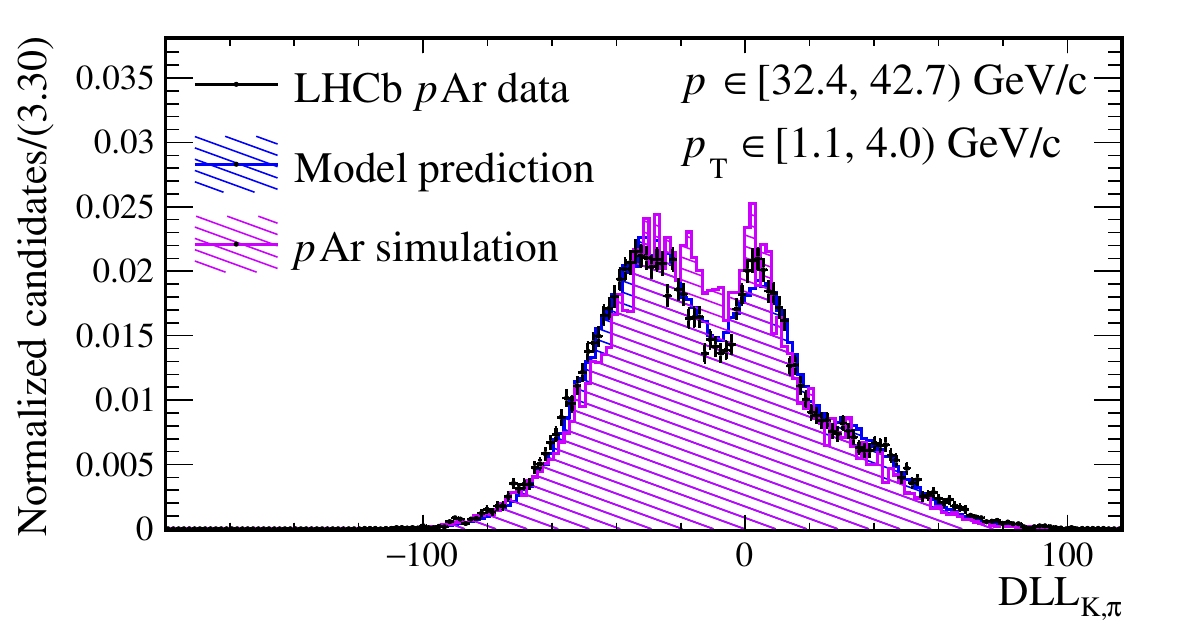} 
\includegraphics[width = 0.480000\textwidth]{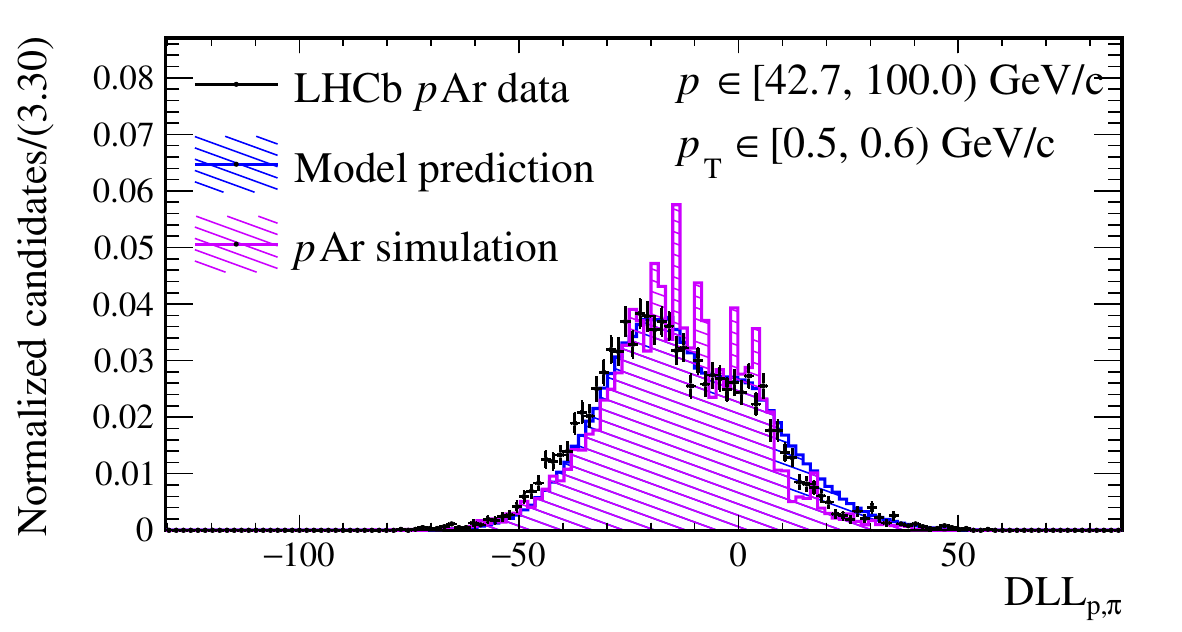} 
\includegraphics[width = 0.480000\textwidth]{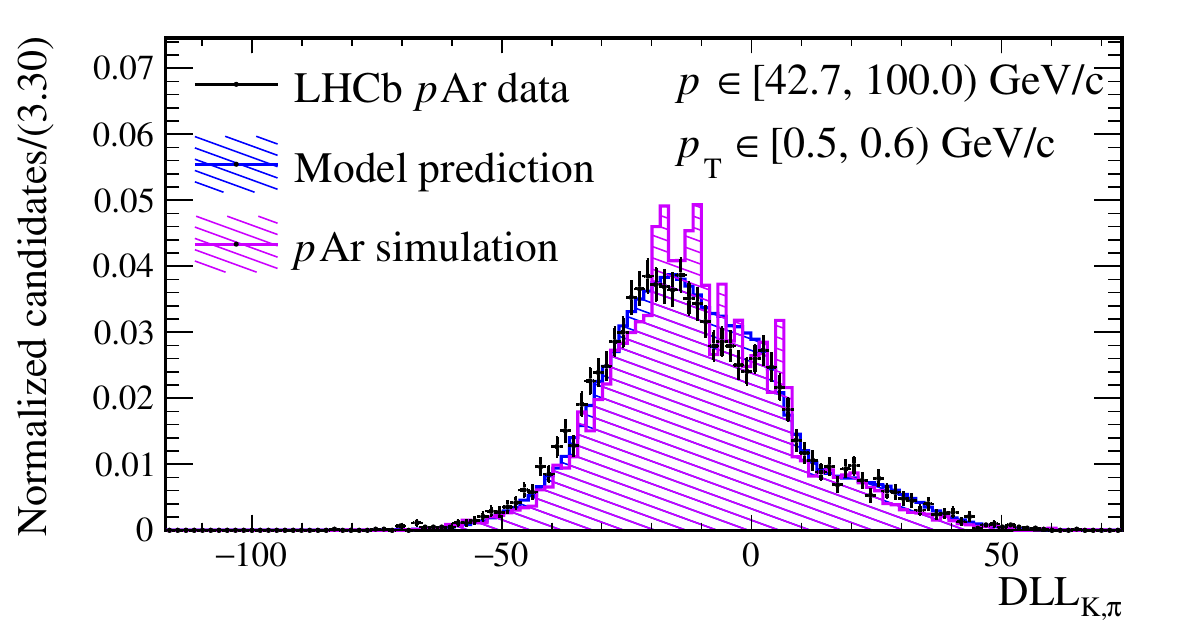} 
\includegraphics[width = 0.480000\textwidth]{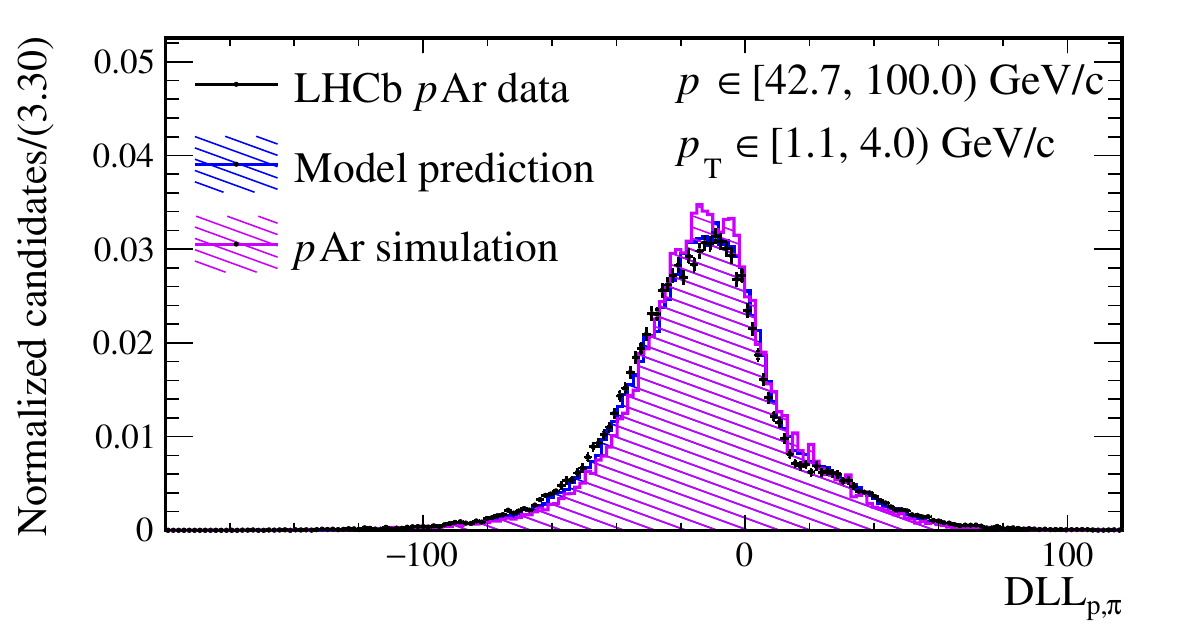} 
\includegraphics[width = 0.480000\textwidth]{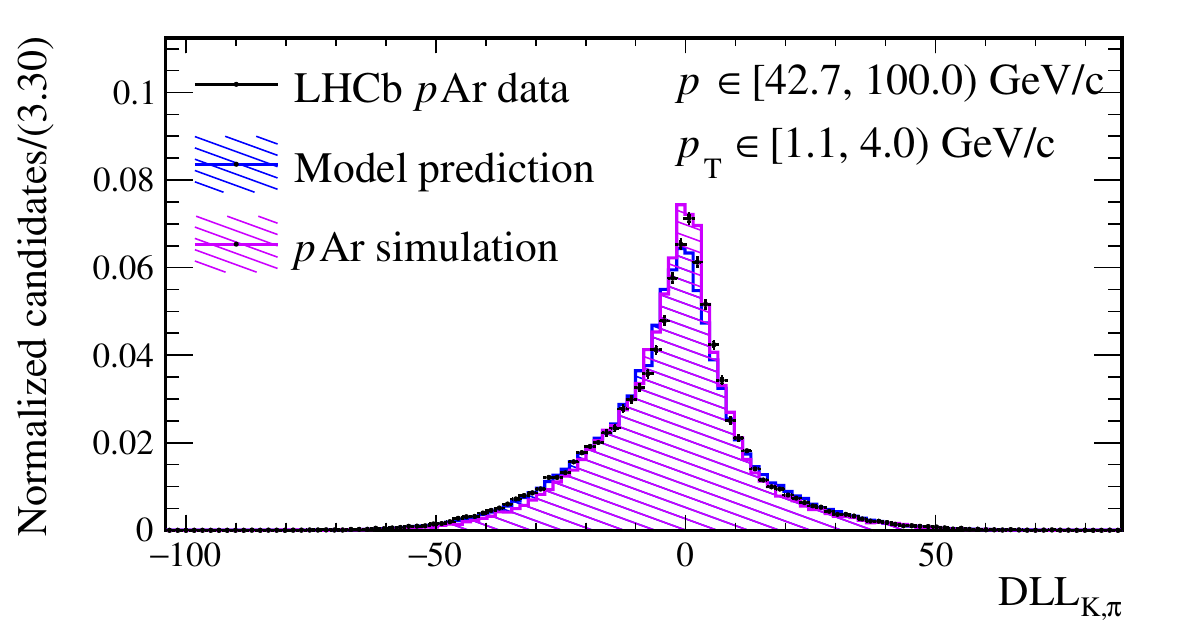} 
\caption{Comparison of the (left) \dllppi and (right) \dllpk distributions in \pAr data modelled with simulation (violet) and data-based (blue) templates.} 
\label{fig_ar:fit_comp_ar} 
\end{figure}

\begin{figure} 
\centering 
\includegraphics[width = 0.900000\textwidth]{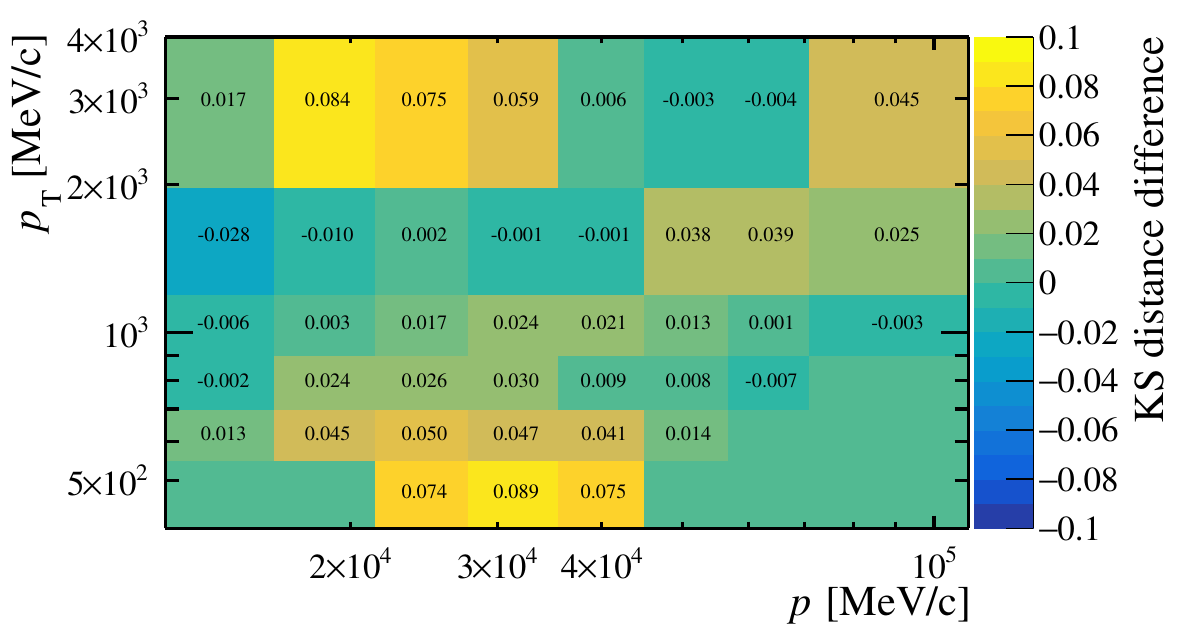}
\includegraphics[width = 0.900000\textwidth]{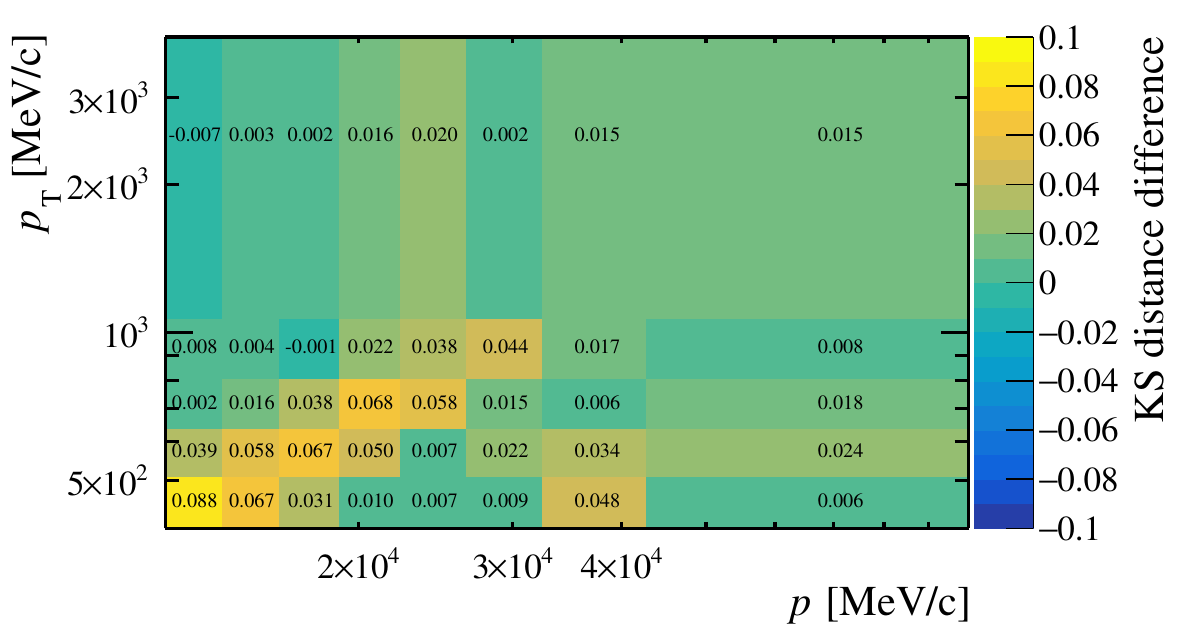} 
\caption{ Difference of the Kolmogorov-Smirnov distance for the modelling of the bidimensional (top) \pHe or (bottom) \pAr target variables distribution composing templates produced with a full simulation approach or with the machine-learning method employing the \pNe calibration data discussed in this document. In all bins presenting a positive value a more precise description of the data is achieved with the latter approach.}
\label{fig_GenpHe:KS_Difference} 
\end{figure}

The PID model, trained separately for each calibration channel selected from the \pNe data sample, is applied to two independent lower-statistics samples of \pHe and proton-Argon (\pAr) collisions collected in 2016 and in 2015, respectively, with a nucleon-nucleon centre-of-mass energy of $\sqrt{s_{NN}} = 110\gev$. The different centre-of-mass energy and target nuclei with respect to the training sample result in significant distortions of the kinematic distributions of the produced particles and of the detector occupancy, while identical detector and data-taking  conditions can be assumed. For each particle species, a template for the classifiers \pdf is produced with the same method adopted for the validation in Sec.~\ref{sec:ML_validation},
but considering the differently-distributed \pHe or \pAr features as an input. 
To evaluate the model performance and the possible improvement over the PID model
adopted so far in antiproton production studies in fixed-target data~\cite{LHCb-PAPER-2018-031}, the data-driven templates for the \dllppi and \dllpk variables are compared with those obtained from the detailed simulation of the detector. For the \pAr sample, the application is performed on the \dllppi and \dllkpi variables. Simulated data samples are generated for fixed-target collisions with EPOS-LHC~\cite{Pierog:2013ria}, the interaction of the generated particles with the detector and its response are implemented using the \geant toolkit~\cite{Allison:2006ve, *Agostinelli:2002hh} as described in Ref.~\cite{LHCb-PROC-2011-006}. 
The simulation provides an overall good description of the input features and the small observed residual disagreements with respect to data can be corrected with reweighting techniques. For the \pHe sample, the same approach of Ref.~\cite{LHCb-PAPER-2018-031} is followed and simulated events are reweighted in two steps, according to the detector occupancy (using $nSPDHits$) and the track transverse momentum. To verify that the simulation-based PID model is not limited by this simple reweighting technique, a more sophisticated approach is adopted for the \pAr sample, where a boosted decision tree algorithm~\cite{hep_ml} is used to determine a single event weight depending on \pt and the number of hits in the \spd and RICH1 subdetectors.
After the reweighting, simulation-based binned templates for the three particle species are obtained. A fourth particle category for tracks reconstructed from hits belonging to different simulated particles and accounting for a few per cent of the candidate tracks, called ghost in the following, is considered. The corresponding  template can only be obtained from simulation and is used also in the data-driven model. Template fits are finally performed on the \pHe and \pAr data, using the simulation-based or the data-driven templates. Binned maximum likelihood fits to the bidimensional target variable distribution are done in kinematic intervals, leaving the relative abundances of the particle species as free parameters of the fit. Examples of the fit projections to the \pHe data are shown in the same kinematic intervals in Fig.~\ref{fig_GenpHe:fit_simulated} for the simulation-based templates and 
in Fig.~\ref{fig_GenpHe:fit_TF} for the data-driven templates and directly compared to data in  Fig.~\ref{fig_GenpHe:fit_Comparison}. Similar plots of fit projections in five other kinematic intervals are shown for the \pAr sample in Figs.~\ref{fig_ar:fit_simulated_ar}-\ref{fig_ar:fit_comp_ar}.
In general, the data-based method is found to provide a more accurate prediction of the 
PID classifiers distribution than the full-simulation one. As expected from the larger size of the training dataset
and from the smooth function assumed in the model, it is less affected by statistical fluctuations and also appears to be less biased.
To better quantify the comparison between the two sets of templates, 
the fit quality is measured from the two-dimensional KS distance between the fitted 
and the actual data distribution, in kinematic bins. 
The difference between the values obtained with the simulation-based and data-driven 
templates is shown in Figs.~\ref{fig_GenpHe:KS_Difference} for the \pHe (top) and \pAr (bottom) data samples, respectively.
In both cases, the observation of a positive difference in most of the bins demonstrates 
that the templates produced with the method presented in this work better describe the data than those based on a detailed simulation.

\subsection{Systematic uncertainties}
\label{sec:syst}
\begin{figure}
\centering
\includegraphics[width = 0.85\textwidth]{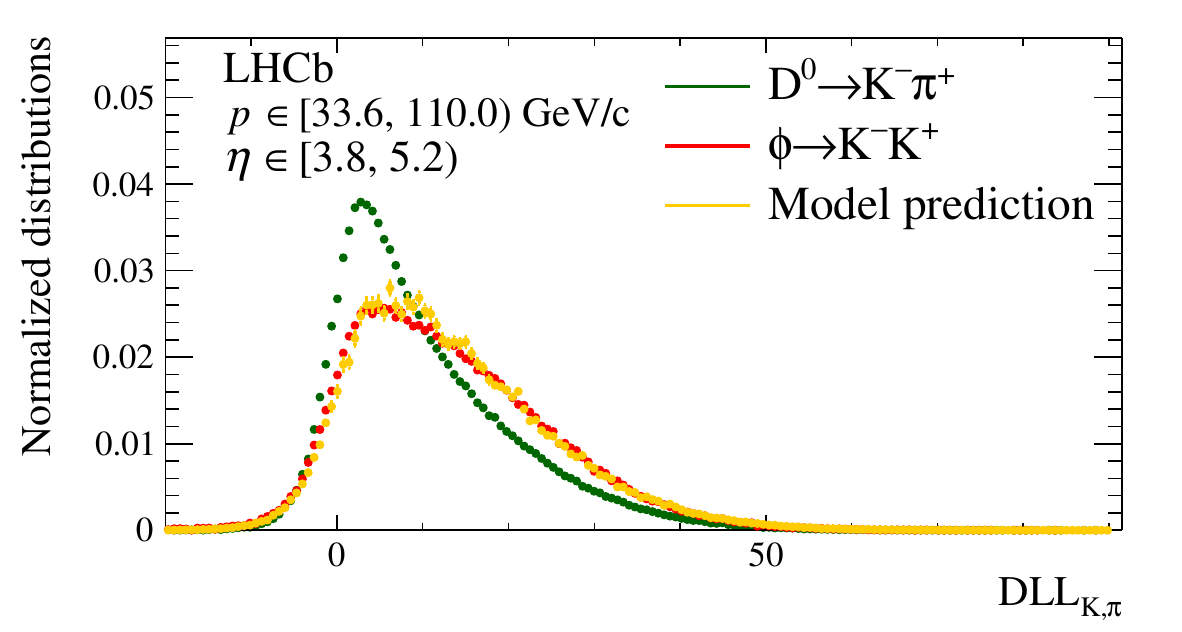}
\includegraphics[width = 0.85\textwidth]{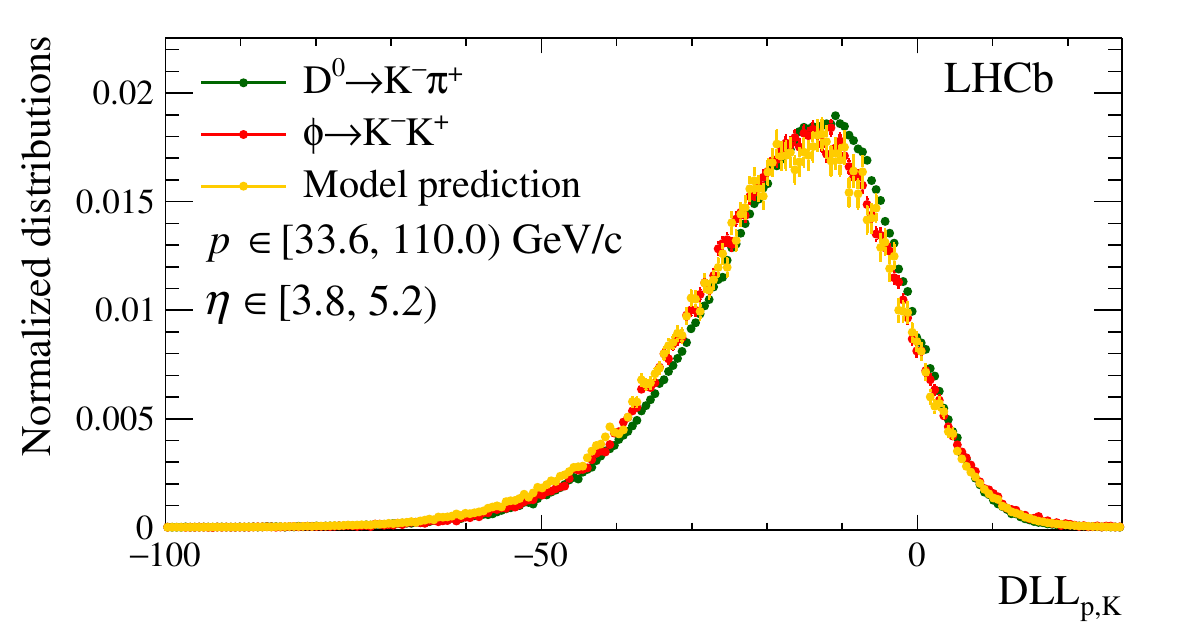}
\caption{Distributions of (top) \dllkpi  and (bottom) \dllpk variables for kaons from the \Dstarp and \phiz calibration channels in 2017 \pp data, in the kinematic bin where the two samples differ most. The prediction of the model trained on \Dstarp and applied to the
\phiz data is also shown.}
\label{fig:systPhiKK}
\end{figure}

Data-driven methods for modelling the detector response such as the one proposed in this paper, while unaffected by biases related to the unavoidable imperfections of the simulation, depend on the selection and peculiarities of the calibration samples employed. We show in this section how the proposed model can be used to investigate these systematic effects. In our benchmark example, the PID response to kaons is the one which is expected to be more difficult to model reliably using the $\decay{\phiz}{\Km\Kp}$ decay because of the sizeable background contamination and the presence of the tag kaon, which is typically produced within a small angle with respect to the probe track and can thus bias the PID response. Indeed, for \pp LHCb data, kaon PID calibration typically relies on the $\decay{\Dstarp}{\Dz(\to \Km\pip)\pip}$ channel~\cite{LHCb-DP-2018-001}, whose statistics is too limited to be useful in fixed-target data. To evaluate the possible bias due to the correlation between the tag and probe kaons in the \phiz channel, a comparison between the PID responses of kaons from the two calibration channels in 2017 \pp data is performed in momentum and pseudorapidity bins.
A clear difference, notably in the \dllkpi variable, is observed in some kinematic bins.
We then train our model for the kaon response using the $\decay{\Dstarp}{\Dz(\to \Km\pip)\pip}$ decay and use the model to predict the templates for kaons in the \phiz channel. The result, shown in Fig.~\ref{fig:systPhiKK} for the kinematic bin featuring the largest difference among the two samples, demonstrates that the difference is mostly explained by our model, taking into account the residual difference in track and event topology between the two decays within the kinematic bin. As the model trained on the \Dstarp channels ignores the presence of the companion kaon and is not affected by the background of the \phiz channel, the systematic effects that have been investigated are found to be minor.
This exercise shows that the proposed model can accurately take into account the 
different distributions of all the relevant experimental features among different data samples, well beyond the simple equalization of kinematic bins. In general, training the model on different calibration channels, also resorting to data recorded at different energy and collision systems, provides a way to evaluate the systematic effects related to the choice of a particular calibration dataset.

\section{Conclusions}
\label{sec:prospects}
In summary, we presented a novel approach based on machine-learning techniques to model particle identification classifiers in high energy physics experiments. The GMM model, fitted to calibration channels using a set of MLP NNs, can be used to identify the relevant experimental features to be considered and to provide a smooth multivariate parametric representation of the PID response, making optimal use of the available training data. State-of-the-art machine-learning software libraries and computing resources provide the possibility to train models with $O(10^5)$ parameters in a relatively short time, depending on the chosen complexity for the NN and the GMM. The method was demonstrated on a concrete case, namely the modelling of the particle identification response in the \lhcb experiment, where it was shown to be able to predict non-trivial correlations among experimental features and to describe data more accurately than a detailed first-principle simulation of the detector.
The method is expected to be employable on a larger variety of use cases dealing with experimental observables, not necessarily related to particle identification, depending on a sizeable number of experimental features.
\newpage





\addcontentsline{toc}{section}{References}
\clearpage
\bibliographystyle{LHCb}
\bibliography{main,standard,LHCb-PAPER,LHCb-CONF,LHCb-DP,LHCb-TDR}

\end{document}